\documentclass[pdflatex,sn-basic]{sn-jnl}
\usepackage{aas_macros}
\usepackage{graphicx}%
\usepackage{multirow}%
\usepackage{amsmath,amssymb,amsfonts}%
\usepackage{amsthm}%
\usepackage{mathrsfs}%
\usepackage[title]{appendix}%
\usepackage{xcolor}%
\usepackage{textcomp}%
\usepackage{manyfoot}%
\usepackage{booktabs}%
\usepackage{caption}%
\usepackage{subcaption}%
\usepackage{algorithm}%
\usepackage{algorithmicx}%
\usepackage{algpseudocode}%
\usepackage{listings}%
\usepackage{cancel}
\usepackage{verbatim}
\usepackage{xspace}
\usepackage{hyperref}
\usepackage{orcidlink}
\begin{document}
\title[The Dark Ages Explorer (DEX)]{The Dark Ages Explorer (DEX): a filled-aperture ultra-long wavelength radio interferometer on the lunar far side}
\author*[1]{\fnm{C.D.} \sur{Brinkerink}} \email{c.brinkerink@astro.ru.nl\orcidlink{0000-0002-2322-0749}}
\author[2,3]{\fnm{M.J.} \sur{Arts}}\email{arts@astron.nl\orcidlink{0000-0003-0084-2526}}
\equalcont{These authors contributed equally to this work.}
\author[3]{\fnm{M.J.} \sur{Bentum}}\email{m.j.bentum@tue.nl\orcidlink{0000-0002-1852-5214}}
\equalcont{These authors contributed equally to this work.}
\author[2]{\fnm{A.J.} \sur{Boonstra}}\email{boonstra@astron.nl\orcidlink{0000-0002-1439-5985}}
\equalcont{These authors contributed equally to this work.}
\author[4]{\fnm{B.} \sur{Cecconi}}\email{baptiste.cecconi@obspm.fr\orcidlink{0000-0001-7915-5571}}
\equalcont{These authors contributed equally to this work.}
\author[11,12]{\fnm{A.}\sur{Fialkov}}\email{afialkov@ast.cam.ac.uk\orcidlink{0000-0002-1369-633X}}
\equalcont{These authors contributed equally to this work.}
\author[5]{\fnm{J.} \sur{Garcia Guti\'errez}} \email{Borja.Garcia.Gutierrez@esa.int\orcidlink{0009-0007-8812-3943}}
\equalcont{These authors contributed equally to this work.}
\author[6]{\fnm{S.} \sur{Ghosh}}\email{soniaghosh@astro.rug.nl\orcidlink{0000-0002-6524-9830}}
\equalcont{These authors contributed equally to this work.}
\author[7]{\fnm{J.} \sur{Grenouilleau}}\email{Jessica.Grenouilleau@esa.int\orcidlink{0000-0002-3700-7228}}
\equalcont{These authors contributed equally to this work.}
\author[8,9]{\fnm{L.I.} \sur{Gurvits}}\email{leonid@gurvits.org\orcidlink{0000-0002-0694-2459}}
\equalcont{These authors contributed equally to this work.}
\author[1]{\fnm{M.} \sur{Klein-Wolt}}\email{M.KleinWolt@astro.ru.nl\orcidlink{0000-0001-7901-9545}}
\equalcont{These authors contributed equally to this work.}
\author[6]{\fnm{L.V.E.} \sur{Koopmans}}\email{koopmans@astro.rug.nl\orcidlink{0000-0003-1840-0312}}
\equalcont{These authors contributed equally to this work.}
\author[3]{\fnm{J.} \sur{Lazendic-Galloway}}\email{j.lazendic.galloway@tue.nl\orcidlink{0000-0003-0166-1647}}
\author[8]{\fnm{Z.} \sur{Paragi}}\email{paragi@jive.eu\orcidlink{0000-0002-5195-335X}}
\equalcont{These authors contributed equally to this work.}
\author[2,3]{\fnm{D.} \sur{Prinsloo}}\email{prinsloo@astron.nl\orcidlink{0000-0001-6052-9692}}
\equalcont{These authors contributed equally to this work.}
\author[9]{\fnm{R.T.} \sur{Rajan}}\email{R.T.Rajan@tudelft.nl\orcidlink{0000-0002-6443-9624}}
\equalcont{These authors contributed equally to this work.}
\author[4]{\fnm{E.} \sur{Rouillé}}\email{erwan.rouille@obspm.fr\orcidlink{0000-0001-8683-2657}}
\equalcont{These authors contributed equally to this work.}
\author[2]{\fnm{M.} \sur{Ruiter}}\email{ruiter@astron.nl}
\equalcont{These authors contributed equally to this work.}
\author[10]{\fnm{J.A.} \sur{Tauber}}\email{tauber@strw.leidenuniv.nl\orcidlink{0000-0002-9005-0495}}
\equalcont{These authors contributed equally to this work.}
\author[2]{\fnm{H.K.} \sur{Vedantham}}\email{vedantham@astron.nl\orcidlink{0000-0002-0872-181X}}
\equalcont{These authors contributed equally to this work.}
\author[1,4]{\fnm{A.} \sur{Vecchio}}\email{a.vecchio@astro.ru.nl\orcidlink{0000-0002-2002-1701}}
\equalcont{These authors contributed equally to this work.}
\author[3]{\fnm{C.J.C.} \sur{Vertegaal}}\email{c.j.c.vertegaal@tue.nl\orcidlink{0000-0001-8071-7922}}
\equalcont{These authors contributed equally to this work.}
\author[3]{\fnm{J.C.F.} \sur{Zandboer}}\email{j.c.f.zandboer@tue.nl\orcidlink{0009-0008-2784-1471}}
\equalcont{These authors contributed equally to this work.}
\author[2]{\fnm{P.} \sur{Zucca}}\email{zucca@astron.nl\orcidlink{0000-0002-6760-797X}}
\equalcont{These authors contributed equally to this work.}
%
%
%
%
%
\affil[1]{\orgname{Radboud University Nijmegen}, \orgaddress{\street{Heyendaalseweg 135}, \city{Nijmegen}, \postcode{6525 AJ}, \country{The Netherlands}}}
\affil[2]{\orgname{ASTRON, the Netherlands Institute for Radio Astronomy}, \orgaddress{\street{Oude Hoogeveensedijk 4}, \city{Dwingeloo}, \postcode{7991 PD}, \country{The Netherlands}}}
\affil[3]{\orgdiv{Department of Electrical Engineering}, \orgname{Eindhoven University of technology}, \orgaddress{\street{Groene Loper 19}, \city{Eindhoven}, \postcode{5612 AP}, \country{The Netherlands}}}
\affil[4]{\orgdiv{LIRA}, \orgname{Observatoire de Paris-PSL, Université PSL, CNRS, Sorbonne Université, Université Paris Cité}, \orgaddress{\street{5 Place Jules Janssen}, \city{Meudon}, \postcode{92190}, \country{France}}}
\affil[5]{\orgdiv{ESA/ESTEC/TEC-SYE}, \orgaddress{\street{P.O. Box 299}, \city{Noordwijk}, \postcode{2200 AG }, \country{The Netherlands}}}
\affil[6]{\orgname{Kapteyn Astronomical Institute, University of Groningen}, \orgaddress{\street{Landleven 12}, \city{Groningen}, \postcode{9747AD}, \country{The Netherlands}}}
\affil[7]{\orgdiv{ESA/ESTEC/HRE-E}, \orgaddress{\street{P.O. Box 299}, \city{Noordwijk}, \postcode{2200 AG }, \country{The Netherlands}}}
\affil[8]{\orgname{Joint Institute for VLBI ERIC}, \orgaddress{\street{Oude Hoogeveensedijk 4}, \city{Dwingeloo}, \postcode{7991 PD}, \country{The Netherlands}}}
\affil[9]{\orgname{Delft University of Technology, Faculty of Aerospace Engineering}, 
\orgaddress{Kluyverweg 1}, \city{Delft}, \postcode{2629 HS},  \country{The Netherlands}}
\affil[10]{\orgname{Leiden Observatory, Leiden University}, \orgaddress{\street{P.O. Box 9513}, \city{Leiden}, \postcode{2300 RA}, \country{The Netherlands}}}
\affil[11]{\orgname{Institute of Astronomy, University of Cambridge}, \orgaddress{\street{Madingley Road}, \city{Cambridge}, \postcode{CB3 0HA}}, \country{UK}}
\affil[12]{\orgname{Kavli Institute for Cosmology}, \orgaddress{\street{Madingley Road}, \city{Cambridge}, \postcode{CB3 0HA}}, \country{UK}}
%
%
%
%
\abstract{The measurement of the spatial fluctuations of the neutral hydrogen 21\,cm signal arising during the Dark Ages and Cosmic Dawn periods of our Universe ($z\sim$\,200-10) holds the potential to resolve these still-unexplored earliest phases of the evolution of matter structures. As these cosmological signals are very weak, large distributed telescopes are required at locations free from terrestrial radio interference and ionospheric disturbances. This paper presents a description of the scientific aims, the instrumental concept, and technological developments of an experiment - dubbed the Dark-ages EXplorer (DEX) - which would allow us to (a) measure the Global Signal and (b) measure the angular density fluctuations and conduct line-of-sight tomography in the Dark Ages and Cosmic Dawn epochs. Additional scientific goals are also briefly described. The experiment consists of a low-frequency radio interferometer, which should ideally be located on the far side of the Moon. The paper presents findings from an ESA Concurrent Design Facility (CDF) study, which was conducted to assess the feasibility of such a system using present-day technologies with a high TRL (Technology Readiness Level). Although the study finds that the number of antennas needed to achieve the primary scientific goals is not yet feasible at the moment, it points to a path of technological development that can lead to a realistic and valuable experiment in the medium-term future (i.e., the next decade(s)), as well as development of multi-purpose use technology that can be applied on Earth, and towards other lunar operations. }
\newcommand\lig[1]{{\color{red!75!black}[LIG: #1]}}
\newcommand\cdb[1]{{\color{red!30!blue}[CDB: #1]}}
\newcommand\bc[1]{{\color{green!70!black}[BC: #1]}}
\newcommand\er[1]{{\color{green!70!black}[ER: #1]}}
\newcommand\jt[1]{{\color{brown!70!black}[JT: #1]}}
\newcommand\zp[1]{{\color{purple!70!black}[ZP: #1]}}
\newcommand\ajb[1]{{\color{cyan!50!black}[AJB: #1]}}
\newcommand\sg[1]{{\color{orange!80!black}[SG: #1]}}
\setlength{\parindent}{20pt}
\maketitle
%
\section{Introduction}\label{sec:Intro}

The high level of interest currently shown by space agencies worldwide in returning to the Moon has prompted the European Space Agency (ESA) to evaluate scientific experiments which could potentially be included in the first wave of lunar missions. Among the topics identified, radio astronomy is prominently featured. Indeed, as the radio environment for Earth-based radio astronomy gradually deteriorates with an increasingly crowded radio spectrum, and as the ionosphere hampers observations below 10 to 30\,MHz, observing low-frequency\footnote{In contrast to the ITU definition, we consider low-frequency radio waves to have a frequency lower than 300\,MHz (a wavelength longer than 1\,m).} radio waves from space is becoming necessary for enabling further scientific discoveries. One of the major science cases that could in particular benefit from observing from space, avoiding the Earth's ionosphere and far from terrestrial interference, is low-frequency neutral hydrogen cosmology.\\

\noindent
 For this reason, ESA convened a Topical Team (TT\footnote{\url{https://scispace.esa.int/topical-teams/}}) on an ``Astronomical Lunar Observatory'' (ALO\footnote{\url{https://alott.astro.ru.nl}}) to develop the maturity of a low-frequency radio telescope concept to be deployed on the lunar far side and to provide advice on this topic. The main concept studied under the ALO umbrella is the Dark Ages Explorer (DEX), a concept that focuses on observing neutral hydrogen from the Dark Ages and Cosmic Dawn in the general frequency range $7 - 50$\,MHz (the specific frequency range depending on implementation). Ancillary science can be supported down to 1\,MHz, albeit with reduced sensitivity as the antenna array is optimized for redshift $z<78$. Based on ALO TT's recommendations, a lunar-based cosmological radio array concept was chosen for an ESA Concurrent Design Facility (CDF) study entitled ``Assessment of an Astrophysical Lunar Observatory on the far side of the Moon'' (\cite{ESA2021}, \cite{KleinWolt2021}), detailing the concept, including mission analysis, implementation, programmatics, risk and cost. The purpose of this paper is to present and discuss the main experimental concept resulting from the activities of the ALO TT, to describe the technical results of the ESA CDF study on DEX, and to discuss the technical requirements needed to bring this experiment into realisation. Throughout this paper, several different array sizes, antenna lengths and antenna spacings will be considered for DEX as these determine which part of the neutral hydrogen spectrum (redshift range) can be detected.\\

\noindent
This paper is organised as follows. \hyperref[sec:History]{Section 2} reviews briefly the history of very low frequency (or ultra-long wavelength, ULW) astronomy instruments, leading up to the possibility of lunar observatories. \hyperref[sec:science]{Section 3} describes the main science cases, which are measuring the global spectrum of redshifted neutral hydrogen from the Dark Ages and Cosmic Dawn, and its spatial power spectrum. The section also presents the resulting requirements for the DEX concept, starting from the type of data we wish to collect and discussing what this implies for the design. In this section, we also include brief descriptions of several ancillary science cases, such as solar and planetary radio emissions, radio transients, and radio measurements on the interstellar medium.\\

\noindent
\hyperref[sec:CDF_Results]{Section 4} describes the outcomes of an ESA CDF study that was performed in collaboration with the ALO TT, which points to what would be technologically feasible with the current state of the art and what is needed to bridge the technological gaps, especially at the lowest frequencies. In \hyperref[sec:technology]{Section 5} we look to the future and lay out the technological developments needed to bring DEX into full realisation. This is followed by a brief discussion on the far-side electromagnetic environment in \hyperref[sec:fs-protect]{Section 6}, and conclusions in \hyperref[sec:conclusion]{Section 7}.
%
%
\section{Low-Frequency Radio Astronomy Facilities}\label{sec:History}

\subsection{Ground-based observatories}

The birth of radio astronomy in 1933 was heralded by Karl Jansky's discovery of emissions originating beyond the Solar System at a wavelength of approximately 14.6\,m (frequency of about 20.5\,MHz) \citep{Jansky-1933N,Jansky-1933P}. Nearly a century later, this spectrum of electromagnetic emissions is recognized in radio astronomy as the low-frequency (or long-wavelength) domain. Initially confined to low-frequency observations, radio astronomy swiftly expanded into higher frequency (shorter wavelength) realms. This transition was driven by both astronomical and technical imperatives. The astronomical motivations were somewhat tied to advancements in other branches of astronomical study, while the technical reasons stemmed from two main factors: the escalating levels of human-produced radio frequency interference (RFI) and the limiting opacity of the Earth's ionosphere at wavelengths exceeding approximately 20\,m. Indeed, Jansky's pioneering observations \citep{Jansky-1933N} were serendipitously made near what can be considered the practical lower frequency limit for terrestrial radio astronomy, or the ionospheric frequency cutoff. The spectral region below this loosely defined cutoff is now referred to as the ultra-long-wavelength (ULW) domain. This domain remains the last largely uncharted band of the cosmic electromagnetic emission spectrum. Concurrently, there is growing evidence suggesting that this wavelength range is extraordinarily rich with potential for astrophysical and cosmological research.\\

\noindent
Very few ground-based radio astronomy facilities have ever conducted observations in the ULW regime below 40\,MHz. Notable facilities are LOFAR \citep{Groeneveld2022, Groeneveld2024}, UTR-2 and other low-frequency radio telescopes in Ukraine \citep[][and references therein]{Konovalenko+2016}, and NenuFAR \citep{Zarka2020}. Although these observations show that it is possible to observe bright astrophysical sources through the interfering layer of the ionosphere up to a certain extent, optimal sensitivity for a variety of sources requires avoiding the ionosphere altogether.

\subsection{Space-based instruments and concepts}

The value of establishing a lunar surface-based or lunar-orbit based ULW observatory for a wide array of astrophysical research was first thoroughly championed by Stanislaw Gorgolewski during his presentation at the First Lunar International Laboratory Symposium on Research in Geosciences and Astronomy, held in Athens, Greece, in 1965 \citep{Gorgolewski-1966}. Although the Moon continued to be a promising site for such future ventures, numerous pioneering observations at ultra-long wavelengths were undertaken from a variety of free-flying platforms and orbiters during the 1960s and 1970s. NASA's Radio Astronomy Explorer satellite, \textit{RAE-1} \citep{Alexander+1969ApJL}, provided crucial data on the cosmic radio emission spectrum ranging from 0.4 to 6.5\,MHz. This work was further extended by experiments conducted on NASA's \textit{IMP-6} (Interplanetary Monitoring Program) satellite, which recorded observations between 130 and 2600\,kHz \citep{Brown1973}. \textit{RAE-2}, the second satellite in the \textit{RAE} series, effectively demonstrated the Moon's ability to act as a shield against Earth-originated RFI at ULW \citep{Alexander+1975}. Moreover, several Soviet planetary probes—\textit{Zond-2, Zond-3, Venera-2, Luna-11} and \textit{Luna-12}—carried out omnidirectional measurements of the intensity of radio emissions at frequencies ranging from 20\,kHz to 2.2\,MHz \citep{Grigorieva+Slysh-1970}. Notably, the lunar orbiters \textit{Luna-11} and \textit{Luna-12} demonstrated the effect of lunar shadowing on radio emissions produced by the interaction between the solar wind and the interplanetary magnetic field at frequencies of 30, 200, and 965 kHz \citep{Grigorieva+Slysh-1970}. These initial spaceborne ULW experiments offered an early view into the radio Universe at frequencies below 10 MHz \citep{Novaco+Brown-1978}.\\

\noindent
Over the next four decades numerous proposals and preliminary design efforts concentrated on diverse concepts for space-borne ultra-low frequency (ULF) astronomical facilities (refer to the citations in Section 1 of \cite{Yan+2023ExpA}). Among these varied initiatives and studies, several have explored the potential for radio astronomy facilities, including ULW versions, on the far side of the Moon (e.g., \cite{Bely-ESA-sci1997, Mimoun+2012}).\\

\noindent
In 2018, the Chinese lunar exploration mission \textit{Chang'E-4} heralded a revival of experimental studies in space-borne ULW astronomy. This mission featured three ULW instruments: 1) the Netherlands-China Low-Frequency Explorer (NCLE) aboard the \textit{Queqiao} relay satellite, strategically stationed at the Earth-Moon system's Lagrangian point L2 \citep{NCLE-2024}; 2) the Very Low Radio Frequency Spectrometer deployed on the Chang'e 4 lunar lander, situated on the Moon’s far side \citep{CE4science-2018}; and 3) the Low-Frequency Interferometer and Spectrometer (LFIS) aboard the satellites \textit{Longjiang-1} and \textit{Longjiang-2}, which orbited the Moon in a selenocentric trajectory as supplementary components of the main spacecraft. Among these, the LFIS on \textit{Longjiang-2} performed valuable measurements of Earth-based RFI as detected from lunar orbit, spanning frequencies from 1.5 to 30\,MHz \citep{Yan+2023ExpA}. These observations provided a novel quantitative analysis of the Moon’s role as a natural barrier against RFI originating from Earth.\\

\noindent
In 2025, the ROLSES-1 (Radiowave Observations on the Lunar Surface of the photo-Electron Sheath instrument) conducted several short ULW observing sessions during the Earth-Moon cruise phase (80~minutes total) and two days on the Moon surface (two runs, 10 and 25 minutes) as a part of the Intuitive Machines’ \textit{Odysseus} lunar lander mission \citep{Hibbard+2025}. The observations were conducted with four monopole stacer antennas that were in a non-ideal deployment at the frequencies between 5\,kHz and 30\,MHz. These short observations confirmed the expectations regarding the ``technosignature'' RFI emission originating from Earth and appeared to be qualitatively consistent with the measurements of the LFIS experiment mentioned above \citep{Yan+2023ExpA}.\\

\noindent
Recently, a concept of a square kilometer radio telescope on the Lunar surface has been presented in China \citep{Chinese-ULWSKA}. The aim of this concept study is to analyse engineering approaches toward construction of such the facility for ultra-sensitive ULW astronomy studies in the ionosphere-free and RFI-free environment. \\

\noindent
Furthermore, ULW radio instruments have been a staple ingredient of Solar and planetary missions as well (WIND, STEREO, Solar Orbiter, Cassini, Parker Solar Probe, JUICE), mostly focusing on the measurement of solar radio bursts, planetary magnetosphere emissions and local plasma effects - in short, emissions generated from within the Solar System. Data from these instruments has been analysed to constrain the Galactic radio emission at low frequencies as well, showing that there is a transition from a bright Galactic plane to bright Galactic poles as we consider lower frequencies, happening around 3\,MHz \citep{Bassett2023} as measured using the FIELDS instrument on Parker Solar Probe.\\

\noindent
Looking ahead, several mission concepts and instruments are under consideration. Project OLFAR proposes a network of small orbiting radio antennas dedicated to ULW astronomical observations \citep{bentum2009olfar,rajan2011AE}. The SunRISE mission aims to elucidate the formation and evolution of solar particle storms through a novel approach using a cluster of small spacecraft as a unified radio telescope. Additionally, the NOIRE (Nanosatellites pour un Observatoire Interférométrique Radio dans l'Espace or Nanosatellites for a Radio Interferometer Observatory in Space) project is exploring the feasibility of a swarm of nanosatellites to form a low-frequency radio observatory in space  \citep{Cecconi:2018ee}. These missions aim to overcome the limitations of poor angular resolution in current space-based radio astronomy instruments by using a distributed network of antennas that operate as an interferometer.\\

\noindent
However, the Moon is of significant interest both as a natural barrier against Earth-originated RFI and as a potential site for ULW astronomy projects, as highlighted in various ongoing studies and strategic roadmaps (e.g., \cite{Burns+2017, Belov+2018, Bentum+2020, Koopmans+2021}). For example, the previously mentioned LFIS experiment was initiated as both a precursor and a technology demonstrator for the ambitious ULW interferometer to be positioned in a selenocentric orbit under the Discovering the Sky at Longest Wavelengths (DSL) project \citep{Boonstra+2016, Chen+2019, Chen2021}.

\section{Science cases and requirements for a Lunar Radio Observatory}
\label{sec:science}

\subsection{Lunar-Based 21-cm Cosmology}

The infant Universe, covering the first billion years, is a crucial but poorly understood era despite significant advancements in observational astronomy. Instruments operating across optical, ultraviolet, infrared and submillimeter spectra, both ground-based and in space, have facilitated remarkable progress. However, the period that stretches from the snapshot provided by the Cosmic Microwave Background (CMB) \citep{Planck2020-I} to when the first stars and galaxies become observable, approximately half a billion to one billion years later, remains observationally almost unreachable. In particular, the redshift interval from $z\sim 14$ to $z \sim 1100$ still lacks direct observational probes of the sources responsible for heating and ionizing the IGM \citep{Carniani2024}. Even with the current capabilities of the James Webb Space Telescope (JWST), direct observations will only marginally extend to $z\sim 15$, and these will be confined to a limited number of very luminous sources across a small field of view.\\

\noindent
As baryons cool inside dark matter halos, they collapse to form the first stars, black holes, and galaxies. Their light spreads throughout the Universe, eventually heating virtually every baryon and coupling the hydrogen spin and gas temperatures. This epoch is referred to as Cosmic Dawn (see \citet{Fialkov2024} for an overview of the processes active in this epoch). The very first galaxies were likely hosted by so-called mini-halos with masses of $10^{6-8}$ M$_\odot$ and were probably populated by massive and bright stars, so-called Population III stars \citep{Klessen2023}. Subsequent generations of more massive galaxies probably have very different ISM properties and different star formation efficiencies compared to later galaxies. They are expected to produce soft UV radiation, which couples the spin temperature to the gas temperature, which would allow us to see the cold intergalactic medium (IGM) against the CMB. By mapping the Cosmic Dawn phase, we will be able to understand the nature of these first stars, galaxies, (super-massive) black holes, the stellar and ISM properties, and possibly even dark matter.\\

\noindent
The redshifted 21-cm line is uniquely suited to study these phases of the evolution of the universe for a number of reasons. Firstly, it probes the bulk of the hydrogen in the IGM and provides us with a global picture of the physical state of the Universe. Secondly, since it comes from a specific wavelength, it allows for a detailed tomographic reconstruction of the history of the universe as a function of redshift. Thirdly, it is one of the very few observables (if not the only one) that is accessible to us in this redshift range, either from the ground or, for the longer wavelengths, from space. The redshifted 21-cm line from the Dark Ages also allows us to detect the IGM in absorption against the CMB \citep{HoganRees1979,ScottRees1990,Madau1997}. Since in this stage of the Universe the matter fluctuations are still mostly well within the linear regime of gravitational instability, they faithfully reflect the initial conditions of the Universe in a very detailed manner. Hence, the redshifted 21-cm line can be used to, for example, constrain models of inflation or its alternatives through the study of non-Gaussianity with high-order statistics. Such data promise to be even more constraining than the CMB, since they will cover a wide range of redshifts and spatial scales.\\

\begin{figure*}[htb]
\centering
   \includegraphics[width=1.1\linewidth]{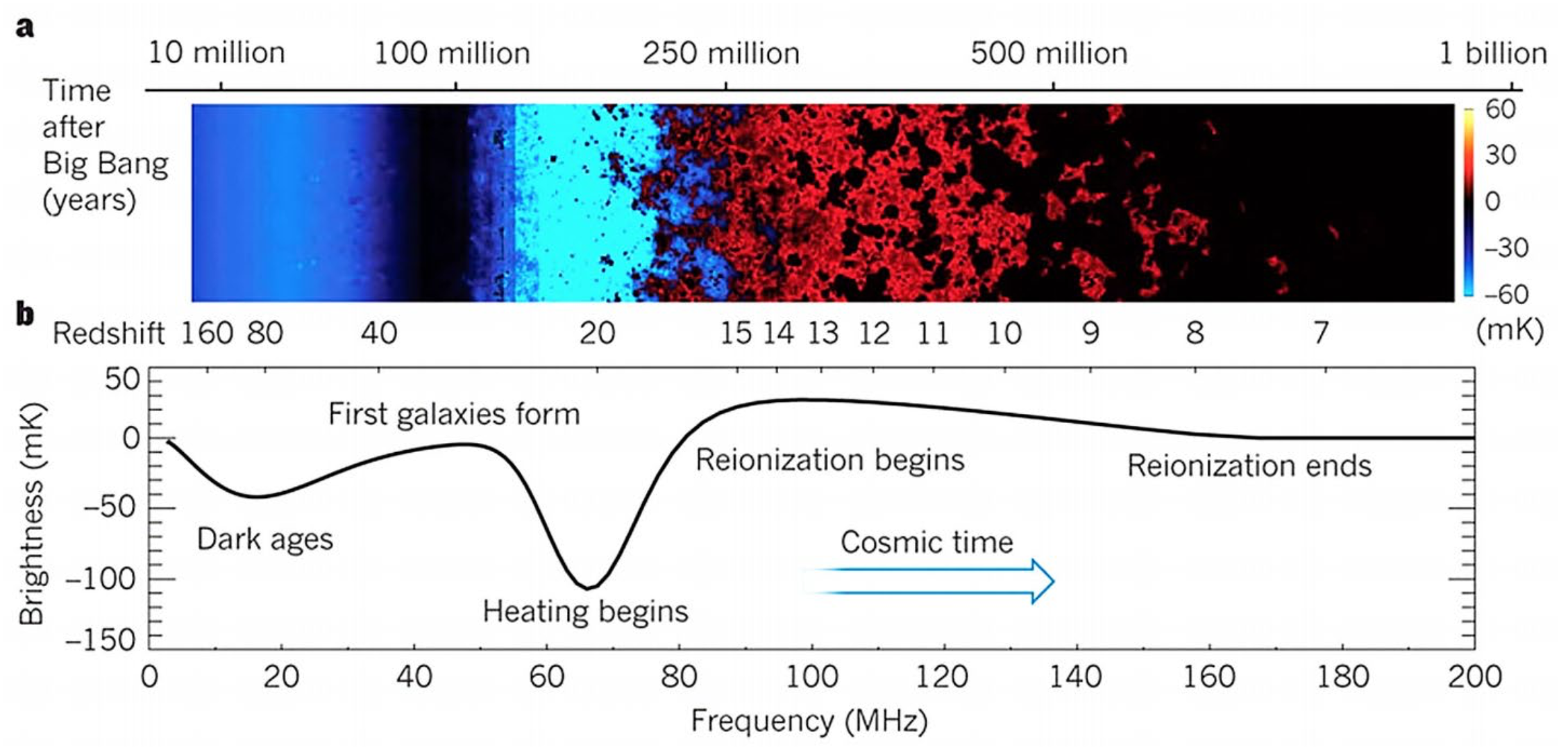}
   \caption{The fiducial model for the Global Signal, depicted together with the corresponding phases of the early Universe. Figure from \citet{Pritchard2012}.}
   \label{Fig:globalsignal}
\end{figure*}

\noindent
The 21-cm line from this era can also be used to study many other topics, such as extreme density peaks as progenitors of supermassive black holes, as well as annihilation and decay of dark matter particle candidates. The first objects in the Universe are expected to form during the Cosmic Dawn phase, which allows us to study the interplay between the Lyman-alpha coupling to the cooler baryons and of the spin temperature \citep{Wouthuysen1952, Field1958, Field1959} and X-ray heating of the IGM \citep{Madau1997,Pritchard2010}. During this era, the IGM remains mostly neutral. Still, the first stars and X-ray sources leave a marked imprint on the fluctuations of the spin temperature in the IGM, most significantly on the contrast between the absorption and emission regions of the 21-cm signal. Figure \ref{Fig:globalsignal} shows an overview of these characteristics.\\

\noindent
The structure and evolution of this spectrum of negative and positive fluctuation sheds light on the processes that formed the first objects, when they formed, and how. This is also the period in which the supermassive black holes formed, that we detect almost routinely now, for example, between redshifts 6 and 7 \citep{Fan2003,Fan2006,Mortlock2011}. Therefore, exploring this era might answer the puzzle of supermassive black hole formation and the rapid build-up of their mass in such a short period in the early history of the Universe. This period might also reveal new physics, as the result of the EDGES telescope tentatively suggests \citep{Bowman2018}. Of course, also here, there is excellent potential to address cosmological questions.\\

\noindent
As described by \citet{Koopmans+2021}, complementary aspects of the 21-cm signal can be studied using two different approaches. The sky-averaged or Global Signal (monopole in CMB parlance) can be measured by a spectrometer connected to a single dipole antenna. Alternatively, angular fluctuations in the 21-cm brightness can be studied at different frequencies via interferometric techniques. The global and spatially fluctuating signals are analogous to the diffuse 2.7\,K blackbody and anisotropies seen in the CMB, respectively. Unlike the CMB, the 21-cm signal, being a spectral line, allows for line-of-sight tomography of the IGM. The measured 21-cm differential brightness temperature primarily depends on the radiation temperature, gas temperature, and gas ionization fraction (assumed to be zero during the early Universe). In the absence of stellar sources of radiation in the Dark Ages, all three quantities are strongly constrained by standard baryonic physics applied to an adiabatically expanding Universe permeated by CMB photons. Any departure from theoretical expectations of the global radio spectrum in the Dark Ages requires an exotic mechanism to alter the physical condition of baryonic matter substantially and/or the cosmic radiation background on cosmic scales (see above). An accurate global radio spectrum measured by a single dipole can, therefore, reveal the presence of a new mechanism to deposit energy into gas on cosmological scales. The predictions are similarly strong for the spatial fluctuations of the 21-cm signal from the Dark Ages, which, in the absence of luminous sources, are largely governed by linear structure formation without any baryonic feedback. The resulting baryonic density fluctuations and peculiar velocity have been computed for the standard CDM paradigm and yield relatively stringent predictions for the Power Spectrum of the 21-cm signal that interferometry can measure. Any measured departures will therefore significantly evolve our understanding of the nature of dark matter and its influence on structure formation.\\

\noindent
Several ground-based low-frequency radio telescopes, some of which are already decommisioned, or are upgraded, have already established upper limits on the 21-cm global signal from the Cosmic Dawn phase and the subsequent Epoch of Reionization (EDGES \citet{Monsalve2017}, SARAS2 \citet{Bevins2022}, SARAS3 \citet{Bevins2024}) or on the Power Spectrum from these epochs, such as LOFAR \citep{LOFAR-2013A&A}, MWA \citep{MWA-2011AAS}, OVRO-LWA \citep{2019AJ....158...84E}, LEDA \citep{OVRO-LWA-2021}, GMRT \citep{2013MNRAS.433..639P}, PAPER \citep{2019ApJ...883..133K},  NenuFAR \citep{NenuFAR-2021}, and HERA \citep{HERA-2023MNRAS}. Future projects like SKA \citep{SKA-2015sci} will have significantly better thermal noise sensitivity than the current generation of instruments and, therefore, can push the detection of spatial fluctuations of the 21-cm line to about $z\sim 25$. However, moving beyond this frontier to higher redshifts is very difficult from the ground due to persistent issues with RFI and ionospheric disturbances, although perspectives of studying the 21-cm line redshifted to $z \sim 85$ by the Ukrainian telescopes UTR-2 and GURT are described by \cite{Konovalenko+2016,UTR-2-GURT2024}.\\

\noindent
Thus, the limitations mentioned so far have spurred interest in space-based solutions. Several ambitious missions are either in operation, in the development phase, or have been proposed as a mission concept. This includes the Dutch-Chinese NCLE instrument aboard the Chang'E-4 relay satellite at lunar L2 point \citep{NCLE-2018P&SS}, which is currently the only radio instrument to study the unexplored low-frequency regime (80\,kHz to 80\,MHz) from the lunar far side.\\

\noindent
Additional experiments that aim to observe the low-frequency radio sky from the lunar far side, using either orbiters or landers, are: LuSEE-Night and Day \citep{LuSEE-2023arXiv}, FARSIDE/FARView \citep{FarSide-2021BAAS,FarView-2024arXiv}, the DSL \citep[China,][]{Chen+2019, Chen2021}, LARAF \citep[China,][]{Chen2024}, PRATUSH \citep[India,][]{PRATUSH-2023} and a CubeSat-based CosmoCube \citep{artuc2024spectrometer}. These experiments are designed either to measure the spatial fluctuations of the 21-cm signal or the sky-averaged Global Signal at redshifts beyond z$\sim$15 and explore the different epochs of the Universe, namely the Epoch of Reionization, the Cosmic Dawn and the Dark Ages. Such initiatives are enabling the discovery of new astrophysics and advancing our understanding of the early Universe.\\

\noindent
In parallel to the above-described future concepts, we identified an opportunity for developing a European-led lunar-based experiment which is focused comprehensively on the 21-cm cosmology. The DEX experiment is envisaged to be a many-element array on the far side of the Moon with a primary science goal to measure the spatial fluctuations in the 21-cm signal from the neutral hydrogen atom during the Dark Ages and Cosmic Dawn periods of the Universe. This experiment is also envisioned to be the main driver for the development of a Lunar Radio Observatory (named Astronomical Lunar Observatory, or ALO), which would also support additional science goals, as described in Section \ref{sec:sci-secondary}. The requirements and current concept for DEX are described in more detail in Section \ref{sec:inst-sci-req}.\\

\subsection{Primary science case for DEX: neutral hydrogen cosmology}
\label{sec:inst-cncpt}

We identify the neutral hydrogen cosmology science case, which includes studying the Dark Ages and Cosmic Dawn epochs of the early Universe, as the primary science case for DEX for the following reasons:

\begin{itemize}
    \item Largest scientific impact: the measurement of redshifted hydrogen emission provides the only known way to study a specific period in the early history of the Universe that is of critical importance to understand the history of structure formation and the role and nature of dark matter, making the characterisation of this epoch one of the major open questions in astrophysics.\\
    \item Unique enabling environment: the lunar far side offers unique conditions (lack of ionospheric interference, shielded from Earth RFI) that make it possible to perform high-sensitivity measurements of the faint redshifted hydrogen signal. Many of the ancillary science cases allow for different locations, antenna configurations and spectral ranges for their radio measurements, but the cosmological case does not.\\
    \item Scalability: Observations focusing on neutral hydrogen cosmology can be done already with a few separate antennas or a modestly-sized array when focusing on the Global Signal. For Power Spectrum measurements, increasing the array sizes would allow us to probe further into redshift space. In contrast, the exoplanet science case can only be meaningfully addressed with a spatially extended array of sufficiently many elements, making it technologically more difficult to implement.
\end{itemize}

\noindent
Measurements of the neutral hydrogen signal from the Dark Ages and Cosmic Dawn can be performed in the following three ways, of which the first two are in scope for DEX:\\

\noindent
{\bf Global Signal from 21-cm line emission}: measurement of spatially averaged redshifted 21-cm hydrogen emission from the Dark Ages and Cosmic Dawn epochs. The global Dark Ages signal peaks at around 20\,MHz with an absorption depth of about 50\,mK. Simple considerations show that even with one dipole antenna and an ideal receiver (only limited by thermal noise) it could be detected with a total integration time of order 100 days, and this could be reduced further by adding more antennas. However, in the presence of foregrounds and receiver non-idealities, this estimate will be significantly increased. Nevertheless, it is clear that in an RFI-free environment such as that envisaged for DEX, the measurement goals can be achieved with a (very) small number of antennas and within a reasonably short time. On the other hand, using only a few antennas implies that the use of cross-correlations is reduced, and therefore the success of the measurement relies on long-term instrument stability and/or extremely accurate calibration.\\

\noindent
{\bf Angular Power Spectrum 21-cm analysis}: performing an analysis of the spatial variations in the distribution of hydrogen gas and, by inference, of dark matter on scales from an arc-minute to a degree. This is the key measurement that DEX aims for. The expected features in the global spectrum are predicted to have depths of several tens of mK to about 100\,mK when integrated in spatial frequency shells over the range of interest (10$^{-3} $ -\ 1\  h\ Mpc$^{-1}$) which implies that the sensitivity reached must be at least around 10\,mK per shell. To achieve these high sensitivities, a very large collecting area is required, implying an interferometric array with a large number of antennas. An analysis of the noise properties of the array, as was performed in the CDF study (Section \ref{sec:CDF_Results}) leads to the conclusion that the most effective topology is a regularly gridded distribution with a very high filling factor, which provides large dwell times (and thus high sampling statistics) per uv-cell. A regular grid also offers the added advantage that it reduces data processing needs dramatically, by moving from a traditional interferometer scheme in which each antenna signal is cross-correlated with all others, to a so-called ``FFT-telescope" in which the correlation step is replaced with a spatial 2D FFT. Detailed simulations show that if the number of antennas in the array is of order 1000, the sensitivity achieved with an integration time of 10000 hours would allow us to perform imaging at redshifts up to z $\sim$ 25 with adequate signal-to-noise ratio. Such an array would yield important new Cosmic Dawn science and may start to give some insights into the Dark Ages physics, but to really increase the redshift horizon to the Dark Ages epoch, it is required to increase the antenna number by orders of magnitude (e.g. to reach z$\sim$50, about $\sim$10$^6$ antennas are required).\\

\noindent
{\bf Tomography of the 21-cm line}: imaging the spatial distribution of hydrogen and dark matter separately for individual redshift slices within the DA and CD epochs, in this way generating a “movie” of the evolution of the infant Universe. This measurement is much more challenging than the average angular Power Spectrum since 1) the use of data from within a single redshift slice restricts the frequency range over which the signal can be integrated, and 2) imaging the spatial variations within a redshift slice directly means that only the emission/absorption from very specific sky directions is involved per resolution element. These factors conspire to greatly increase the number of antennas needed (by orders of magnitude) to perform this measurement compared to what is needed for the Power Spectrum measurements. Given the considerations for the former, this objective remains for the moment beyond the scope of DEX.\\

\subsection{Secondary science cases}
\label{sec:sci-secondary}

\subsubsection{Heliospheric physics}
\noindent
The low-frequency capabilities of DEX make it particularly well-suited for heliospheric observations from its unique vantage point on the lunar far side, free from terrestrial interference and ionospheric effects. The instrument will provide groundbreaking measurements of solar and heliospheric phenomena by operating in a pristine radio environment. The array will enable detailed studies of Type II and Type III radio bursts, providing direct measurements of solar energetic particle events and their propagation through the heliosphere \citep{Zucca2018, Morosan2019}. Type II bursts are generated by shock waves driven by Coronal Mass Ejections (CMEs), while Type III bursts are produced by electron beams accelerated during solar flares, typically with energies of 2-20\,keV \citep{Zhang2020}. The interferometric capabilities of DEX will revolutionize our understanding of CME dynamics by enabling direct imaging of CME radio emissions, which has so far been achieved only in a handful of cases \citep{Bastian2001, Maia2000}. The array's high spatial and temporal resolution allows tracking of CME-driven shocks through the inner heliosphere and investigation of particle acceleration at quasi-perpendicular shock regions along CME flanks. This will provide unprecedented insights into the magnetic field structure of CMEs through radio emission mechanisms \citep{Mondal2020}. The DEX array configuration provides comprehensive measurements of the solar wind structure and dynamics through interplanetary scintillation patterns, Faraday rotation measurement, and detection of large-scale density structures and turbulence in the solar wind. The measurements of Faraday rotation, in particular, can enable the determination of magnetic field strengths with unprecedented accuracy due to the absence of ionospheric corruption \citep{Fallows2022b}. The combination of DEX observations with in-situ measurements from spacecraft like Parker Solar Probe and Solar Orbiter will enable unprecedented studies of solar wind evolution from the corona through the inner heliosphere. This multi-messenger approach will be particularly powerful for understanding the initiation and evolution of space weather events and their potential impacts both at Earth \citep{Dresing2023} and for crewed spaceflight, where advance warning for enhanced energetic particle flux events is important from a safety point of view.\\

\subsubsection{Planetary magnetosphere interactions}


\noindent
Besides the Sun itself, the brightest objects in the low-frequency radio sky observable from the Moon are the planets of the Solar System and, more specifically, their magnetospheres. They emit radio emissions below $1\,$MHz ($40\,$MHz for Jupiter) which is below the Earth's ionospheric cutoff ($\sim10\,$MHz). Hence, there is a strong motivation to observe them from space. Contrary to in-situ probes, a radio interferometer can take advantage of its higher angular resolution to improve its sensitivity, preventing the need to fly to the vicinity of the target.\\


\noindent
The planetary magnetospheres emit cyclotron-maser radio emission \citep{1979ApJ...230..621W,1996P&SS...44..211L}. This emission is induced by charged particles trapped along the magnetic field lines of the planet, releasing their free energy with electromagnetic radiation through the so-called Cyclotron-Maser Instability (CMI) . The emission is coherent, elliptically, or circularly, polarized, and occurs at the ambient cyclotron frequency which is proportional to the magnetic field strength.
It is also strongly beamed and, therefore, may appear variable depending on the relative geometric configuration between the observer, the sources and the local Noon. The plasma sources in the planetary magnetospheres are the solar wind and the planet’s ionosphere, but they can also be the planetary moons, like Io in the case of Jupiter \citep{zarka_GM119_00}. By taking advantage of the beaming pattern and polarization of the emission, the precise modulation pattern can be exploited to determine the rough magnetic topology, magnetic obliquity and planetary rotation rate \citep{hess_2011_mag_sig}.



\begin{figure}
    \centering
    \includegraphics[width=0.99\linewidth]{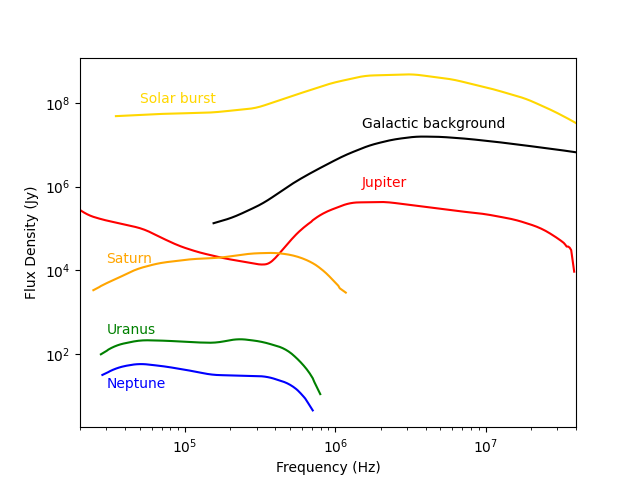}
    \caption{The average flux density spectra (in Jansky) of the radio emissions observable from the Moon. Data from \citet{MASER_zarka2022solar}.}
    \label{fig:pkr_spectra}
\end{figure}

\noindent
The spectra of the planetary radio emissions, as seen from the Moon, are depicted in Fig.~\ref{fig:pkr_spectra}. These spectra represent the average brightness observed when radio emissions occur. They compile what has been observed by previous probe flybys, with Voyager 2 being the only probe to provide radio measurements of Uranus and Neptune. The galactic background appears to be the brightest source 
when no solar bursts are occurring. The galactic spectrum plotted in Fig.~\ref{fig:pkr_spectra} is based on measurements performed in the direction of the galactic poles \citep{Dulk2001}. However, in the direction of the galactic plane, the galaxy may appear fainter because of the ISM causing notably electron free-free absorption \citep{Jester-Falcke-2009}. 

\noindent
The spectra shown in Fig.~\ref{fig:pkr_spectra} are averaged and hide complex time-
frequency structures.
On the one hand, some planetary radio components are narrow-band, continuous, and last for about an hour, like Jupiter's narrow-band kilometric radiation (nKOM). Their observation can thus benefit from large time integration. On the other hand, some emissions are much more sporadic, like Jupiter's broadband kilometric radiation (bKOM), and may overlap with continuous components. Identifying and discriminating the sporadic emission from the continuous ones is essential for the phenomenological studies, and depends on the time-frequency resolutions and the duration of the observation.
Finally, some emissions have a fine ``arc'' structure, like Jupiter decameter emission (DAM) that requires greater time-frequency resolutions to be detected. With even finer time resolution it is possible to observe millisecond bursts in Jupiter decameter emissions (S-burst). Saturn, Uranus and Neptune have their own phenomenology of radio components, yet, their emissions seem to have similar types of time-frequency structure as Jupiter.


\noindent
These spectra are averages of what has been observed by previous probe flybys. Jupiter is widely studied because it functions as a test lab for the radio observation of exoplanets as presented in section \ref{sec:exoplanets}. Yet, joined observations of a probe with a distant static observer are highly valuable as they allow to disentangle between temporal variability and source beaming effects. Providing new systematic observations of Jupiter and Saturn would deepen our understanding of these planets, especially if they complement simulataneous in-situ observations  (with, e.g., future planetary exploration missions, like JUICE for Jupiter in the 2030's or the possible mission to Uranus planned for the 20240's). While Jupiter benefited from multiple missions observing its radio emissions, Uranus and Neptune have yet to be observed again since Voyager 2 flybys in the 80s. The only dynamic spectra ever recorded give an insight into the flux and structures of the emissions we may expect from these planets, as compiled by \citet{lamy2020auroral}. However, they were recorded under quite specific conditions as studied by \citet{solarwind+magnetosphere_donaldson2024}. These planets lack long-term observation of their kilometric radiation. Notably, Neptune's internal rotation period has yet to be measured with precision.\\


\subsubsection{Exoplanets}
\label{sec:exoplanets} 

%
\noindent
As mentioned before, all magnetized Solar System planets emit CMI radio emission, which can reach very high brightness temperatures ($T_b \sim 10^{20-22}$\,K), and only depends on the magnetic field strength: $\nu_c \approx 2.8(B/$Gauss)\,MHz. This allows for a direct measurement of the planetary magnetic field strength, which may not be possible by other means for objects with cold molecular envelopes such as exoplanets and cold brown dwarfs \citep{2012P&SS...74..156Z,2024NatAs...8.1359C}. CMI emission is also beamed and, therefore, appears rotationally modulated. The precise modulation patterns depend on the magnetic topology and viewing geometry. The beaming can be exploited to determine the rough magnetic topology, magnetic obliquity and planetary rotation rate \citep{hess_2011_mag_sig}. Finally, a component of an object's cyclotron maser emission may be induced by the interaction of its magnetic field with an orbiting moon or a companion. This phenomenon is widely studied in case of the Jupiter–Io interaction \citep{zarka_GM119_00}. The temporal modulation in this case occurs at the synodic period, i.e., the beat period between Jupiter's rotation and Io's orbital motion. Similar emission from exoplanets could also be used to detect the presence of exomoons and evaluate the energetics of the exoplanet-moon interaction. Finally, the mere presence of a magnetic field necessarily requires the planet to have a conductive interior with sufficiently vigorous convection. A radio detection can therefore provide indirect information on the internal composition and heat flow of an exoplanet.\\

\noindent
The maximum frequency at which an exoplanet can emit CMI radio emission depends on the planetary magnetic-field strength at the top layer of its polar atmosphere. This value is around 14\,Gauss on Jupiter which corresponds to a maximum cyclotron frequency of around 40\,MHz. In the absence of magnetic field measurement of exoplanets, dynamo scaling laws can be employed to predict the peak emission frequency. Although there is no universally accepted dynamo scaling law, literature predictions of peak cyclotron frequency in gas-giant planets fall within the DEX spectral range  \citep{2019RAA....19...23Z}. The cyclotron frequencies of ice-giants and telluric planets are likely to be below $\sim10$\,MHz and will require a telescope design optimized for much lower frequencies than those considered here.\\

\noindent
Once the peak frequency is known, the flux must be predicted for comparison with DEX’s target sensitivity. Here we encounter a problem inherent to all coherent emission mechanisms: the flux density is highly variable and can only be estimated to within an order of magnitude, even in the best case. Several authors have done so with varying assumptions. The most widely accepted flux scaling relationship is the so-called Radio-Magnetic Scaling Law that scales the radio flux of the exoplanet with the stellar wind’s Poynting flux on the planet’s magnetosphere \citep{Zarka2007,Zarka2018}. This scaling is inspired by the theoretical expectation that the Poynting flux is the energy-source for the radio emission and by empirical agreement of Solar System planetary radio emission fluxes. From such analysis it can be inferred that only planets whose radio fluxes are at least three orders of magnitude larger than Jupiter can be detected for telescope collecting areas that can be reasonably expected from upcoming proposals, DEX included. The wide extrapolation of the scaling law from the Solar System values raises questions of whether some saturation mechanism can prevent exoplanets from reaching a detectable level of emission. Saturation mechanisms have been considered \citep{Turnpenney2020} with a conclusion that saturation must be anticipated at radio powers of order $10^{22}$\,erg\,s$^{-1}$. This corresponds to a flux density of $\sim 10$\,mJy for a planet at a distance of 5\,pc with a surface field of 15\,Gauss. For comparison, the Jovian radio power is around $10^{18}$\,erg\,s$^{-1}$. It is also possible that the rotation of the planet itself is the source of radio power, with the stellar wind pressure only determining the size of its closed magnetosphere where rotational energy can be converted to radio energy \citep{Nichols2011}. Regardless of the driving engine, one can anticipate saturation due to the emission mechanism itself. A conservative saturation value is around a brightness temperature of $\sim 1020$\,K. Adopting Jovian values for emission bandwidth, this corresponds to a flux density of around 50\,mJy at 5\,pc.\\

\noindent
The conclusion of the above analysis is that a suitable instrument for gas-giant exoplanet detection, or for ruling out existing models in case of non-detection, must access frequencies below $\sim 30$\,MHz and be able to reach a 5-sigma detection threshold in Stokes V of $10^4\times$ Jovian radio flux value. With this sensitivity, we can either achieve a detection, or rule out existing models for exoplanet radio flux scaling. The brightest Jovian bursts reach $10^9$\,Jy (at a sub-second scale) and $10^8$\,Jy (up to 1 hour long) when viewed from 1 AU distance. Let us conservatively adopt $10^4 \times 10^8$\,Jy as our minimum 5-sigma detection threshold in 1 hr integration which is 50\,mJy for an exoplanet at the distance of 5 pc. A futuristic expansion of DEX to achieve 5\,$\mu$Jy noise (at 5-sigma) in a 1-hr-long observation will ‘guarantee’ a detection as no extrapolation of Jovian radio flux values is necessary in that case.\\

\subsubsection{Fast Radio Bursts}

\noindent
An emerging application of ultra-low-frequency observatories in space involves the investigation of coherent astrophysical phenomena. A prime example are Fast Radio Bursts (FRBs), millisecond-scale emissions of uncertain origin. Current research suggests that the most probable progenitors of FRBs are magnetars, highly magnetized neutron stars, or high-energy binary systems \citep{Petroff+2022A&ARv}. However, the mechanisms behind their generation, whether magnetospheric or due to relativistic shocks, remain to be clarified \citep{Zhang-2020Nature}. Because FRBs traverse the host galaxy and intergalactic plasma, they experience significant dispersion. The dispersion measure (DM), after adjustment for contributions from the local environment, host galaxy and the Milky Way, can be employed to delineate the baryonic content of the IGM. When coupled with the redshifts of the host galaxies, the DM values from a set of FRBs can facilitate cosmological studies and help constrain theories of galactic feedback and cosmic baryon redistribution \citep{Macquart+2020Nature}.\\

\noindent
While FRBs are a broadband phenomenon, they exhibit a limited bandwidth in individual bursts, predominantly detected within the GHz range (cm band) from 400\,MHz to 5\,GHz \citep{Petroff+2022A&ARv}. Notably, LOFAR recorded emissions from FRB 20180916B down to a minimum frequency of 110\,MHz, indicating that FRB spectra may extend below this threshold \citep{Pleunis+2021ApJ}. The highest redshift FRB detected so far, FRB 160102, originated from a galaxy with redshift between $z=2.1$ and  $z=2.6$ and exhibited a DM of 2596.1\,pc/cm$^3$ \citep{Bhandari2018}.\\

\noindent
The DM values are expected to increase with redshift \citep{Macquart+2020Nature}, and addressing this challenge requires sophisticated coherent de-dispersion techniques \citep{Bassa+2017AC}. Extending the detection of FRBs to redshifts well beyond $z=1$ is pivotal for cosmological applications. For instance, a typical CHIME-detected FRB peaking around 400\,MHz in the local Universe would manifest at 66, 36 and 19\,MHz at redshifts of 5, 10 and 20, respectively. The epochs during which the first magnetars or high-mass binaries formed remain unknown, but given the abundance of massive stars in the early Universe, these structures could have emerged rapidly on cosmological timescales. The upper limits of the FRB luminosity function have yet to be determined, which affects the feasibility of detecting extraordinarily powerful bursts. Nevertheless, the potential for observing gravitationally-lensed FRBs exists, with galaxy clusters possibly providing a magnification factor up to 100 times.\\

\subsection{Science requirements for DEX}
\label{sec:inst-sci-req}

\noindent
Resulting from the above high-level considerations, a configuration for DEX is proposed which consists of two separate elements, an antenna configuration for Global Signal detection, and an imaging array for 21-cm line tomography.\\

\noindent
{\bf Global Signal detection experiment}: for the Global Signal measurement of the redshifted hydrogen spectrum from the early Universe, a small number of carefully characterised and calibrated antennas are needed. These antennas need to be placed well away (at least several hundreds of meters) from other antennas and systems. The individual antennas should have a well-behaved and thoroughly understood antenna pattern for the frequency range of interest. For a $10\sigma$ detection of a 50\,mK Global Signal, the required integration time can be approximated by
\begin{equation}\label{eq:tint}
    t_{int} = 17\; \rm{h} \left(\frac{\nu}{70\;\rm{MHz}}\right)^{-5.1}\left(\frac{\Delta\nu}{1\;\rm{MHz}}\right)^{-1}\left(\frac{\Delta T}{10\;\rm{mK}}\right)^{-2}
\end{equation}
which means that if the width of the signal exceeds 1\,MHz, one can reach 10\,mK in less than one day at 70\,MHz. Towards a lower frequency range this quickly increases, e.g. to little over a year at 20\,MHz. However, given the required spectral dynamic range ($\approx10^7$) this would require extremely accurate band-pass calibration.\\

\noindent
{\bf Power Spectrum detection experiment}: for measurement of the Power Spectrum of the Cosmic Dawn era, a minimum goal is to implement a compact and regularly spaced array of $32\times32$ elements. The compactness ensures that the most effective use of uv-coverage is achieved, and the regularity enables greatly simplified data processing at a fraction of the capacity that would be needed for an irregularly spaced array.\\

\noindent
The sensitivity estimates of the array are based on the following three assumptions: 1) the array geometry is circularly symmetrical and densely packed (we neglect the minor difference in performance caused by using a square versus a circular array geometry), 2) the system temperature is taken as constant across frequency for a single slice in redshift, and 3) the Power Spectrum is assumed to be static in time over the range of redshifts we use for each slice. For each chosen redshift, we calculate the co-moving volume we see, which is a function of our available bandwidth for that redshift (we pick a 'slice' of redshifts). We hence sample a 'cylinder' where our volume has two dimensions on the plane of the sky (the circular base of the cylinder) and one dimension in frequency (the height of the cylinder). Because we are interested in measuring the amplitude of variations as a function of scalar length, all three of these dimensions play into this measurement. We therefore convert all three axes into the same units (k-space) and then integrate over shells in this space to obtain the signal for every spatial scale separately.\\


 \begin{figure*}[htb]
\centering
   \includegraphics[width=1.1\linewidth]{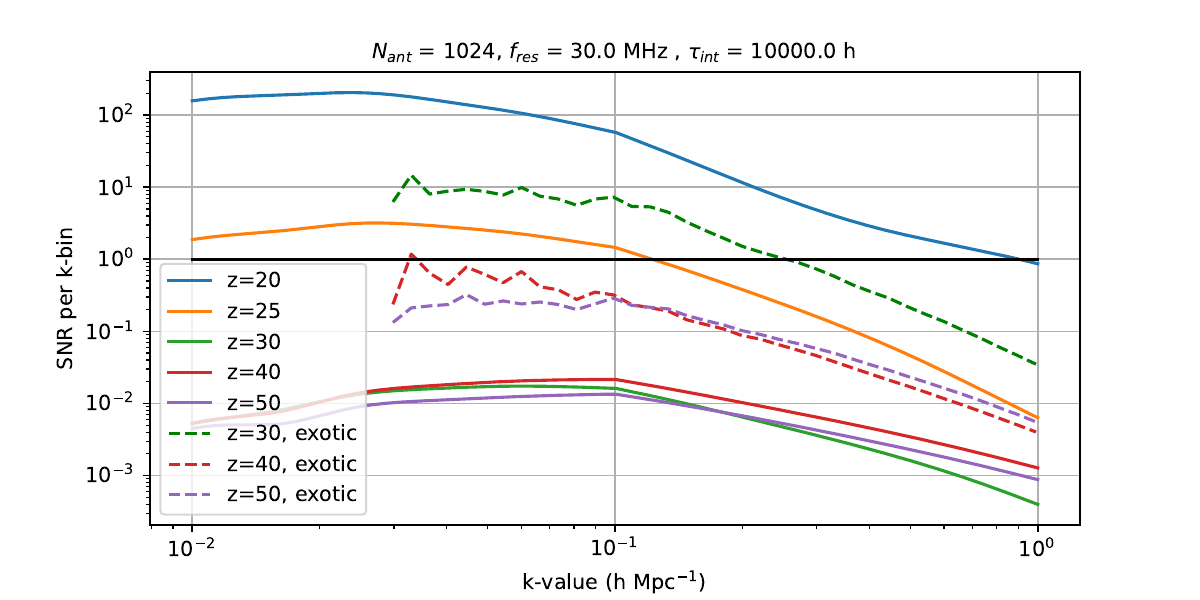}
   \caption{Redshifted 21-cm Power Spectrum array performance for redshifts in the range $z=20-50$.}
   \label{Fig:arraysnr}
\end{figure*}

\noindent
Figure~\ref{Fig:arraysnr} shows the Power Spectrum signal-to-noise (SNR) for a $32\times32$ array for five redshifts, and for three redshifts of power spectra influenced by exotic physics (e.g., dark matter or primordial black holes). As SNR scales linearly with (dense) aperture size, the curves in the figure would shift $+12$\,dB and $-18$\,dB for array sizes of respectively $128\times128$ and $4\times4$ elements. This means that measuring the Power Spectrum at low redshifts (around $z=20$) is already possible with a modest-size $4\times4$ array, albeit over a limited range of co-moving distances around $k=2\times 10^{-2}$\,h\,Mpc$^{-1}$. The $32\times32$ array would cover co-moving distances up to about 0.1\,h\,Mpc$^{-1}$.\\

\noindent
{\bf Planetary magnetosphere interactions}: The basic configuration of DEX ($32\times32$ elements) can provide continuous observations over $\sim10\,$days with high spectral and temporal resolution for the brightest planets (Jupiter and Saturn). The large field of view of DEX, coupled with its lower angular resolution and temporal variability of foregrounds and secondary radio sources from one observation to the other, may make it difficult to isolate a source target in the measured data, thus lowering the instrument's sensitivity. This should be the topic of a dedicated study. If the resulting sensitivity allows for the confident detection of Uranus' or Neptune's kilometric radiation, DEX observations would be the first observations of such kind in decades and would give access to unique planetary magnetosphere dynamics.\\

\begin{figure}[htb]
\centering
   \includegraphics[width=0.7\linewidth]{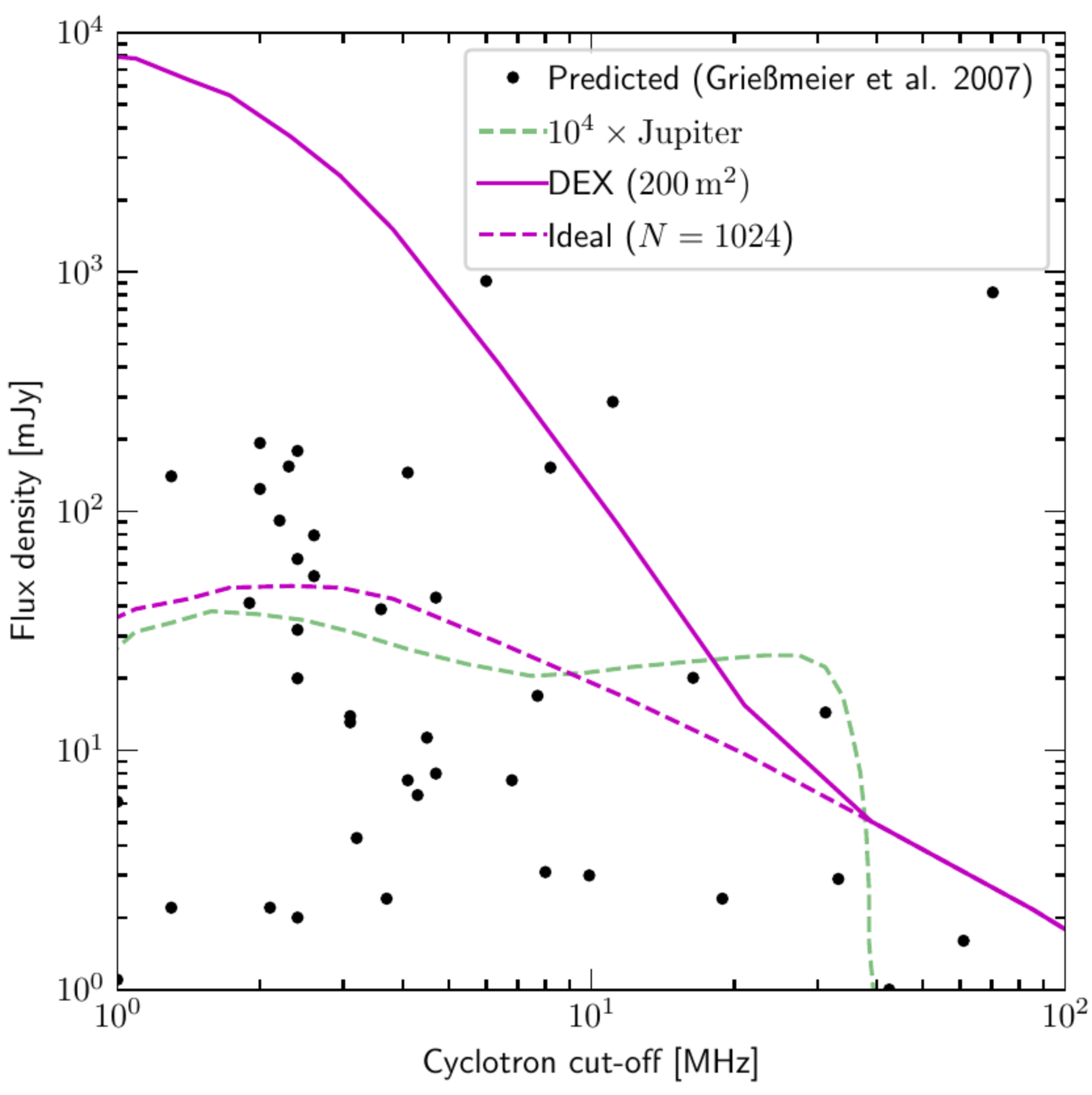}
   \caption{Comparison of DEX (32 × 32) dipole array sensitivity with predicted exoplanet flux densities. The solid and broken magenta lines show sensitivities with and without aperture oversampling loss.}
   \label{Fig:exoplanets}
\end{figure}

\noindent
{\bf Exoplanet detection}: The predicted exoplanet radio flux densities from Section~\ref{sec:exoplanets} are compared with the sensitivity of a DEX (32 $\times$ 32) array contained within a $200\times 200$\,m$^2$ area in Fig.~\ref{Fig:exoplanets}. To compute the sensitivity, we assumed an integration bandwidth of $\nu_c$ / at center frequency $\nu$, background sky temperature of $T_{sky} = 60(\lambda/{\rm meter})^{2.55}$ Kelvin \citep{Roger+1999}, sky-noise dominated receiver and 1\,hr of integration. The predicted exoplanet flux densities are from \cite{Griessmeier2007}. The broken magenta sensitivity line shows the ideal sensitivity of a 32 $\times$ 32 array. The actual achieved sensitivity (solid magenta line) is worse due to oversampling of the electric field below $\approx$ 30 MHz by the close-packed dipoles. We assumed that each dipole has an effective collecting area of $\lambda^2/3$. From the plot, we can see that there are a few exoplanet systems above $\approx$ 5 MHz that could be detected by such an array.\\

\subsection{System requirements for DEX}

{\bf{Spectral:}} For the science measurements presented in Section~\ref{sec:inst-sci-req}, the top-level system requirements are defined and listed in Table~\ref{Table:telparam}. Concerning the frequency range, observing the Dark Ages should ideally start at 1\,MHz, however, given the sensitivity that can be achieved with a limited-size first generation of DEX, it cannot yet provide a Power Spectrum detection in this frequency range. The frequency range $10 - 30$\,MHz is accessible partially from Earth, but the ionospheric disturbances and the RFI situation make observing and calibration near impossible for the Dark Ages observations. The frequency range $30 - 45$\,MHz for the Dark Ages is included to be able to cross-reference scientific results with Earth-based telescopes. The Cosmic Dawn phase in the frequency range $45 - 80$\,MHz is included because of the favourable ionospheric and RFI situation and because the observation system is only a modest extension of the Dark Ages observation system. Optimization is done for frequencies below 30\,MHz. Bandwidths are chosen as a compromise between minimizing compute load and maximizing sensitivity.  The frequency resolution of 12\,kHz was chosen to meet the narrow band condition, $\Delta f_{nbc} \ll c / (2\pi D)$, implying a channel width much smaller than 78\,kHz for the $128 \times 128$ array.\\

\newcommand{\pbA}{\parbox{3.5cm}}
\newcommand{\pbB}{\parbox{1.8cm}}
\begin{table}[htb]
	\begin{centering}
		\caption{Basic characteristics of the Global Signal and the Power Spectrum experiments. The columns refer to the frequency/redshift ranges corresponding to the Dark Ages and the Cosmic Dawn epochs.}
		\begin{tabular}{lllll}
		
            \toprule
	       & \multicolumn{2}{c}{Global Signal}  &\multicolumn{2}{c}{Power Spectrum}\\ 
            \cmidrule(l){2-3} \cmidrule(l){4-5}
		& \pbB{Dark Ages} &\pbB{Cosmic Dawn} &\pbB{Dark Ages} &
           \pbB{Cosmic Dawn}\\
		
            \midrule
		\pbA{\emph{Spectral}} & & & &\\
            \midrule
		  \pbA{Frequency range (MHz)} & \pbB{18-45}& \pbB{45-80} & \pbB{18 - 45} & \pbB{45 - 80}  \\
		  \pbA{Bandwidth (MHz)} & \pbB{27}& \pbB{35} & \pbB{27} & \pbB{35}  \\
		\pbA{Frequency resolution (kHz)} & \pbB{12}& \pbB{12} & \pbB{12} & \pbB{12}  \\ \midrule
     	\pbA{\emph{Temporal}} & & & &\\ \midrule
			  \pbA{Total integration time (h)} & \pbB{2000\\$(35\leq z\leq80$)}& \pbB{2000\\$(z\leq35)$} & {\color{red}\pbB{$10^4$ {($z\leq50$)}}} & \pbB{$10^4$ $(z\leq25$)}  \\
		\pbA{Time resolution (s)} & \pbB{1}& \pbB{1} & \pbB{1760} & \pbB{1760}  \\

            \midrule
		\pbA{\emph{Spatial, antennas}} & & & &\\
            \midrule
            $N_{ant}$ & \pbB{$\geq$1} & \pbB{$\geq$1} & {\color{red}\pbB{512$\times$512 }} & \pbB{32$\times$32} \\ 
            \pbA{\vspace{2mm}Antenna system\vspace{4mm}} & \pbB{Separate antennas} & \pbB{Separate antennas} & \pbB{Regular array} & \pbB{Regular array} \\
            \pbA{Spatial resolution ($^\circ)$} & \rm{all-sky} & \rm{all-sky} & 0.7 - 0.23 & 7 - 4 \\
            \pbA{\vspace{2mm}Location accuracy, devia-\\tion from planarity, per\\ antenna} &\pbB{-}&\pbB{-}&\pbB{$\leq$ 30 cm} & \pbB{$\leq$ 19 cm} \\ 
            
            \midrule
		\pbA{\emph{Data handling}} & & & &\\
            \midrule
            \pbA{Data rate per\\ antenna, Gbps\vspace{2mm}} &1.6&1.6&1.6&1.6\\
            \pbA{Processing power,\\ hub, Gflops\vspace{2mm}} &-&-&{\color{red}2.4$\times10^7$}&1.15$\times10^4$\\
            \pbA{Data rate to Earth, Mbps} &\pbB{0.5 per\\ antenna}&\pbB{0.5 per\\ antenna}&{\color{red}80}&0.3\vspace{2mm}\\
            \pbA{Clock stability} &-&-&$10^{-11}$ & $10^{-11}$\\ \midrule
		\pbA{\emph{Power}} & & & &\\ \midrule
		\pbA{Power per antenna (W)}& ~0.30 & ~0.30 & ~0.30 & 0.30 \\	   
            \pbA{Power to hub (W)}& - & - & {\color{red}$\geq 80\cdot10^3$} & $\geq$250 \\ 
           
            \bottomrule
		\end{tabular}\label{Table:telparam}
	\end{centering}
\end{table}

\noindent
{\bf{Temporal:}} From the Global Signal equation~\ref{eq:tint}, detecting a 10\,mK signal in a 10\,MHz bandwidth at $z=80$ ($f=18$\,MHz) should be possible within 2000 hours. This is reaching far beyond what is possible from Earth. As mentioned above, Power Spectrum imaging would be possible at redshifts up to $z\approx 20$ and $z\approx 25$ with adequate signal-to-noise ratio when observing 10000 hours  for array sizes of respectively $4\times4$ and $32\times32$. The maximum integration time before re-calibration is needed, given in \cite{rajan2016}, is for practical clocks about half an hour, or 1760 seconds. This is related to the fact that most available clocks have Allan variances better than $10^{-11}$. However, as the Moon rotates 0.54 degrees per hour and the array beam width is larger than 0.34 degrees for $f<80$\,MHz for the $128\times128$ array, the integration interval needs to be split into a few bins to avoid spatial smearing. Each analog/digital converter (ADC) could be given its own clock, or the reference signal could be supplied centrally. In the latter case, a lower-cost clock could be considered, but this would require a stable signal transport network. Both scenarios are possible, the choice depends on system level design choices, for example relating to the physical transport layer of the data and control signals (coax, fibre, RF, or free space optical).\\

\noindent
{\bf{Spatial, antennas}}: The size of the Power Spectrum array ($D$) is chosen to match the expected spatial structures. For the $128\times128$ configuration the spatial resolution is about 1.5 degrees at 18\,MHz to 0.6 degrees at 45\,MHz, covering the Dark Ages frequency range, and for the Cosmic Dawn period it comes to a few tenths of degrees. The resolution increases linearly with the size of the array. The antenna element size was initially chosen to be 5\,m tip-to-tip, a value balancing high sensitivity at the lowest frequencies and placing antenna resonances at relatively high frequencies, but subject to change according to the eventual frequency range that DEX is to be optimised for. Fat dipoles give an overall better performance compared to thin ones (a flatter frequency response), and also are easier to integrate with solar panels and electronic circuitry. No ground plane is foreseen. The 15\, cm antenna terminal separation distance is a small fraction of the antenna size. The number of antenna elements follows from the desired antenna array size and antenna length.\\

\begin{figure}[htb]
    \centering
    \includegraphics[height=0.43\linewidth]{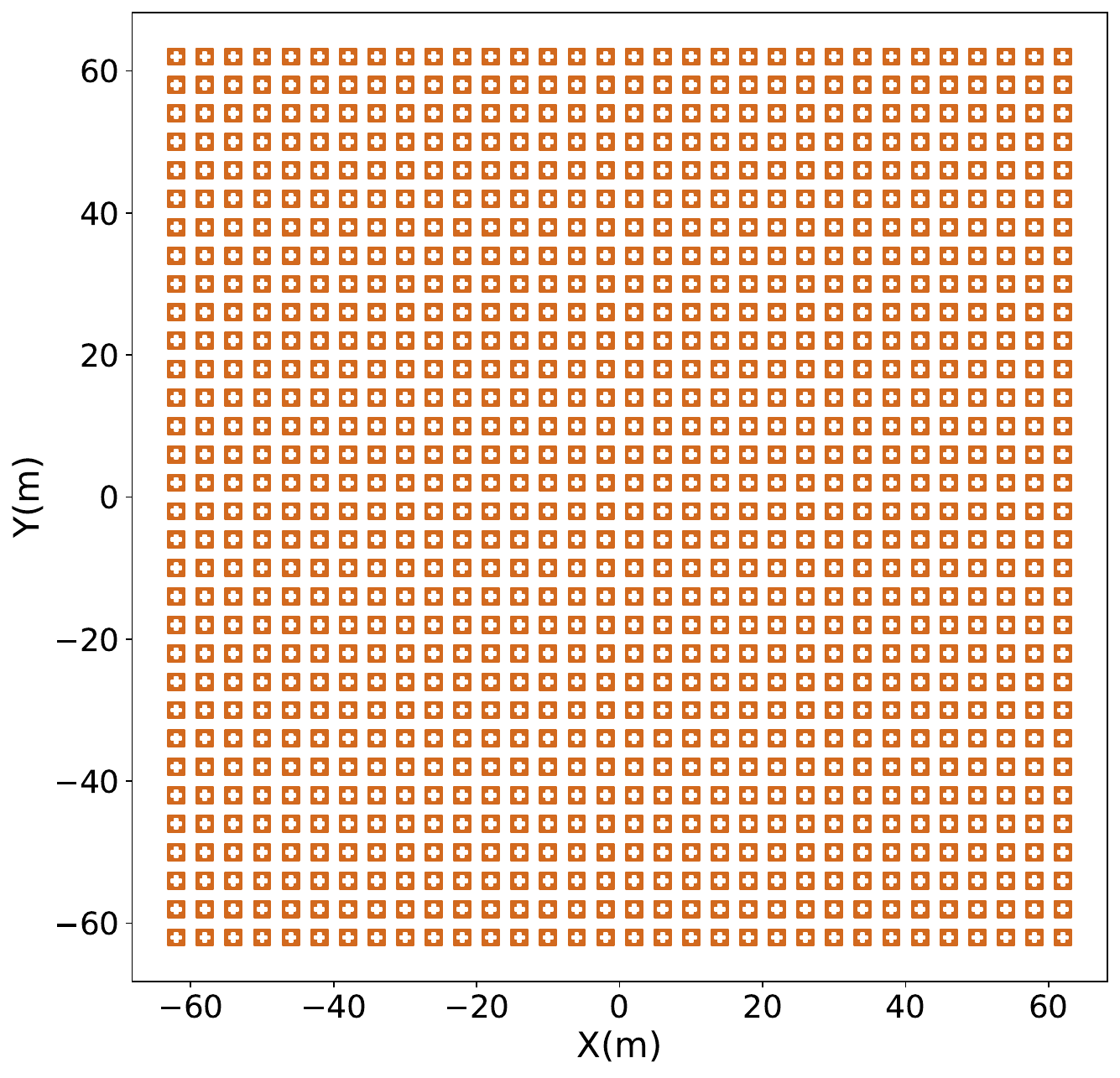}\hspace{1em}
    \includegraphics[height=0.43\linewidth]{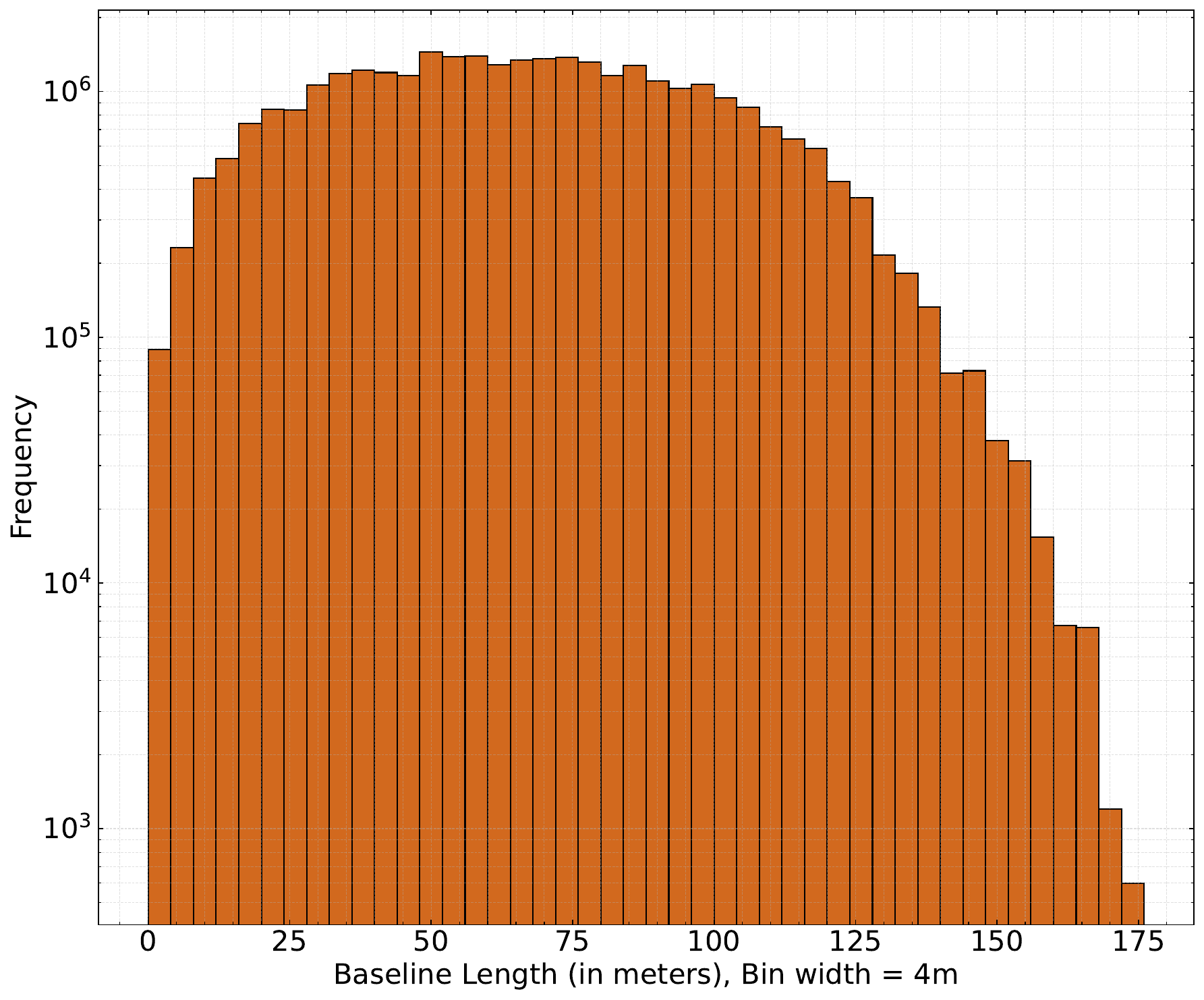}\vspace{1em}
    \includegraphics[height=0.4\linewidth]{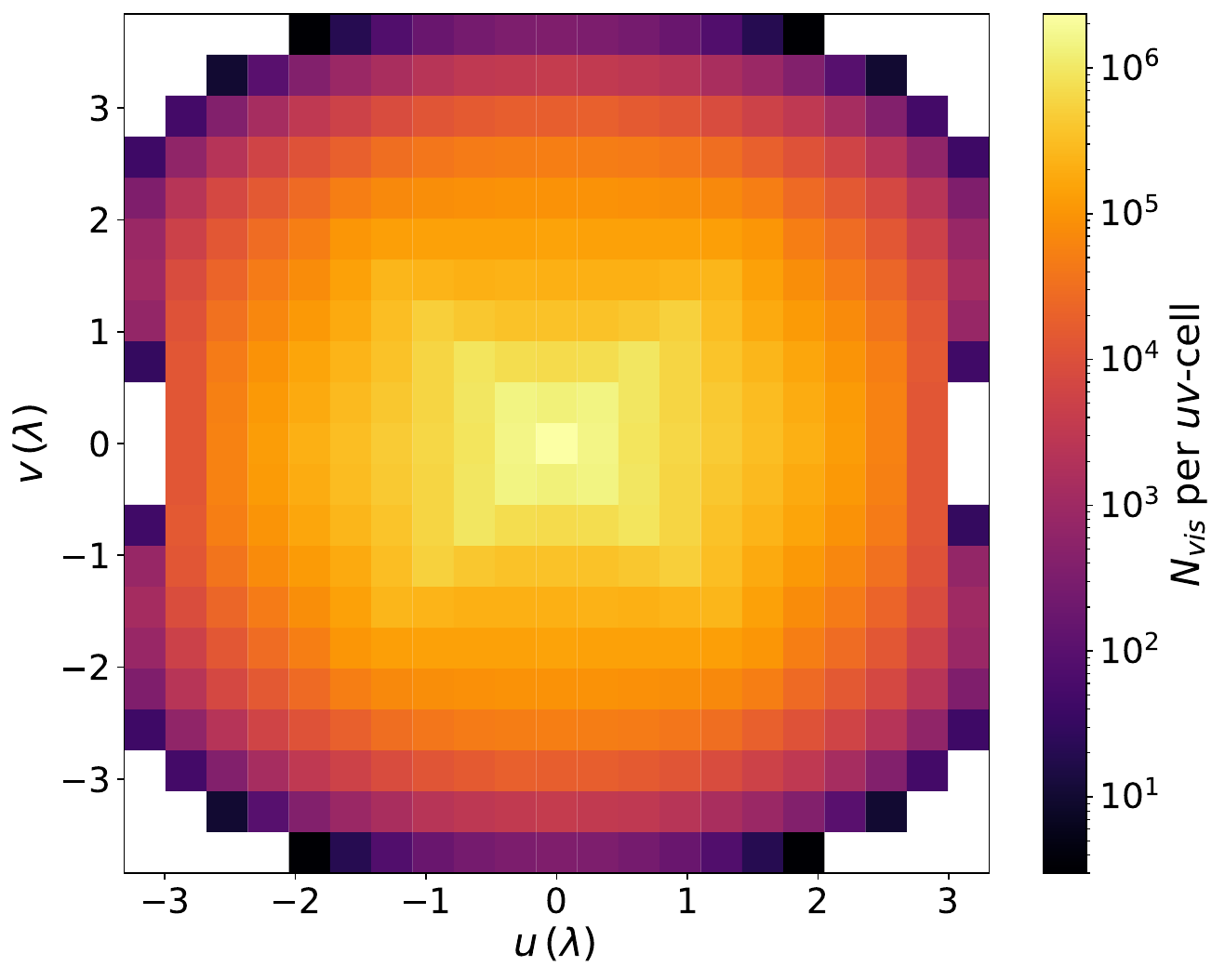}\hspace{0.5em}
    \includegraphics[height=0.4\linewidth]{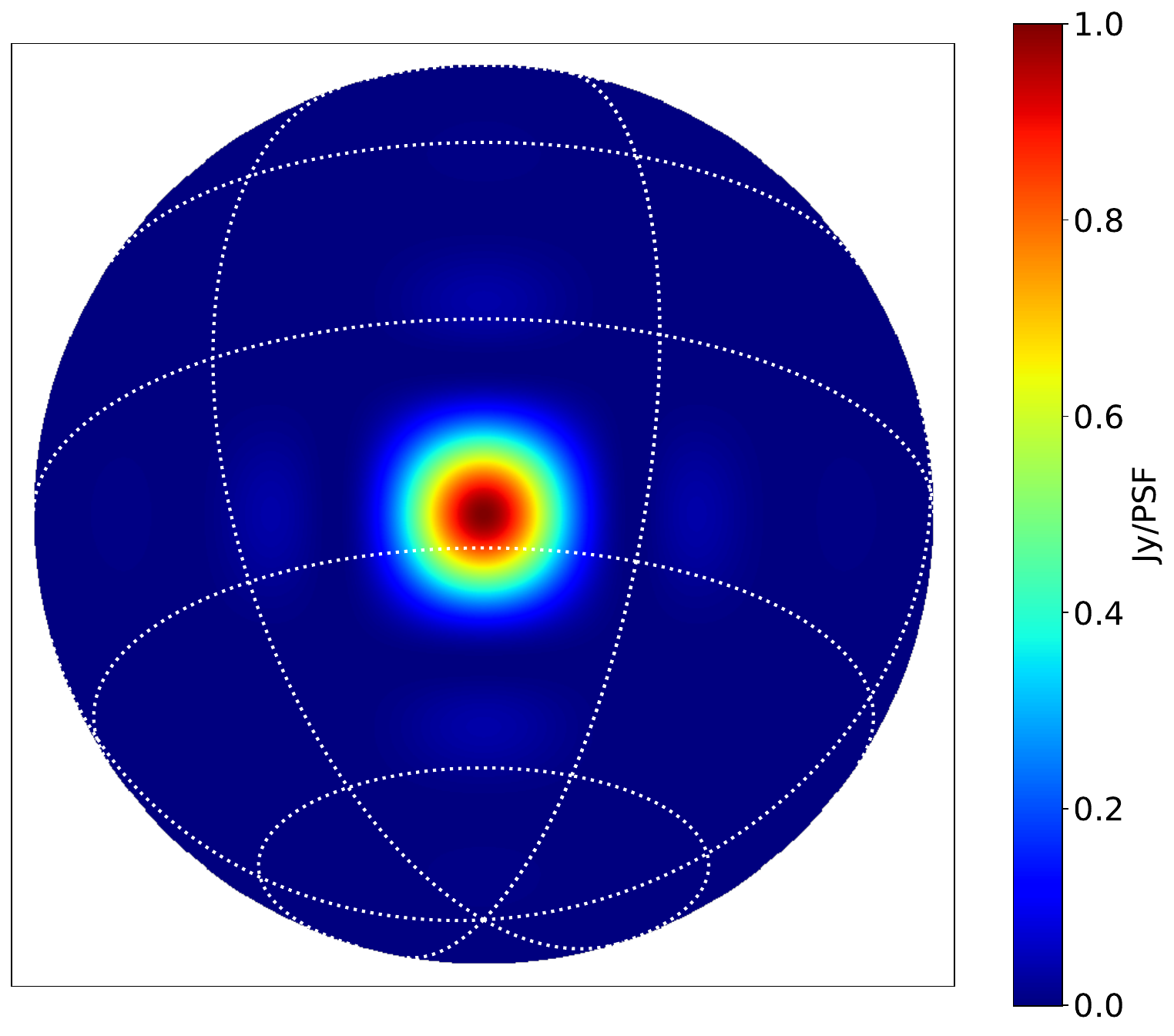}
    \caption{\textbf{Upper left}: The current proposed configuration of DEX consists of $32\times32$ array (1024 elements), arranged on a regular grid, with possibility of an expansion to more number of antennas. Each of the adjacent elements is spaced 4 meter center-to-center apart. \textbf{Upper right}: The 1D histogram of baseline length ($\textrm{L} = \sqrt{u^{2}+v^{2}}$) distribution and \textbf{Lower left}: a 1 lunar day ($\sim$30 Earth days) synthesis $uv$-coverage of DEX at 7\,MHz demonstrates the redundancy of the array highlighting the desire to maximize sensitivity on short baselines. This is a simulated observation for 30 Earth days (January 1st to 29th, 2040), corresponding to 1 lunar day. The observation has been conducted at 56 discrete times steps, each representing a 12-hour long time interval. \textbf{Lower right}: Naturally-weighted point spread function (PSF) for a source at zenith, imaged at 7\,MHz.} 
    \label{fig:dex-uv}
\end{figure}

\noindent
The antennas will be located on the far side of the Moon at a mid-range latitude because the locations close to the lunar poles suffer from two drawbacks. The first drawback is the susceptibility to Earth RFI leakage that diffracts around the lunar limb and which may still be able to reach a polar/high-latitude site as a result (an effect that is avoided for locations further away from the poles on the lunar far side). The second drawback is the fact that an array close to the pole does not sample a large fraction of the sky, meaning that cosmic variance will affect the statistics of the Power Spectrum measurements more strongly. Locations close to the lunar equator sample the sky more completely, but suffer from a reduction in quality of $uv$-coverage, as well as a more extreme temperature range compared to higher latitudes. An example of the $uv$-coverage attained by a $32\times32$ version of DEX is illustrated in Fig.~\ref{fig:dex-uv}.\\

\noindent
{\bf{Data handling}}: Given the mass and power constraints, the number of (re-coded) bits from each sample of the antenna signal is limited to four, and low levels of RFI are hence of vital importance to avoid saturation of the ADCs. Combined with a signal bandwidth of 100\,MHz, two polarizations and Nyquist sampling this yields a data stream of 1.6\,Gbps per antenna element. The data processing load for each hub, combining multiple antennas, is modest: taking $N_{\textrm{ant}}$ antenna elements per hub, it involves channelising the data stream from each antenna element into $N_{\textrm{chan}}$ channels ($N_{\textrm{pol}} \times N_{\textrm{ant}} \times N_{\textrm{chan}} \times \log{N_{\textrm{chan}}}$ operations) and performing a spatial 2D FFT using all antennas for each frequency channel ($N_{\textrm{pol}} \times N_{\textrm{chan}} \times N_{\textrm{dir}} \times \sqrt{N_{\textrm{ant}}} \times \log{\sqrt{N_{\textrm{ant}}}}$ steps) before sending the intermediate data products to the central processing module. There, all partial spatial FFTs are completed for the full array for each frequency channel and polarisation, for $N_{\textrm{hubs}}$ hubs this involves $N_{\textrm{pol}} \times N_{\textrm{chan}} \times N_{\textrm{dir}} \times \sqrt{N_{\textrm{hubs}}} \times \log{\sqrt{N_{\textrm{hubs}}}}$ operations per sampled spectrum. Once the full spatial FFTs have been calculated, the power per Fourier pixel can be accumulated over the integration period, which greatly reduces the aggregate data rate for storage and transmission.\\

\noindent
{\bf{Power}}: The power budget for operation of the array can be split up into antenna-based power usage, hub processing power usage and central processing module power usage (including communications). Due to the large number of antennas in the fiducial DEX layout, power consumption per antenna element is a critical quantity to be minimised. According to current estimates, a power consumption of 200\,mW per dual-polarisation antenna may be feasible for operation of the low-noise-amplifiers (LNAs) and filters. Additional power will be needed for signal transmission and hub-level processing. More extensive estimates as to the full power consumption of the DEX array are included in the results from the CDF study, linked in the next Section.\\

\noindent
In the following Section, we describe the results of the 2021 CDF study  which attempted a realistic implementation of this configuration, whose basic characteristics are summarized in Table~\ref{Table:telparam} and shown in Fig.~\ref{fig:dex-uv}.

\clearpage
\section{Outcomes of 2021 ALO-DEX CDF study}
\label{sec:CDF_Results}

\subsection{Setup and scope of CDF study}
\label{ss:cdf-study}

ESA's Concurrent Design Facility (CDF) is designed to bring together a multidisciplinary team of experts in an environment which facilitates efficient communications and provides common engineering tools, with the objective of designing a future space mission and covering all relevant engineering aspects within a short period of time. It is meant to lay a solid foundation for the development of more classical engineering project. In the case of DEX\footnote{While the CDF study was named using the ALO name, it actually concentrated on the cosmological radio-interferometer element of ALO, which is DEX.}, ESA called for the CDF study to identify technical options and concepts for the deployment and operation of DEX, and to put forward a baseline mission concept with an assessment of its technology needs, as well as risks and constraints. The study took place in the period June-July 2021, and was carried out by a team of around 40 ESA engineers, supported by the ALO scientific Topical Team external to ESA\footnote{\url{https://alott.astro.ru.nl}}. A detailed report of the results of the study is available at \url{https://alott.astro.ru.nl/index.php/our-projects}.\\

\noindent
The CDF study was carried out in the context of ESA's lunar exploration programme, which includes the development of a vehicle capable of carrying a large cargo to the surface of the Moon. This vehicle, dubbed Argonaut\footnote{\url{https://www.esa.int/Science_Exploration/Human_and_Robotic_Exploration/Exploration/Argonaut}} (previously known as the European Large Logistic Lander, or EL3) has three main components: the lunar descent element (LDE) that takes care of flying to the Moon and landing on target, the cargo platform element (CPE) that is the interface between the lander and its payload, and a payload. The CDF study assumed the use of Argonaut to carry DEX to the surface of the Moon in the mid-2030's. While Argonaut was (and still is) in a design phase, its notional payload capability at the time of the study (1800\,kg including CPE, volume\footnote{In the context of payload volume designations Ø represents diameter.} of Ø$4\times 2$\,m) was considered to be the envelope for DEX, and DEX was not allowed to make any requirements on the LDE. Furthermore, the CDF was tasked with identifying the ``smallest reasonable observatory'' which would provide enough new science based on a single Argonaut mission. Since it was clear from the outset that only one Argonaut mission would be insufficient to carry a payload to achieve the full objectives of DEX experiment, a key requirement was imposed on the baseline design that it should be {\it scalable}, namely that additional landing events would allow to expand the performance of the baseline system up to the desired capabilities.

\subsection{Landing site}
\label{ss:landing-site}

The location of DEX is driven by: (a) the uv-coverage on the sky achievable by a fixed antenna array, which improves with higher lunar latitudes; and (b) the need to reduce RFI from Earth (which improves at lower latitudes as the distance to the limb increases), as well as local man-made disturbances (which are likely to be concentrated near the South Pole in the near future). The simple compromise is to select a site at mid-latitudes. The array requires a fairly large ($\sim 1 \times 1$\,km) flat topography, which can often be found in the interior of large craters. Available elevation maps show that favourable regions (average slopes $<5$ degrees) can be found inside Tsiolkovsky and von Karman craters, and for this reason the former was selected as a baseline landing site. It should be noted however, that small-scale surface features (e.g. boulders) are also important to comply with the array alignment requirements, and such irregularities are not yet mapped to the level required.  One critical parameter for the performance of the array is the dielectric behavior of the (sub)surface regolith on which the antennas will be placed, which however, was not known in great detail at the time of the study, and therefore its impact on the design was not taken into account. However, studies that investigate the impact of the local properties of the lunar regolith on antenna performance are currently underway, making use of the radar measurements performed as part of the Chang'e 3 mission \citep{Ding2020, Ding2024}. Other characteristics of the lunar environment (particles, plasma, magnetic, etc) are important, but are not currently expected to differ significantly from site to site.

\subsection{Deployment}
\label{ss:deployment}

From the outset of the study it was assumed that the deployment of the DEX elements from Argonaut would not include astronaut operations\footnote{Since DEX should be deployed at a site distant from any human occupation, including astronauts' operations does not seem advisable at this stage. However, this aspect could be reconsidered at a later time, when more is known about the extent of human activities on the Moon.}. Instead, a rover would be carried by Argonaut and lowered from the CPE to the lunar surface. The rover would then offload DEX and deploy the array onto the lunar surface. Furthermore, it was assumed in the study that the LDE/CPE can be used to provide some operational services to DEX.\\

\noindent
Although the rover itself was not designed in detail in the study, a number of considerations and/or assumptions arise from the above:
\begin{itemize}
\item deployment of the rover to the surface (vertical difference $\sim 3$\,m) requires a mechanism, which could be fixed on the CPE or could be part of a robotic arm on the rover itself. A mechanism is also needed to deploy DEX itself from the CPE to the rover. This could be done using the same mechanism used to deploy the rover (e.g. a robotic arm on the rover).\\
\item The rover needs to be able to transport parts of the array (``sub-arrays") to assigned locations, and place them on the surface with a location and orientation accuracy compatible with the requirements of the array, and with whatever umbilical connections are required with the LDE or with other sub-arrays. This is a repetitive operation, with the rover needing to return to the LDE to retrieve more sub-arrays and place them at their assigned locations without disturbing the already-deployed parts.\\
\end{itemize}

The baseline selected is to have a large robotic arm mounted on the rover, which would be used both to retrieve DEX sub-arrays from the CPE and to deploy them on the surface.\\

\noindent
With the above considerations and the current state of rover development, a concept-level design was made which yields a characteristic envelope for the rover shown in Table \ref{Tab:Rover}.

\begin{table}[htb]
\caption{Characteristics of the deployment rover}
\begin{centering}
\begin{tabular}{ll}
\toprule
Rover property & Value\\
\midrule
Rover mass & 420\,kg\cr
Payload capacity & 160\,kg \cr
Volume, width$\times$ depth$\times$height & $2.20\times2.20\times1.00$\,m \cr
Driving power consumption & 120\,W\cr
Robotic arm power consumption & 614\,W \cr
\bottomrule
\end{tabular}
\end{centering}
\label{Tab:Rover}
\end{table}

\subsection{Architecture}
\label{ss:architect}

\begin{figure*}[htb]
    \centering
    \includegraphics[width=1\linewidth]{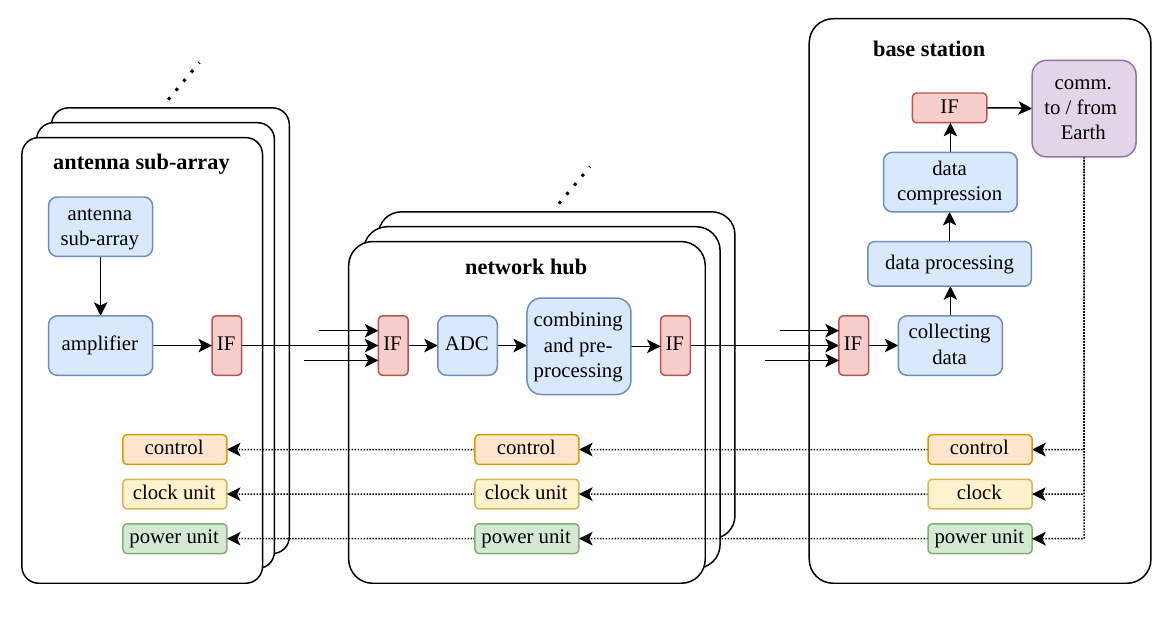}
    \caption{The DEX array baseline data architecture.}
    \label{Fig:Architecture}
\end{figure*}

The architecture of DEX is based on the following considerations:
\begin{itemize}
\item Power and signal need to be distributed to each of the antennas in the array, and the harness needed for this purpose is a major mass driver. A simple architecture where each antenna is independently fed from the base station quickly requires an infeasible level of harness mass (e.g. $\sim 87$ tons for 1000 antennas). Considerable harness mass savings can be achieved with an architecture based on hubs which serve antenna sub-arrays and this is assumed to be the baseline for the study. The optimal number of antennas served per hub depends on the total number of antennas in the array.\\
\item The signal acquired by each antenna in the array must be amplified and digitized and all signals must eventually be combined, processed and transmitted to Earth. At each step, electronics are required and data transmitted to the next step in the processing chain. The mass required for this is, as for power, a major consideration.  A trade-off analysis was made to define what is the optimal location for each processing step within an architecture including individual antennas, hubs and base station. The result of this trade-off indicates that the analog signals should be transported from individual antennas and digitized at each hub, from where the digital signals can be transported to the base station for combination and further processing, as shown in Fig.~\ref{Fig:Architecture}.\\
\end{itemize}

\subsection{Power generation and distribution}
\label{ss:power}

Power needs are clearly a key driver for the design of DEX. One of the main challenges is to manage the lunar day/night cycle\footnote{At near-equatorial locations, the lunar daylight and night period have about the same duration of $\sim 354$ hours. At the selected location of the Tsiolkovsky crater, the day surface temperature gradually increases to a maximum above 90 deg C, and during the night it falls to a minimum of -170 deg C).}. From a scientific point of view there are many advantages to being able to operate DEX during the lunar night, but a detailed trade off analysis of power generation options led to the conclusion that, currently, there is no viable solution that allows this. The baseline adopted is to operate DEX only during the lunar day using solar panels mounted on the base station (LDE/CPE). It still remains necessary to power DEX electronics during the lunar night to ensure their survival – the most promising approach for this identified during the study is a Regenerative Fuel Cell system currently in development by ESA. Regarding distribution of power to each antenna from its parent hub, a tradeoff indicated that Series AC feed would be most suited in terms of mass, but it remains to be verified whether the presence of transformers/rectifiers is acceptable in terms of EMC.
 
\subsection{Antenna considerations}
\label{ss:antenna}

The basic design of individual elements of the array consists of a pair of orthogonally-oriented dipoles (a ``crossed-dipole"), each with a length of 5\,m tip-to-tip. This size implies immediately that they will have to be compressed into a small volume for transport on Argonaut, and re-expanded (either individually or as sub-arrays) on the surface of the Moon. Whatever method is used for packing and re-expanding constitutes a key challenge for DEX. During the CDF study, a number of existing and future technologies were investigated, including: 1) antennas made of Shaped Memory Alloy (SMA) tapes; 2) antennas printed on a film and stored as a roll; 3) tubular booms; and 4) sail-like structures with motorised and non-motorised deployment. Taking the development status and applicability for DEX into consideration, the only technology which was found to have high enough TRL is tubular boom antenna, which is currently being used on ESA's Jupiter Icy Moons Explorer \cite[JUICE,][]{JUICE-2018cospar}, and NASA's Parker Solar Probe \cite{Fox+2016}, and this is the solution that was selected for the CDF study. The estimated mass of this (single crossed-dipole) unit is 1.9\,kg.\\

\noindent
An important consideration for the antenna design is the inclusion of a ground plane. All low-frequency arrays deployed on Earth use ground planes to control the beam patterns and stray radiation. However, a ground plane would need to be at a distance of at least several decimetres from the antenna itself, and would convert the antennas into 3D structures which are much more difficult to deploy onto the lunar surface than 2D antennas lying directly on the lunar regolith. An extra complication is that a ground plane would add significant mass to the instrument. During the CDF study a set of electromagnetic simulations of simple dipoles were carried out geared to evaluate the effect of lunar regolith lying directly under them. The simulations used regolith properties derived from recent radar measurements by the Chang'e 3 rover \citep{Ding2024}. These simulations provided preliminary reassurance that it is possible to obtain good performance for 2D dipole-like antennas placed directly on the lunar surface. However, this needs to be confirmed taking into account better measurements of the lunar surface, in particular of the variability of its properties across the array extent and several metres into the lunar regolith.

\subsection{Array considerations}
\label{ss:array-config}

From a scientific point of view, the number of antennas in the array is the most important performance driver. The considerations made in the CDF study presented so far allow to estimate the mass required for a given number of crossed-dipoles (including power and signal harness and data processing electronics). When adding everything needed (antenna units, hubs, harness, base station, rover), the mass that would need to be carried by Argonaut is $\sim 1.1$\,t for a $4\times4$ array (1 hub), and 1.8\,t for an $8\times8$ array (4 hubs). It becomes immediately clear that with the baseline choices made, the ``largest reasonable observatory” that can be delivered by one Argonaut launch is a $4\times4$ array. 

 \subsection{Data production and handling}
 \label{ss:data-handl}

Interferometers traditionally require very high amounts of data processing power, due to the need to correlate all baselines separately. However, the fact that DEX requires a very compact array can also lead to a significant reduction in data processing. The trick consists in arranging all antennas into a completely filled and regularly spaced rectangular array. \citet{Zheng2014} have shown that in such a situation, all the useful information can be retrieved by carrying out 2D (spatial) FFTs on successive chunks of the data stream. This approach reduces the data processing requirements significantly and is therefore adopted as the baseline for DEX. A further potential reduction may be achieved by distributing the computation such that individual sub-array hubs pre-process part of the FFT, at the cost of some additional noise.\\

\begin{figure*}[htb]
    \centering
    \includegraphics[width=1\linewidth]{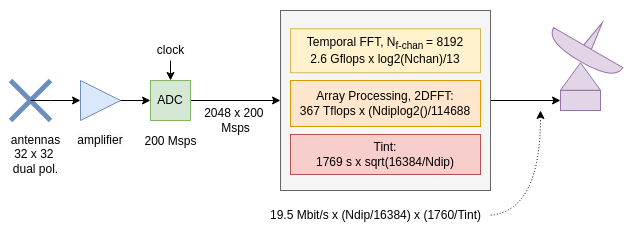}\\
    \caption{Signal transmission scheme, and data production rates at different stages of the data chain.}
    \label{Fig:Signal}
\end{figure*}

\noindent
The basic concept then is that ``snapshot'' images will be produced by FFT on the lunar surface and transmitted to Earth for further processing. During the CDF study, a detailed assessment of the data production yields rates are shown in Fig.~\ref{Fig:Signal}, including the processing power required.

For the minimum ($4\times4$ antenna) scenario, the requirements are:
\begin{itemize}
    \item 720 MB of science data storage per day.
    \item 102 GFlops of computational power.
    \item 0.6 kbit/s science data transmission to Earth.
\end{itemize}

\begin{figure*}[htb]
    \begin{center}
    \includegraphics[width=0.95\linewidth]{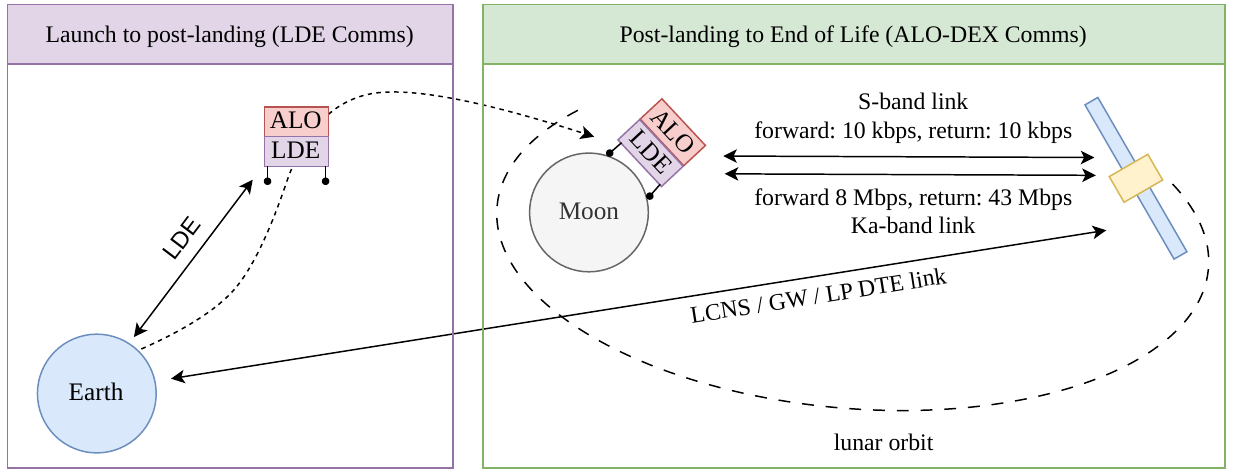}
    \caption{Scheme for communications between DEX and Earth, where LDE is the lunar descent element of Argonaut, LCNS stands for Lunar Communications and Navigation Services, GW stands for Gateway and LP stands for Lunar Pathfinder.}
    \label{Fig:Comms}
    \end{center}
\end{figure*}

These requirements are relatively straightforward to comply with using state-of-the-art equipment\footnote{The baseline to transmit data from each hub to the base station is to use current wireless transmission technology, though for the minimum 16-antenna system the hub is itself the base station. }. In particular, the scheme for communications with Earth could use one of three possible relay satellites which are already in development stages (Gateway, Lunar Pathfinder or Lunar Communications and Navigation Services), as illustrated in Fig.~\ref{Fig:Comms}.  

\subsection{CDF study conclusions}
\label{ss:study-concl}

The CDF study was extremely useful to identify all major requirements and constraints of a future DEX, and serves as a stepping stone towards future studies. An important second objective was to identify a ``minimum reasonable observatory" to be deployed with a single Argonaut launch. The result is a system comprising a $4\times4$ antenna array based on technology with high TRL. A summary of the configuration of this minimum system is shown in Tables~\ref{Table:4x4a} and \ref{Table:4x4b}.\\

\renewcommand{\pbA}{\parbox{3.9cm}}
\renewcommand{\pbB}{\parbox{5.9cm}}
\begin{center}
 \begin{table}[htb]
		\caption{Basic characteristics of the minimum $4\times4$ antenna system.}
		\begin{tabular}{ll}
		
            \toprule
		Characteristic &Value\\
  
            \midrule
		\emph{Mass}&\\
            \midrule
            \pbA{ALO Payload (incl. CPE$^1$)} & 984.6\,kg\\
            Antenna subsystem & 29.6\,kg\\
            Network subsystem & 58\,kg incl. 7.6\,kg of harness\\
            Base station subsystem & 503.5\,kg\\
            Rover & \pbB{393.4\,kg incl. 43.5\,kg antenna deployment arm}\\
            
            \midrule
		\pbA{\emph{Power}} & \\
            \midrule
            Night survival system demand & $\approx$225\,W\\
            Day operation system demand & $\approx$250\,W\\
            Power source & 8.1 m$^2$ Solar array, RFC$^2$, Battery, PCDU$^3$\\
            
            \midrule
            \emph{Observatory}&\\
           \midrule
            Antenna hubs & 16 items\\
            Magazine hub & 1 item\\
            Science Data Processor Unit & 1 item\\
            
            \midrule
            \emph{Data Handling}&\\
            \midrule
            Computing and Interfaces & \pbB{OBC$^4$+ RTU$^5$}\\
            Data Storage & SSMM$^6$ \\
            
            \midrule
            \emph{Communications}&\\
            \midrule
            Ka-band & 1$\times$HGA$^7$, 2$\times$Rx/Tx, 2$\times$TWTA$^8$, 1$\times$RFDN$^9$\\
            S-band & 2$\times$LGA$^{10}$, 2$\times$Rx/Tx, 2$\times$TWTA, 1$\times$RFDN\\
            
            On-ground wireless network &\\
            \bottomrule
		\end{tabular}\label{Table:4x4a}
  $^1$CPE: Cargo Platform Element, $^2$RFC: Regenerative Fuel Cell, $^3$PCDU: Power Conditioning and Distribution Unit, $^4$OBC: On-board Computer, $^5$RTU: Remote Terminal Unit, $^6$SSMM: Solid State Mass Memory, $^7$HGA: High Gain Antenna, $^8$TWTA: Traveling Wave Tube Amplifier, $^9$RFDN: Radio Frequency Distribution Network, $^{10}$LGA: Low Gain Antenna

\end{table}
\end{center}

\renewcommand{\pbA}{\parbox{2.8cm}}
\renewcommand{\pbB}{\parbox{8.0cm}}
\begin{center}
 \begin{table}[htb]

		\caption{Basic characteristics of the minimum $4\times4$ antenna system, mechanical and thermal systems.}
		\begin{tabular}{ll}
		
            \toprule
		Characteristic &Value\\
            \midrule
            \emph{Mechanisms and Robotics} &\\
            \midrule
            Mechanisms & \pbB{Solar array HDRMs$^1$ and hinges, HGA$^2$ Pointing Mechanism, Rover Unloading Mechanism}\vspace{2mm}\\
            Robotics & \pbB{Hub deployment arm}\\
            \midrule
            \emph{Structures}&\\
            \midrule
            \midrule
            Primary Structure (LDE$^3$ I/F)& \pbB{$\varnothing$2.121 m $\times$ 0.2 m CFRP$^4$ core / Al skin honeycomb sandwich, central cylinder with Al ring, CFRP core / Al skin honeycomb sandwich rover deck}\vspace{2mm}\\
            \pbA{Secondary structure (Rover garage)} & \pbB{CFRP core / AL skin honeycomb sandwich side walls Al core / Al skin honeycomb sandwich roof with shear support, NOTE: Deck extensions serves as rover driveway}\vspace{2mm}\\
            Support structures & \pbB{Robotic arm support, equipment I/Fs, brackets and inserts}\\
            \midrule
            \emph{Thermal Control} &\\
            \midrule
            Antennas & \pbB{MLI$^5$, heaters, washers, white paint/SSM$^6$/OSR$^7$}\vspace{1mm}\\
            Hub & \pbB{louvers, MLI, heaters, washers, white paint/SSM/OSR}\vspace{1mm}\\
            Bases Station & \pbB{louvers, MLI, heaters, washerss, white paint/SSM/OSR, heat pipes, loop heat pipes (TBC), RHU$^8$ (TBC)}\\
            \bottomrule
		\end{tabular}\label{Table:4x4b}
  $^1$HDRM: Hold-Down and Release Mechanism, $^2$HGA: High Gain Antenna, $^3$LDE: Lunar Descent Element, $^4$CFRP: Carbon Fibre Reinforced Polymers, $^5$MLI: Multi-Layer Insulation , $^6$SSM: Secondary Surface Mirrors, $^7$OSR: Optical Solar Reflectors, $^8$RHU: Radioisotopic Heater Unit

\end{table}	
\end{center}

\noindent
{However, this minimum system is clearly only capable of achieving the scientific objectives of one of the two experiments included in DEX, namely the Global Signal measurement. As shown in Section~\ref{sec:inst-cncpt}, the more ambitious Angular Power Spectrum experiment requires at least an order of 1000 antennas to become cosmologically interesting. An exercise to understand the consequences of scaling up the CDF baseline to a much larger number of antennas is summarized in Table~\ref{table:scaling}}.\\

\renewcommand{\pbA}{\parbox{2.0cm}}
\renewcommand{\pbB}{\parbox{1.7cm}}
\newcommand{\pbC}{\parbox{1.2cm}}
\begin{table}[htb]
\begin{centering}
\caption{Summary of the budget for scaling the DEX array size, as determined by the CDF study.}
    \begin{tabular}{lll lll}
        \toprule
	\pbA{Array size} & \pbB{Mass\\($\times 10^3$ kg)} & \pbB{Max. Power\\(kW)} & \pbB{Compute (Mflops)} & \pbB{Data Rate\\(kbits/s)} & \pbC{Number\\of EL3s}\\
        \midrule
        4$\times$4 (16)        & 0.98  & 0.25 & 102     & 0.6    & 1 \\
        8$\times$8 (64)        & 1.43  & 0.7  & 614     & 0.5    & $\approx$1 \\
        16$\times$16 (256)     & 3.2   & 2.7  & 3,277   & 40     & $\approx$2-3 \\
        32$\times$32 (1024)    & 10.5  & 11   & 16,384  & 300    & $\approx$7 \\
        64$\times$64 (4096)    & 39.7  & 42   & 78,643  & 2,400  & $\approx$25 \\
        128$\times$128 (16384) & 157.8 & 179  & 367,000 & 19,500 & $\approx$90\\
        \bottomrule
    \end{tabular}\label{table:scaling}
\end{centering}
\end{table}

\noindent
Table~\ref{table:scaling} shows clearly that the Angular Power Spectrum experiment will not be achievable if we assume the use of the same technologies as baselined for the minimum CDF system. Significant advances are required in all payload resources (mass, volume, power, and the rest), which we discuss in the following Section.

\clearpage
\section{Technology development needed for DEX}
\label{sec:technology}

The conclusions of the CDF study described in the previous Section highlight the fact that significant advances must be made in several key technologies to enable an effective version of DEX to be deployed to the surface of the Moon. The critical properties of each subsystem that require technological development are listed below:

\begin{itemize}
\item{The individual antennas need to have very low mass, be packaged in a compact configuration for transport, be deployed in a reliable manner onto the lunar surface, and have a good RF performance without the use of a ground plane on the lunar surface.}\\
\item{The low-noise receivers need to have very low power consumption during operation, to survive the lunar night with as little power draw as possible (preferably none), and have a very low mass and a form factor that allows for their integration into the packaged and deployed antennas.}\\
\item{The power distribution and data transmission systems need to minimise the use and mass of electrical harnessing, for instance by making use of optical or RF solutions.}\\
\item{The data processing units must have the capability to carry out the analog-to-digital conversion and computing needs of the system at low power consumption, both at sub-array and array levels.}\\
\end{itemize}

\noindent
The following subsections describe in more detail some of the most promising areas of technological development which are being investigated.

\subsection{Low-mass antennas}
\label{ss:low-mass-antennas}

Due to the large number of antennas needed to achieve the sensitivity requirements for the scientific goals, large but low-mass antennas are needed. The antennas must occupy a small volume during launch, but must expand to full size on the lunar surface. Multiple promising directions of research are identified, from shape memory alloy to inflatable antennas.\\

\noindent
Shape memory alloy (SMA), as the name implies, is an alloy that has the ability to be trained to a certain shape such as an antenna. After training, the alloy can be manipulated such that it is folded down for storage or in preparation for launch. By adding extra energy (e.g. by solar heating) and using the limited resistivity of the material, the SMA will return to its trained shape. Recent research conducted by \cite{Vertegaal2021} has shown the possibilities and limitations of using SMA as antenna material. However, SMA can only be used as a deployment mechanism as using it for the antennas themselves would make them too massive.\\

\noindent
The second identified type of antenna potentially suitable for DEX is the use of inflatable antennas. These antennas are made of lightweight material such as kapton, on which antennas are printed/deposited. They can be deployed as part of the inflation process. The use of inflatable antennas comes with multiple challenges, such as the inflation mechanism and limited lifetime. Due to the limited gravity and therefore also air density, only a small pressure difference is needed to inflate a structure. Using a sublimation process, a powder-efficient gas generation can be achieved. Due to the limited air supply, considerations should be made in terms of design and rigidization options such as UV-curing or SMA, as mentioned before. A specific challenge here is that the antennas need to be deployed over a particular surface area, i.e. in two dimensions. This makes any unrolling or unfolding process associated with inflatable structures a non-trivial exercise. In this context, the deployment of a full $32\times32$ array will likely need to be separated into the deployment of multiple sub-arrays. These sub-arrays may have a linear character, or may be smaller two-dimensional arrays of antennas themselves. The choice of how best to define these sub-arrays ties in closely with the topology of the power and data distribution systems, as well as the properties of the local lunar surface: because the deployment is particularly sensitive to any obstructions and terrain geometry variations, the deployment architecture will need to be adapted to these factors.\\

\noindent
In the CDF study, it was assumed that the basic antenna element considered for the ALO-DEX instrument is a cross-dipole. This is easily justified: not only is the dipole concept compact, light, has a high TRL, it is also very well understood and proven in theory, it is also very affordable. This all together makes it suitable for the deployment of large arrays in space.\\

\begin{figure}[htb]
    \centering
    \includegraphics[width=0.75\linewidth]{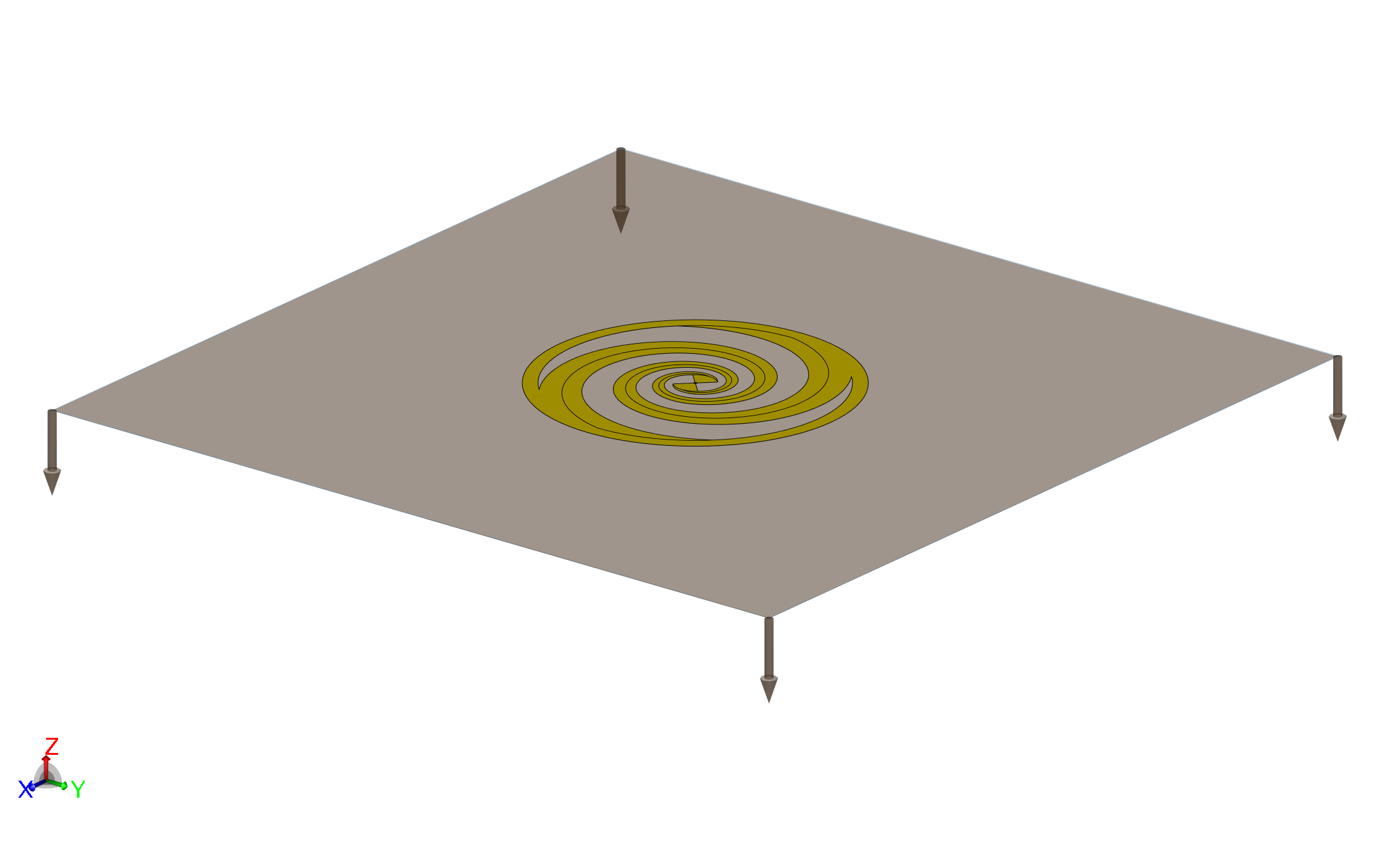}
    \caption{Equiangular spiral plate or log-spiral antenna simulation model on top of a dielectric lunar regolith as proposed in \cite{Zandboer2023}, representing outrigger stations aimed at observing the Global Signal.}
    \label{Fig:LogSpiralAntennaModel}
\end{figure}

The dipole, however, also has its downsides. Not only does it have a relatively narrow resonance bandwidth, it also becomes very long when used in the low MHz range. Therefore, the alternative is often to use non-resonant, electrically small antennas as, for example, presented in \cite{Arts2019} for the NCLE mission. However, the stand-alone global Dark Ages observation only requires a single antenna instead of an array. This reduces the requirements on size, weight and cost, opening up the possibility of non-dipole designs.\\

\noindent
With this idea in mind, internal research of \cite{Zandboer2023} and published in \cite{Zandboer2024} discusses the design considerations of an active antenna (i.e., a combined amplifier and antenna) that satisfies the requirements on frequency invariant characteristics and stability required to perform the stand-alone DEX science mission. For the array specifically, four antennas are considered in more detail based on a list of 21 options. This includes, for example, the antenna type used by EDGES (\cite{Mozdzen2016}), as well as the one proposed for the SKA (\cite{DeLeraAcedoRazaviGhods2015}).\\

\noindent
Due to its extremely broadband and frequency-invariant response, the equiangular spiral plate antenna (also called log-spiral antenna) was selected as the most promising antenna type, as shown in figure \ref{Fig:LogSpiralAntennaModel}. This is evident both in the simulations for radiation pattern and for impedance characteristics conducted by \cite{Zandboer2023}. Next to that, a parameter sweep shows that the log-spiral antenna is relatively insensitive to the unknown relative permittivity of the lunar regolith. Combined with its 2D planar shape, this means the antenna can be placed directly on the lunar regolith, allowing for easier deployment compared to other proposed (3D) structures.\\

\subsection{Power generation and distribution}

In the CDF study, the assumption was made that all power is made available at the location of the lander, and is to be distributed across the antenna array using electrical wiring. The mass associated with this harnessing was found to be problematically high (taking up a significant part of the total system mass budget), and the efficiency losses induced by the distances involved were found to be large as well. Two fundamental approaches can be considered to sidestep these issues: non-conductive power distribution and distributed power generation.\\

\noindent
Electrical harnessing requires insulation, a return path, and sufficient conductivity (i.e., conductor cross-section) to keep losses limited. This means that the mass of electrical harnessing quickly gets out of control for a DEX-sized array when power distribution from a central point is considered, as inferred from the CDF study. Alternative possibilities for electrical wiring are power-over-fiber solutions, where the electrical power needed at all LNAs is delivered through optical fiber and converted into electrical power locally, and free-space power distribution using RF or optical systems. The imperfect efficiency of any power conversion occurring close to the LNAs may actually be beneficial, since it can provide heating power during periods of low environmental temperature.\\

\noindent
The second way to limit electrical harnessing mass is to consider a system where electrical power is generated in a distributed fashion, at the sub-station level or even at the individual antenna level. For example, if antennas are printed on foils, it could be considered whether solar cells could also be printed on the remaining foil space. Here, the term sub-station refers to a set of $\sim16$ antennas that are deployed as a unit and in each others' vicinity. With this architecture, power distribution only happens on the small scales within a sub-station which also limits resistive losses. While harnessing mass can be dramatically reduced this way, this option also means that power conversion occurs much more close to the antennas. Therefore, care must be taken in the design to avoid having the power conversion subsystems generate RFI that may interfere with the measurements of the weak science signals.

\subsection{Electronics survivability and night operations}

A lunar day/night cycle lasts for approximately 29.5 Earth days. This means that any system located on the lunar surface is exposed to both high-temperature and low-temperature conditions, each for approximately two weeks at a time. While in free space heat transport can be facilitated between the warm side and the cold side of a 3-axis stabilised spacecraft in order to stabilise thermal conditions for the payload, the design of a lunar surface system needs to consider the use of thermal reservoirs and high-quality insulation to keep the payload temperature in the desired temperature range for operations or survival during these extended periods of extreme temperatures. In particular, the LNA hardware, being located close to the antenna terminals on the lunar surface and thus more exposed than any component in the body of the lander, needs to have a thermal management system that protects it from thermal shock (through insulation) and from the most extreme temperatures that the environment can take (through the use of a thermal reservoir). Even taking these measures into account, the temperature range that the electronics will reach is still much larger than is typically encountered in spacecraft payloads, and the selection and design of the involved components will need to account for this.\\

\noindent
Furthermore, the lunar night presents the most favourable conditions for taking radio measurements: the Sun (being one of the most prominent sources of radio emissions) is not present in the local sky and the low temperatures mean the noise levels contributed by the electronics may be reduced as well. Sustaining operations during lunar night-time requires energy storage with sufficient capacity to power DEX for at least part of the night, and to heat those components that need it. Batteries quickly inflate the mass budget for the system side of DEX, and so this is another reason that power consumption should be kept to the absolute minimum. Assuming a power draw per (dual-polarisation) antenna element of 0.2\,W, the operation of the full DEX array throughout the lunar night requires approximately 70\,kWh of capacity (excluding digital data processing). With a fiducial battery energy density of 200\,Wh/kg, this means that 350\,kg of batteries are needed across the array, which is a significant fraction of the system mass budget. Alternatively, battery capacity can be scaled so that array operation is possible only during the first part of the night.\\

\subsection{Road map for the nearby future: PRE-DEX and other stepping stones toward DEX}
\label{s:road-map}

Following the ESA CDF study, a pre-phase A concept study dubbed ``PRE-DEX"\footnote{PRE-DEX - EL3 Polar Explorer Radio Antenna Payload Pre-Phase A Study, ESA-ESTEC contract 4000135880/21/NL/AT}, was carried out aiming at a top-level design for a precursor antenna system for ALO-DEX. The baseline design was based on NCLE (Karapakula et al (2024)), a 10\,kg low-frequency instrument, deployed on the Chang’E 4 relay satellite {\it Queqiao} at the Earth-Moon L2 point. The design of PRE-DEX is adapted to meet the requirements of the Argonaut (previously called EL3) lander.\\

\noindent
The scientific objective of PRE-DEX is to conduct astrophysical studies from the lunar surface as a precursor to a cosmological Dark Ages and Cosmic Dawn mission, including measuring and characterising astrophysical radio sources at the low frequencies. It will also characterise the influence of the local lunar ionosphere on low-frequency radio measurements, and will monitor the lunar radio background spectrum over short (minutes to hours) and long (days to months) timescales to assess its temporal variability characteristics. This includes effects from cosmic rays, dust impacts, RFI, and Auroral kilometric radiation.\\

\subsubsection{High-TRL option}

\noindent
A high-TRL option for PRE-DEX includes two orthogonal 5\,m tip-to-tip horizontal dipoles and a 2.5 m length vertical monopole, to be placed on top of the lander. As PRE-DEX will be subject to a larger temperature range than NCLE, the carbon fiber reinforced polymer antennas used for NCLE will need to be replaced for example by beryllium-copper alloy antennas. The targeted frequency range is that of NCLE, 80\,kHz to 80\,MHz. Simulated antenna performance indicate the system is limited by the sky noise in the range $7 - 70$\,MHz. The fundamental mode of operations for PRE-DEX is envisioned to consist of performing cross-correlation measurements between the three antennas to offer all-sky sensitivity to all modes of incoming polarised radio waves. In addition to this, a separate recording mode records time-domain voltages for all antennas. This mode, while addressing specific science cases, is also invaluable as a diagnostic tool to fully understand the instrument response in the lunar environment. Specifically, the influence of possible dust buildup on the antenna elements from electrostatic effects and the influence of cosmic ray interactions in the regolith are phenomena that are considered of interest to future surface radio instruments. These effects are yet to be quantified.\\

\noindent
This PRE-DEX architecture involves an external subsystem-enclosure supporting antennas and LNA, and an enclosure hosting the analogue to digital converters,  subject to a $-110$ to 150$^\circ$C temperature range.\\

\noindent
System characteristics include a mass of 10.1\,kg (allowing 20\% margin), science-mode power consumption under 27\,W (allowing 20\% margin), and a fiducial data budget of 1.8\,GByte per 24 hours. The standard volume of housekeeping data is estimated to be 1\,kByte per minute, and is stored alongside the science data. Initial designs require allowable temperature ranges of $-110$ to 150\,$^\circ$C for external instrument components.

\subsubsection{Lower-TRL option}

\begin{figure*}[htb]
    \centering
    \includegraphics[width=1\linewidth]{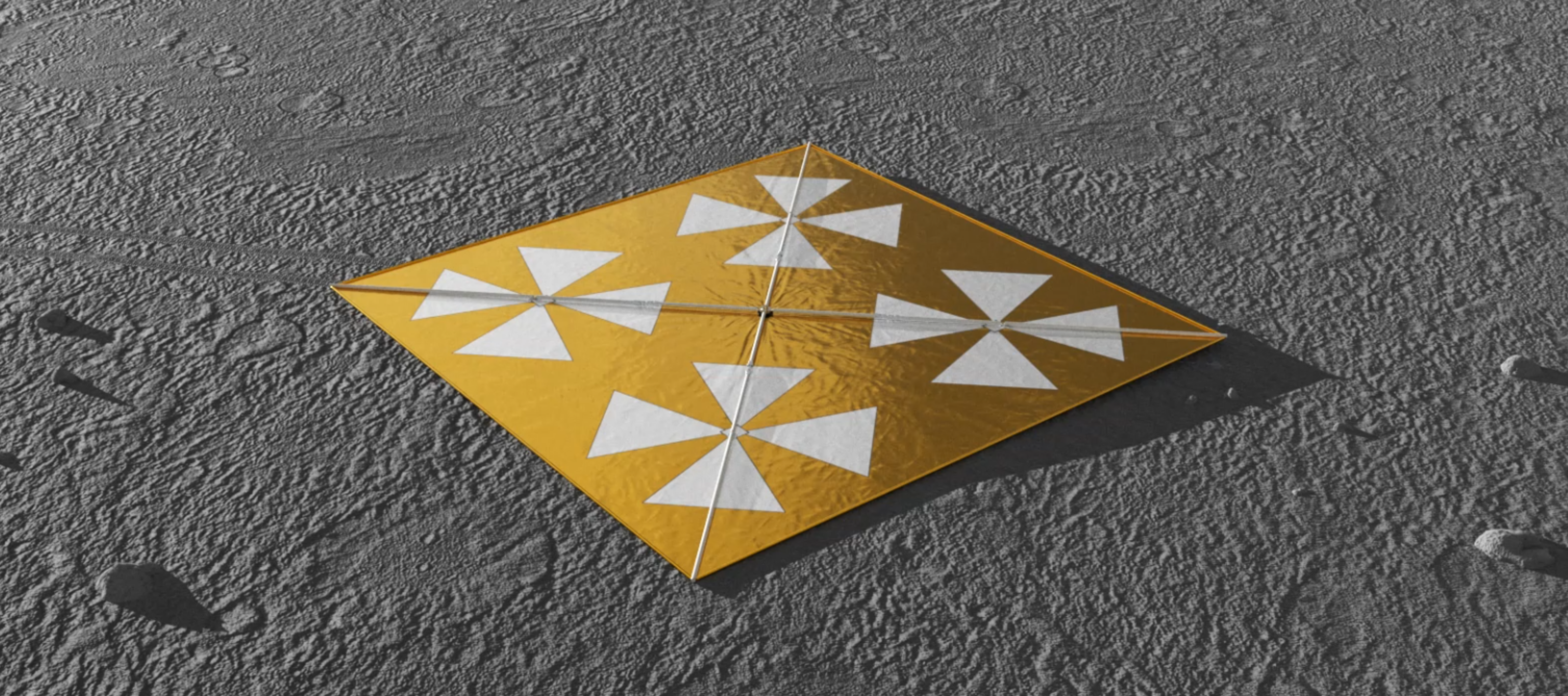}
    \caption{One PRE-DEX four-dual-polarisation antenna cluster deployed on the lunar surface, envisioned to be scaled to $16\times16$ of these modules to form a $32\times32$ array with dual-polarization for the Global Signal measurement research. Image credit: ATG Europe.}
    \label{Fig:PreDEX}
\end{figure*}

\begin{figure*}[htb]
    \centering
    \includegraphics[width=1\linewidth]{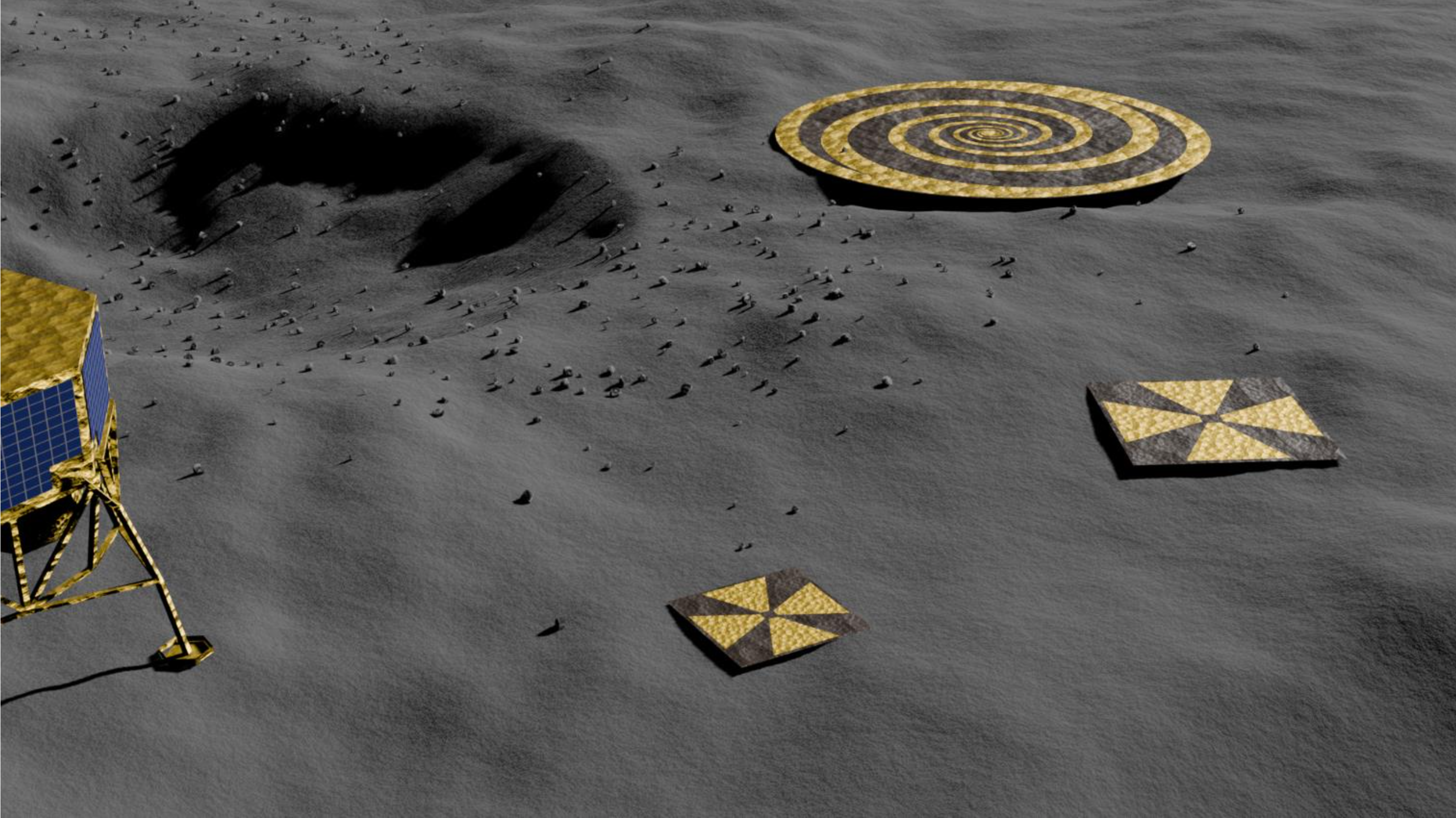}
    \caption{A simplified representation of the GloDEX instrument, showing multiple antennas of different designs are used in conjunction. Distance scales have been compressed in this image.}
    \label{Fig:GloDEX}
\end{figure*}

The high-TRL option for PRE-DEX, described above, can be developed relatively quickly as the technology is mature (including flight heritage for several of its subsystems). However, it does not scale well to a full DEX array, because the instrument is situated on top of the lander and the mass per set of antennas is several kilograms. For this reason, a lower-TRL but more scaleable option for PRE-DEX was studied as well. This option makes use of a low-mass inflatable deployment system and foil-based antennas, to be deployed directly onto the lunar surface instead being mounted atop the lander platform. This option uses a small array of $2\times2$ dual-polarisation antenna elements, as illustrated in Fig.~\ref{Fig:PreDEX}. With this configuration, cross-correlation of the signals from antennas in different locations is included as an observational mode, and both the observing mode and the type of data product of the full DEX array can be tested. This design does require more development effort for the thermal management of the LNAs, located at the antenna terminals, as they will be subject to a large range of environmental temperatures without the benefit of being shielded in a larger structure. Furthermore, an instrument concept dedicated to studying the global neutral hydrogen signal (GloDEX, see Figure \ref{Fig:GloDEX}) has been formulated, which makes use of technologies that are relevant to DEX as well. This concept consists of multiple spatially isolated foil-based antennas of different design, of which the measurements are to be cross-checked to help separate antenna and instrument response from astrophysical spectral signatures.\\

\subsection{DEX spin-off technologies for multi-purpose applications}


The technological developments envisioned for DEX experiment hold significant potential for spin-offs in other lunar-surface activities and terrestrial applications. Advances in low-mass, deployable antennas, such as inflatable structures and shape memory alloys, provide highly compact solutions suitable for rapid terrestrial deployment scenarios, including disaster-relief communications, remote environmental monitoring, and portable satellite ground stations. Additionally, robust temperature-tolerant electronics designed to withstand extreme thermal variations of the lunar surface can directly enhance the durability and reliability of electronics in terrestrial harsh environments, benefiting deep-sea exploration equipment, polar research instrumentation, and critical technologies operating in extreme climates. The low-power distributed data-processing architectures required for efficient lunar operation can similarly advance terrestrial edge computing networks, IoT sensor arrays, and autonomous robotic platforms, enabling effective data handling under stringent power constraints. Furthermore, innovative approaches in communication architectures, including RF-based wireless solutions designed to minimize wiring mass and power consumption, are highly applicable to terrestrial communication systems, particularly in remote, rugged or constrained environments, such as underground mining operations, deep-space ground communication arrays, and rural broadband deployment. Together, these technological innovations to be developed for DEX represent valuable opportunities to drive multi-purpose advancements across lunar and Earth-based applications.

\subsection{Lunar far side electromagnetic environment}

\noindent
The attractiveness of the Moon's far side for astronomical research, in particular in the radio domain of the electromagnetic spectrum, is essentially based on shielding from Earth-originated RFI. However, a massive drive toward exploration of the Moon by many space agencies and private enterprises might bring emitters of RFI to the Moon's far side or in the translunar space above the far side. Making the lunar far side or a reasonable volume above its surface completely radio quiet is unrealistic and in fact impractical. Even radio astronomy facilities themselves on or above the far side will need to communicate their data to the processing center and further to Earth. However, a well thoughtout approach to the use of EM spectrum on and above the lunar far side is possible if implemented along guidelines that have been developed by the International Astronomical Union (IAU) and the International Telecommunications Union (ITU) in the past decades.\\
\begin{figure*}[htb]
    \centering
    \includegraphics[width=0.9\linewidth]{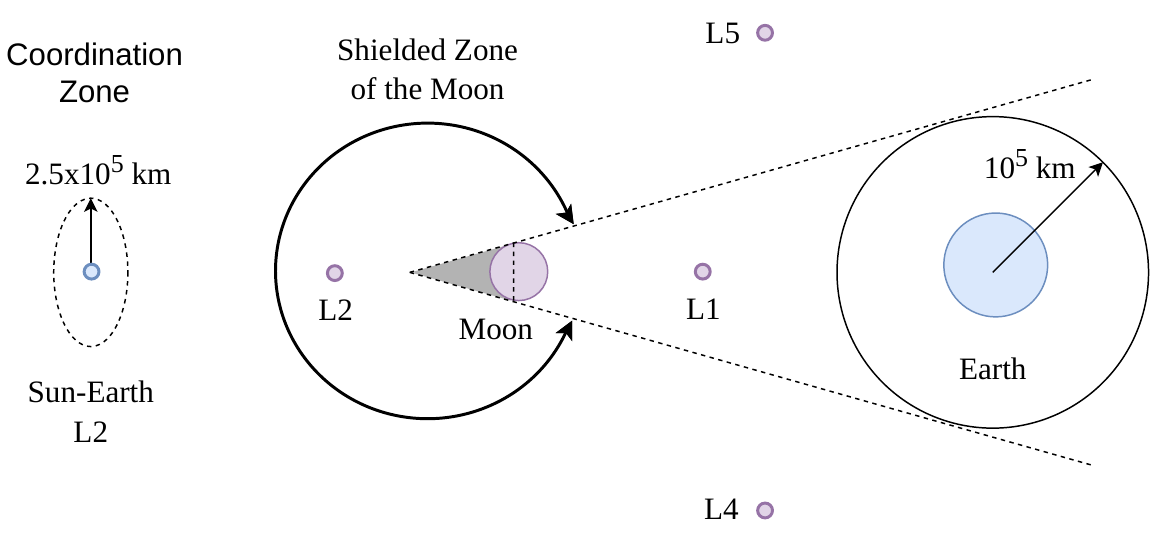}
    \caption{Schematic depiction of the Shielded Zone of the Moon (SZM), once a month aligned with the Sun-Earth L2 point (not to scale).}
    \label{Fig:SZM}
\end{figure*}

\noindent
Concerning the Shielded Zone of the Moon (SZM), the ITU Radio Regulations \cite{ITU-RR} state in article 22, section V \cite{ITU-R-479}: ``The SZM comprises the area of the Moon's surface and an adjacent  volume of space which are shielded from emissions originating within a distance of 100\,000\,km from the center of the Earth''. It also states that emissions causing harmful interference to passive services in the SZM shall be prohibited in the entire frequency spectrum except in certain bands related to space research and Earth exploration satellite services. This includes using active sensors, radio location, radio communications, and space research transmissions. Figure~\ref{Fig:SZM} schematically shows the SZM together with the Earth-Moon Lagrangian points depicted relative to the SZM, and with the Sun-Earth L2 point \cite{ITU-R-1417} with which it is aligned once a month.\\

\noindent
The IAU recommends necessary transmissions avoiding bands of great astronomical importance \cite{IAU-B15}, including all frequencies below 2\,GHz. It also recommends two alternative band(s) allocated on a time-coordination basis to retain access by the passive services to the entire radio spectrum. The ITU recommends in the Radio Regulations and Recommendation RA.479-5 ``Protection of frequencies for radio astronomical measurements in the shielded zone of the Moon'' that in radio spectrum planning ``account be taken of the need to provide for radio astronomy observations in the SZM''. The recommendation details preliminary guidelines for this for frequency bands up to 1\,THz.\\

\noindent
In summary, the IAU and ITU recommendations and regulation show the need to protect the bands below 2\,GHz in the SZM. In addition, guidelines are provided for agencies and industries to make sure provisions are taken into account regarding the SZM radio astronomy protection for bands up to 1\,THz. However, there are two aspects not covered by the Radio Regulations that may hamper radio astronomy observations: blocking and unintended electromagnetic radiation. Blocking is the effect that a receiver system is driven into non-linear behaviour due to a strong spatially-nearby transmitter in an adjacent band. Typically, receivers are expected to be designed with sufficient steep filters to avoid this from happening. However, when a new unexpected allocation and/or usage occurs, this may happen, and given the remoteness of the location, would incur loss of observational bandwidth or incur great cost in remedying it. A second effect is (cumulative) interference from unintended electromagnetic radiation from satellites and satellite constellations, such as observed with terrestrial low-frequency instruments such as LOFAR \cite{DiVruno2023}.\\

\noindent
As generic regulations and recommendations are in place, governmental agencies, and active and passive services need to participate in consultation processes, nationally and internationally, when planning the use of radio spectrum, as recommended in \cite{ITU-R-479} and confirmed in its latest edition of 2023.\\ 

\noindent
Considering that Moon-to-Earth and lunar Moon-to-space communication may be needed for scientific or technology research conducted on the lunar surface, and noting SZM protection as mentioned by ITU-R RA.479-5, Resolution 680 of the 2023 World Radiocommunication Conference (WRC-23) has been brought forward. It invites the ITU radio communications sector to conduct compatibility studies in preparation for WRC-27. These studies include spectrum needs, system characteristics, and SZM protection. Also the impact of unintended electromagnetic radiation (cf. Question ITU-R 243/1) is considered for inclusion. Currently, the ITU Working Party 7D (WP-7D) is drafting a report on unintended electromagnetic radiation from space systems into Radio Astronomy Service (RAS) frequency bands.\\

\clearpage
\section{Conclusion}
\label{sec:conclusion}

Active exploration of the Moon is high on the agenda of major space agencies and many leading private enterprises. It is certain that in the next several decades a set of human and robotic bases will appear in various locations on the lunar surface. They will address a broad range of operations, among which scientific studies of the Moon and from the Moon will be very prominent. We argue that in the wide spectrum of research activities on the Moon, radio astronomy, more specifically ultra-long-wavelength  astronomy (ULWA) must occupy a significant place defined by the scientific significance of the expected science return. We also argue that ULWA studies from the Moon's far side are very likely to deliver breakthrough scientific results.\\

\noindent
For the Dark Ages and the Cosmic Dawn science cases, the ESA CDF study \cite{ESA2021} assessed the feasibility of a low-frequency array radio telescope on the lunar far side using available high-TRL technologies. Given the extreme lunar environment and limited mass and power budgets, it was inferred that a limited $4\times4$ antenna array would fit the constraints. These constraints include the assumption that the mission should fit within one Argonaut lander mission. Scaling up the array size would require developing lightweight antennas, low-power and low-mass integrated receiver electronics, and data transport and synchronisation systems, able to survive extreme environmental conditions on the Moon. Some of these developments can have a multi-purpose use, such as the development of low-power and low-mass technologies, which would be also beneficial for terrestrial applications.

\bibliography{DEX-ALO-references}  

\begin{thebibliography}{117}
\providecommand{\natexlab}[1]{#1}
\providecommand{\url}[1]{{#1}}
\providecommand{\urlprefix}{URL }
\providecommand{\doi}[1]{\url{https://doi.org/#1}}
\providecommand{\eprint}[2][]{\url{#2}}
 \bibcommenthead

\bibitem[{{Alexander} et~al(1969){Alexander}, {Brown}, {Clark}, {Stone}, and {Weber}}]{Alexander+1969ApJL}
{Alexander} JK, {Brown} LW, {Clark} TA, et~al (1969) {The Spectrum of the Cosmic Radio Background Between 0.4 and 6.5 MHz}. ApJL 157:L163. \doi{10.1086/180411}

\bibitem[{{Alexander} et~al(1975){Alexander}, {Kaiser}, {Novaco}, {Grena}, and {Weber}}]{Alexander+1975}
{Alexander} JK, {Kaiser} ML, {Novaco} JC, et~al (1975) {Scientific instrumentation of the Radio-Astronomy-Explorer-2 satellite.} Astron and Astrophys 40(4):365--371

\bibitem[{Arts et~al(2019)Arts, Prinsloo, Ruiter, and Boonstra}]{Arts2019}
Arts MJ, Prinsloo DS, Ruiter M, et~al (2019) {Design of a reconfigurable array of monopoles for the Netherlands China Low-frequency Explorer}. European Conference on Antennas and Propagation

\bibitem[{Artuc and de~Lera~Acedo(2024)}]{artuc2024spectrometer}
Artuc K, de~Lera~Acedo E (2024) The spectrometer development of cosmocube, lunar orbiting satellite to detect 21-cm hydrogen signal from cosmic dark ages. RASTI {\href{https://arxiv.org/abs/2406.10096}{{arXiv:2406.10096}}}

\bibitem[{{Bale} et~al(2023){Bale}, {Bassett}, {Burns}, {Dorigo Jones}, {Goetz}, {Hellum-Bye}, {Hermann}, {Hibbard}, {Maksimovic}, {McLean}, {Monsalve}, {O'Connor}, {Parsons}, {Pulupa}, {Pund}, {Rapetti}, {Rotermund}, {Saliwanchik}, {Slosar}, {Sundkvist}, and {Suzuki}}]{LuSEE-2023arXiv}
{Bale} SD, {Bassett} N, {Burns} JO, et~al (2023) {LuSEE 'Night': The Lunar Surface Electromagnetics Experiment}. arXiv e-prints arXiv:2301.10345. \doi{10.48550/arXiv.2301.10345}, {\href{https://arxiv.org/abs/2301.10345}{{arXiv:2301.10345}}} {[astro-ph.IM]}

\bibitem[{{Bassa} et~al(2017){Bassa}, {Pleunis}, and {Hessels}}]{Bassa+2017AC}
{Bassa} CG, {Pleunis} Z, {Hessels} JWT (2017) {Enabling pulsar and fast transient searches using coherent dedispersion}. Astronomy and Computing 18:40--46. \doi{10.1016/j.ascom.2017.01.004}, {\href{https://arxiv.org/abs/1607.00909}{{arXiv:1607.00909}}} {[astro-ph.IM]}

\bibitem[{{Bassett} et~al(2023){Bassett}, {Rapetti}, {Nhan}, {Page}, {Burns}, {Pulupa}, and {Bale}}]{Bassett2023}
{Bassett} N, {Rapetti} D, {Nhan} BD, et~al (2023) {Constraining a Model of the Radio Sky below 6 MHz Using the Parker Solar Probe/FIELDS Instrument in Preparation for Upcoming Lunar-based Experiments}. \apj 945(2):134. \doi{10.3847/1538-4357/acbc76}, {\href{https://arxiv.org/abs/2301.09612}{{arXiv:2301.09612}}} {[astro-ph.GA]}

\bibitem[{{Bastian} et~al(2001){Bastian}, {Pick}, {Kerdraon}, {Maia}, and {Vourlidas}}]{Bastian2001}
{Bastian} TS, {Pick} M, {Kerdraon} A, et~al (2001) The coronal mass ejection of 1998 april 20: Direct imaging at radio wavelengths. The Astrophysical Journal Letters 558:L65--L69

\bibitem[{{Belov} et~al(2018){Belov}, {Branch}, {Broschart}, {Castillo-Rogez}, {Chien}, {Clare}, {Dengler}, {Gao}, {Garza}, {Hegedus}, {Hernandez}, {Herzig}, {Imken}, {Kim}, {Mandutianu}, {Romero-Wolf}, {Schaffer}, {Troesch}, {Wyatt}, and {Lazio}}]{Belov+2018}
{Belov} K, {Branch} A, {Broschart} S, et~al (2018) {A space-based decametric wavelength radio telescope concept}. Experimental Astronomy 46(2):241--284. \doi{10.1007/s10686-018-9601-6}, {\href{https://arxiv.org/abs/1904.00119}{{arXiv:1904.00119}}} {[astro-ph.IM]}

\bibitem[{{Bely} et~al(1997){Bely}, {Laurance}, {Volonte}, and {et al.}}]{Bely-ESA-sci1997}
{Bely} PY, {Laurance} RJ, {Volonte} S, et~al (1997) {Very Low Frequency Array on the Lunar Far Side. Report by the Very Low Frequency Astronomy Study Team.} Tech. Rep. ESA SCI(97)2, {European Space Agency}

\bibitem[{Bentum et~al(2009)Bentum, Verhoeven, and Boonstra}]{bentum2009olfar}
Bentum MJ, Verhoeven C, Boonstra AJ (2009) Olfar-orbiting low frequency antennas for radio astronomy. In: Proceedings of the ProRISC 2009, Annual Workshop on Circuits, Systems and Signal Processing, Veldhoven, pp 1--6

\bibitem[{{Bentum} et~al(2020){Bentum}, {Verma}, {Rajan}, {Boonstra}, {Verhoeven}, {Gill}, {van der Veen}, {Falcke}, {Wolt}, {Monna}, {Engelen}, {Rotteveel}, and {Gurvits}}]{Bentum+2020}
{Bentum} MJ, {Verma} MK, {Rajan} RT, et~al (2020) {A roadmap towards a space-based radio telescope for ultra-low frequency radio astronomy}. Advances in Space Research 65(2):856--867. \doi{10.1016/j.asr.2019.09.007}, {\href{https://arxiv.org/abs/1909.08951}{{arXiv:1909.08951}}} {[astro-ph.IM]}

\bibitem[{Bevins et~al(2022)Bevins, de Lera Acedo, Fialkov, Handley, Singh, Subrahmanyan, and Barkana}]{Bevins2022}
Bevins HTJ, de Lera Acedo E, Fialkov A, et~al (2022) A comprehensive bayesian reanalysis of the saras2 data from the epoch of reionization. Monthly Notices of the Royal Astronomical Society 513(3):4507–4526. \doi{10.1093/mnras/stac1158}, \urlprefix\url{http://dx.doi.org/10.1093/mnras/stac1158}

\bibitem[{{Bevins} et~al(2024){Bevins}, {Heimersheim}, {Abril-Cabezas}, {Fialkov}, {de Lera Acedo}, {Handley}, {Singh}, and {Barkana}}]{Bevins2024}
{Bevins} HTJ, {Heimersheim} S, {Abril-Cabezas} I, et~al (2024) {Joint analysis constraints on the physics of the first galaxies with low-frequency radio astronomy data}. \mnras 527(1):813--827. \doi{10.1093/mnras/stad3194}, {\href{https://arxiv.org/abs/2301.03298}{{arXiv:2301.03298}}} {[astro-ph.CO]}

\bibitem[{{Bhandari} et~al(2018){Bhandari}, {Keane}, {Barr}, {Jameson}, {Petroff}, {Johnston}, {Bailes}, {Bhat}, {Burgay}, {Burke-Spolaor}, {Caleb}, {Eatough}, {Flynn}, {Green}, {Jankowski}, {Kramer}, {Krishnan}, {Morello}, {Possenti}, {Stappers}, {Tiburzi}, {van Straten}, {Andreoni}, {Butterley}, {Chandra}, {Cooke}, {Corongiu}, {Coward}, {Dhillon}, {Dodson}, {Hardy}, {Howell}, {Jaroenjittichai}, {Klotz}, {Littlefair}, {Marsh}, {Mickaliger}, {Muxlow}, {Perrodin}, {Pritchard}, {Sawangwit}, {Terai}, {Tominaga}, {Torne}, {Totani}, {Trois}, {Turpin}, {Niino}, {Wilson}, {Albert}, {Andr{\'e}}, {Anghinolfi}, {Anton}, {Ardid}, {Aubert}, {Avgitas}, {Baret}, {Barrios-Mart{\'\i}}, {Basa}, {Belhorma}, {Bertin}, {Biagi}, {Bormuth}, {Bourret}, {Bouwhuis}, {Br{\^a}nza{\c{s}}}, {Bruijn}, {Brunner}, {Busto}, {Capone}, {Caramete}, {Carr}, {Celli}, {Moursli}, {Chiarusi}, {Circella}, {Coelho}, {Coleiro}, {Coniglione}, {Costantini}, {Coyle}, {Creusot}, {D{\'\i}az}, {Deschamps}, {De Bonis}, {Distefano}, {Palma}, {Domi}, {Donzaud},
  {Dornic}, {Drouhin}, {Eberl}, {Bojaddaini}, {Khayati}, {Els{\"a}sser}, {Enzenh{\"o}fer}, {Ettahiri}, {Fassi}, {Felis}, {Fusco}, {Gay}, {Giordano}, {Glotin}, {Gregoire}, {Gracia-Ruiz}, {Graf}, {Hallmann}, {van Haren}, {Heijboer}, {Hello}, {Hern{\'a}ndez-Rey}, {H{\"o}{\ss}l}, {Hofest{\"a}dt}, {Hugon}, {Illuminati}, {James}, {de Jong}, {Jongen}, {Kadler}, {Kalekin}, {Katz}, {Kie{\ss}ling}, {Kouchner}, {Kreter}, {Kreykenbohm}, {Kulikovskiy}, {Lachaud}, {Lahmann}, {Lef{\`e}vre}, {Leonora}, {Loucatos}, {Marcelin}, {Margiotta}, {Marinelli}, {Mart{\'\i}nez-Mora}, {Mele}, {Melis}, {Michael}, {Migliozzi}, {Moussa}, {Navas}, {Nezri}, {Organokov}, {P{\v{a}}v{\v{a}}la{\c{s}}}, {Pellegrino}, {Perrina}, {Piattelli}, {Popa}, {Pradier}, {Quinn}, {Racca}, {Riccobene}, {S{\'a}nchez-Losa}, {Salda{\~n}a}, {Salvadori}, {Samtleben}, {Sanguineti}, {Sapienza}, {Sch{\"u}ssler}, {Sieger}, {Spurio}, {Stolarczyk}, {Taiuti}, {Tayalati}, {Trovato}, {Turpin}, {T{\"o}nnis}, {Vallage}, {Van Elewyck}, {Versari}, {Vivolo}, {Vizzocca},
  {Wilms}, {Zornoza}, and {Z{\'u}{\~n}iga}}]{Bhandari2018}
{Bhandari} S, {Keane} EF, {Barr} ED, et~al (2018) {The SUrvey for Pulsars and Extragalactic Radio Bursts - II. New FRB discoveries and their follow-up}. \mnras 475(2):1427--1446. \doi{10.1093/mnras/stx3074}, {\href{https://arxiv.org/abs/1711.08110}{{arXiv:1711.08110}}} {[astro-ph.HE]}

\bibitem[{{Boonstra} et~al(2016){Boonstra}, {Garrett}, {Kruithof}, {Wise}, {van Ardenne}, {Yan}, {Wu}, {Zheng}, Eberhard K. A.~{Gill}, {Guo}, {Bentum}, {Girard}, {Hong}, {An}, {Falcke}, {Klein-Wolt}, {Wu}, {Chen}, {Koopmans}, {Rothkaehl}, {Chen}, {Huang}, {Chen}, {Gurvits}, {Zarka}, {Cecconi}, and {de Haan}}]{Boonstra+2016}
{Boonstra} AJ, {Garrett} M, {Kruithof} G, et~al (2016) {Discovering the Sky at the Longest Wavelengths (DSL)}. {IAAA Aerospace Conference} 1:20. \doi{10.1109/AERO.2016.7500678}

\bibitem[{{Bowman} et~al(2018){Bowman}, {Rogers}, {Monsalve}, {Mozdzen}, and {Mahesh}}]{Bowman2018}
{Bowman} JD, {Rogers} AEE, {Monsalve} RA, et~al (2018) {An absorption profile centred at 78 megahertz in the sky-averaged spectrum}. \nat 555(7694):67--70. \doi{10.1038/nature25792}, {\href{https://arxiv.org/abs/1810.05912}{{arXiv:1810.05912}}} {[astro-ph.CO]}

\bibitem[{{Brown}(1973)}]{Brown1973}
{Brown} LW (1973) {The Galactic Radio Spectrum Between 130 and 2600 kHz}. Astrophys J 180:359--370. \doi{10.1086/151968}

\bibitem[{{Burns} et~al(2017){Burns}, {Bradley}, {Tauscher}, {Furlanetto}, {Mirocha}, {Monsalve}, {Rapetti}, {Purcell}, {Newell}, {Draper}, {MacDowall}, {Bowman}, {Nhan}, {Wollack}, {Fialkov}, {Jones}, {Kasper}, {Loeb}, {Datta}, {Pritchard}, {Switzer}, and {Bicay}}]{Burns+2017}
{Burns} JO, {Bradley} R, {Tauscher} K, et~al (2017) {A Space-based Observational Strategy for Characterizing the First Stars and Galaxies Using the Redshifted 21 cm Global Spectrum}. The Astrophysical Journal 844(1):33. \doi{10.3847/1538-4357/aa77f4}, {\href{https://arxiv.org/abs/1704.02651}{{arXiv:1704.02651}}} {[astro-ph.IM]}

\bibitem[{{Callingham} et~al(2024){Callingham}, {Pope}, {Kavanagh}, {Bellotti}, {Daley-Yates}, {Damasso}, {Grie{\ss}meier}, {G{\"u}del}, {G{\"u}nther}, {Kao}, {Klein}, {Mahadevan}, {Morin}, {Nichols}, {Osten}, {P{\'e}rez-Torres}, {Pineda}, {Rigney}, {Saur}, {Stef{\'a}nsson}, {Turner}, {Vedantham}, {Vidotto}, {Villadsen}, and {Zarka}}]{2024NatAs...8.1359C}
{Callingham} JR, {Pope} BJS, {Kavanagh} RD, et~al (2024) {Radio signatures of star-planet interactions, exoplanets and space weather}. Nature Astronomy 8:1359--1372. \doi{10.1038/s41550-024-02405-6}, {\href{https://arxiv.org/abs/2409.15507}{{arXiv:2409.15507}}} {[astro-ph.EP]}

\bibitem[{{Carniani} et~al(2024){Carniani}, {Hainline}, {D'Eugenio}, {Eisenstein}, {Jakobsen}, {Witstok}, {Johnson}, {Chevallard}, {Maiolino}, {Helton}, {Willott}, {Robertson}, {Alberts}, {Arribas}, {Baker}, {Bhatawdekar}, {Boyett}, {Bunker}, {Cameron}, {Cargile}, {Charlot}, {Curti}, {Curtis-Lake}, {Egami}, {Giardino}, {Isaak}, {Ji}, {Jones}, {Kumari}, {Maseda}, {Parlanti}, {P{\'e}rez-Gonz{\'a}lez}, {Rawle}, {Rieke}, {Rieke}, {Del Pino}, {Saxena}, {Scholtz}, {Smit}, {Sun}, {Tacchella}, {{\"U}bler}, {Venturi}, {Williams}, and {Willmer}}]{Carniani2024}
{Carniani} S, {Hainline} K, {D'Eugenio} F, et~al (2024) {Spectroscopic confirmation of two luminous galaxies at a redshift of 14}. \nat 633(8029):318--322. \doi{10.1038/s41586-024-07860-9}, {\href{https://arxiv.org/abs/2405.18485}{{arXiv:2405.18485}}} {[astro-ph.GA]}

\bibitem[{Cecconi et~al(2018)Cecconi, Dekkali, Briand, Segret, Girard, Laurens, Lamy, Valat, Delpech, Bruno, Gelard, Bucher, Nenon, Griesmeier, Boonstra, and Bentum}]{Cecconi:2018ee}
Cecconi B, Dekkali M, Briand C, et~al (2018) {NOIRE study report: Towards a low frequency radio interferometer in space}. 2018 IEEE Aerospace Conference pp 1 -- 19. \doi{10.1109/aero.2018.8396742}

\bibitem[{{Chen} et~al(2019){Chen}, {Burns}, {Koopmans}, {Rothkaehl}, {Silk}, {Wu}, {Boonstra}, {Cecconi}, {Chiang}, {Chen}, {Deng}, {Falanga}, {Falcke}, {Fan}, {Fang}, {Fialkov}, {Gurvits}, {Ji}, {Kasper}, {Li}, {Mao}, {Mckinley}, {Monsalve}, {Peterson}, {Ping}, {Subrahmanyan}, {Vedantham}, {Klein Wolt}, {Wu}, {Xu}, {Yan}, and {Yue}}]{Chen+2019}
{Chen} X, {Burns} J, {Koopmans} L, et~al (2019) {Discovering the Sky at the Longest Wavelengths with Small Satellite Constellations}. arXiv e-prints arXiv:1907.10853. {\href{https://arxiv.org/abs/1907.10853}{{arXiv:1907.10853}}} {[astro-ph.IM]}

\bibitem[{{Chen} et~al(2021){Chen}, {Yan}, {Deng}, {Wu}, {Wu}, {Xu}, and {Zhou}}]{Chen2021}
{Chen} X, {Yan} J, {Deng} L, et~al (2021) {Discovering the sky at the longest wavelengths with a lunar orbit array}. Philosophical Transactions of the Royal Society of London Series A 379(2188):20190566. \doi{10.1098/rsta.2019.0566}, {\href{https://arxiv.org/abs/2007.15794}{{arXiv:2007.15794}}} {[astro-ph.IM]}

\bibitem[{{Chen} et~al(2024){Chen}, {Gao}, {Wu}, {Zhang}, {Wang}, {Liu}, {Zou}, {Deng}, {Gong}, {He}, {Li}, {Sun}, {Suo}, {Wang}, {Wu}, {Xu}, {Xu}, {Yue}, {Zhang}, {Zhou}, {Zhou}, {Zhu}, and {Zhu}}]{Chen2024}
{Chen} X, {Gao} F, {Wu} F, et~al (2024) {Large-scale Array for Radio Astronomy on the Farside}. arXiv e-prints arXiv:2403.16409. \doi{10.48550/arXiv.2403.16409}, {\href{https://arxiv.org/abs/2403.16409}{{arXiv:2403.16409}}} {[astro-ph.IM]}

\bibitem[{De~Lera~Acedo et~al(2015)De~Lera~Acedo, Razavi-Ghods, Troop, Drought, and Faulkner}]{DeLeraAcedoRazaviGhods2015}
De~Lera~Acedo E, Razavi-Ghods N, Troop N, et~al (2015) {SKALA, a log-periodic array antenna for the SKA-low instrument: design, simulations, tests and system considerations}. Experimental Astronomy 39(3):567--594. \doi{10.1007/S10686-015-9439-0/FIGURES/29}

\bibitem[{{Di Vruno} et~al(2023){Di Vruno}, {Winkel}, {Bassa}, {J{\'o}zsa}, {Brentjens}, {Jessner}, and {Garrington}}]{DiVruno2023}
{Di Vruno} F, {Winkel} B, {Bassa} CG, et~al (2023) {Unintended electromagnetic radiation from Starlink satellites detected with LOFAR between 110 and 188 MHz}. \aap 676:A75. \doi{10.1051/0004-6361/202346374}, {\href{https://arxiv.org/abs/2307.02316}{{arXiv:2307.02316}}} {[astro-ph.IM]}

\bibitem[{{Ding} et~al(2020){Ding}, {Xiao}, {Su}, and {Cui}}]{Ding2020}
{Ding} C, {Xiao} Z, {Su} Ya, et~al (2020) Compositional variations along the route of chang’e-3 yutu rover revealed by the lunar penetrating radar. Prog Earth Planet Sci 32(7). \doi{10.1186/s40645-020-00340-4}

\bibitem[{Ding et~al(2024)Ding, Chang, Su, Wang, and Xie}]{Ding2024}
Ding C, Chang Y, Su Y, et~al (2024) Rover-mounted radar observation of discrete layers within the top 4 meters of regolith at the chang’e-3 landing site, the moon. IEEE Transactions on Geoscience and Remote Sensing 62:1--16. \doi{10.1109/TGRS.2024.3365130}

\bibitem[{{Donaldson} et~al(2024){Donaldson}, {Olsen}, {Paty}, and {Caggiano}}]{solarwind+magnetosphere_donaldson2024}
{Donaldson} K, {Olsen} AJ, {Paty} CS, et~al (2024) {Characterizing the Solar Wind-Magnetosphere Viscous Interaction at Uranus and Neptune}. Journal of Geophysical Research (Space Physics) 129(8):e2024JA032518. \doi{10.1029/2024JA032518}

\bibitem[{{Dresing} et~al(2023){Dresing}, {Rodríguez-García}, {Jebaraj}, {Warmuth}, {Wallace}, {Balmaceda}, {Podladchikova}, {Strauss}, {Kouloumvakos}, {Palmroos}, {Krupar}, {Gieseler}, {Xu}, {Mitchell}, {Cohen}, {de Nolfo}, {Palmerio}, {Carcaboso}, {Kilpua}, {Trotta}, {Auster}, {Asvestari}, {da Silva}, {Dröge}, {Getachew}, {Gómez-Herrero}, {Grande}, {Heyner}, {Holmström}, {Huovelin}, {Kartavykh}, {Laurenza}, {Lee}, {Mason}, {Maksimovic}, {Mieth}, {Murakami}, {Oleynik}, {Pinto}, {Pulupa}, {Richter}, {Rodríguez-Pacheco}, {Sánchez-Cano}, {Schuller}, {Ueno}, {Vainio}, {Vecchio}, {Veronig}, and {Wijsen}}]{Dresing2023}
{Dresing} N, {Rodríguez-García} L, {Jebaraj} IC, et~al (2023) {The 17 April 2021 widespread solar energetic particle event}. \aap 674:A105. \doi{10.1051/0004-6361/202345938}, {\href{https://arxiv.org/abs/2303.10969}{{arXiv:2303.10969}}}

\bibitem[{{Dulk} et~al(2001){Dulk}, {Erickson}, {Manning}, and {Bougeret}}]{Dulk2001}
{Dulk} GA, {Erickson} WC, {Manning} R, et~al (2001) Calibration of low-frequency radio telescopes using the galactic background radiation. A\&A 365(2):294--300. \doi{10.1051/0004-6361:20000006}, \urlprefix\url{https://doi.org/10.1051/0004-6361:20000006}

\bibitem[{{Eastwood} et~al(2019){Eastwood}, {Anderson}, {Monroe}, {Hallinan}, {Catha}, {Dowell}, {Garsden}, {Greenhill}, {Hicks}, {Kocz}, {Price}, {Schinzel}, {Vedantham}, and {Wang}}]{2019AJ....158...84E}
{Eastwood} MW, {Anderson} MM, {Monroe} RM, et~al (2019) {The 21 cm Power Spectrum from the Cosmic Dawn: First Results from the OVRO-LWA}. The Astronomical Journal 158(2):84. \doi{10.3847/1538-3881/ab2629}, {\href{https://arxiv.org/abs/1906.08943}{{arXiv:1906.08943}}} {[astro-ph.CO]}

\bibitem[{ESA(2021)}]{ESA2021}
ESA (2021) {Astrophysical Lunar Observatory, Assessment of an Astrophysical Lunar Observatory on the farside of the Moon, CDF Study Report, J. Grenouilleau, HRE-E, Study Manager and B. García Gutiérrez, TEC-SYE, Team Lead}. Tech. rep., ESA, ESTEC Concurrent Design Facility 219(A)

\bibitem[{{Fallows} et~al(2022){Fallows}, {Iwai}, {Jackson}, {Zhang}, {Bisi}, and {Zucca}}]{Fallows2022b}
{Fallows} RA, {Iwai} K, {Jackson} BV, et~al (2022) Application of novel interplanetary scintillation visualisations using lofar: A case study of merged cmes from september 2017. arXiv e-prints p arXiv:2210.02135

\bibitem[{{Fan} et~al(2003){Fan}, {Strauss}, {Schneider}, {Becker}, {White}, {Haiman}, {Gregg}, {Pentericci}, {Grebel}, {Narayanan}, {Loh}, {Richards}, {Gunn}, {Lupton}, {Knapp}, {Ivezi{\'c}}, {Brandt}, {Collinge}, {Hao}, {Harbeck}, {Prada}, {Schaye}, {Strateva}, {Zakamska}, {Anderson}, {Brinkmann}, {Bahcall}, {Lamb}, {Okamura}, {Szalay}, and {York}}]{Fan2003}
{Fan} X, {Strauss} MA, {Schneider} DP, et~al (2003) {A Survey of z>5.7 Quasars in the Sloan Digital Sky Survey. II. Discovery of Three Additional Quasars at z>6}. \aj 125(4):1649--1659. \doi{10.1086/368246}, {\href{https://arxiv.org/abs/astro-ph/0301135}{{arXiv:astro-ph/0301135}}} {[astro-ph]}

\bibitem[{{Fan} et~al(2006){Fan}, {Strauss}, {Becker}, {White}, {Gunn}, {Knapp}, {Richards}, {Schneider}, {Brinkmann}, and {Fukugita}}]{Fan2006}
{Fan} X, {Strauss} MA, {Becker} RH, et~al (2006) {Constraining the Evolution of the Ionizing Background and the Epoch of Reionization with z\raisebox{-0.5ex}\textasciitilde6 Quasars. II. A Sample of 19 Quasars}. \aj 132(1):117--136. \doi{10.1086/504836}, {\href{https://arxiv.org/abs/astro-ph/0512082}{{arXiv:astro-ph/0512082}}} {[astro-ph]}

\bibitem[{{Fialkov} et~al(2024){Fialkov}, {Gessey-Jones}, and {Dhandha}}]{Fialkov2024}
{Fialkov} A, {Gessey-Jones} T, {Dhandha} J (2024) {Cosmic mysteries and the hydrogen 21-cm line: bridging the gap with lunar observations}. Philosophical Transactions of the Royal Society of London Series A 382(2271):20230068. \doi{10.1098/rsta.2023.0068}, {\href{https://arxiv.org/abs/2311.05366}{{arXiv:2311.05366}}} {[astro-ph.CO]}

\bibitem[{{Field}(1958)}]{Field1958}
{Field} GB (1958) {Excitation of the Hydrogen 21-CM Line}. Proceedings of the IRE 46:240--250. \doi{10.1109/JRPROC.1958.286741}

\bibitem[{{Field}(1959)}]{Field1959}
{Field} GB (1959) {The Spin Temperature of Intergalactic Neutral Hydrogen.} \apj 129:536. \doi{10.1086/146653}

\bibitem[{{Fox} et~al(2016){Fox}, {Velli}, {Bale}, {Decker}, {Driesman}, {Howard}, {Kasper}, {Kinnison}, {Kusterer}, {Lario}, {Lockwood}, {McComas}, {Raouafi}, and {Szabo}}]{Fox+2016}
{Fox} NJ, {Velli} MC, {Bale} SD, et~al (2016) {The Solar Probe Plus Mission: Humanity's First Visit to Our Star}. \ssr 204(1-4):7--48. \doi{10.1007/s11214-015-0211-6}

\bibitem[{{Garsden} et~al(2021){Garsden}, {Greenhill}, {Bernardi}, {Fialkov}, {Price}, {Mitchell}, {Dowell}, {Spinelli}, and {Schinzel}}]{OVRO-LWA-2021}
{Garsden} H, {Greenhill} L, {Bernardi} G, et~al (2021) {A 21-cm power spectrum at 48 MHz, using the Owens Valley Long Wavelength Array}. {MNRAS} 506(4):5802--5817. \doi{10.1093/mnras/stab1671}, {\href{https://arxiv.org/abs/2102.09596}{{arXiv:2102.09596}}} {[astro-ph.CO]}

\bibitem[{{Gorgolewski}(1966)}]{Gorgolewski-1966}
{Gorgolewski} S (1966) {Lunar Radio Astronomy Observatory}. In: {Malina} FJ (ed) LIL Symposium on Research in Geosciences and Astronomy, pp 78--84

\bibitem[{Griessmeier et~al(2007)Griessmeier, Zarka, and Spreeuw}]{Griessmeier2007}
Griessmeier JM, Zarka P, Spreeuw H (2007) Predicting low-frequency radio fluxes of known extrasolar planets. \aap 475(1):359–368. \doi{10.1051/0004-6361:20077397}, \urlprefix\url{http://dx.doi.org/10.1051/0004-6361:20077397}

\bibitem[{{Grigor'eva} and {Slysh}(1970)}]{Grigorieva+Slysh-1970}
{Grigor'eva} VP, {Slysh} VI (1970) {Long-Wave Cosmic Radio Radiation in Circumlunar Space}. Cosmic Research 8:260--264

\bibitem[{Groeneveld et~al(2022)Groeneveld, van Weeren, Miley, Morabito, de~Gasperin, Callingham, Sweijen, Brüggen, Botteon, Offringa, Brunetti, Moldon, Bondi, Kappes, and Röttgering}]{Groeneveld2022}
Groeneveld C, van Weeren RJ, Miley GK, et~al (2022) Pushing sub-arcsecond resolution imaging down to 30 mhz with the trans-european international lofar telescope. \aap 658:A9. \doi{10.1051/0004-6361/202141352}, \urlprefix\url{http://dx.doi.org/10.1051/0004-6361/202141352}

\bibitem[{Groeneveld et~al(2024)Groeneveld, van Weeren, Osinga, Williams, Callingham, de~Gasperin, Botteon, Shimwell, Sweijen, de~Jong, Jansen, Miley, Brunetti, Brüggen, and Röttgering}]{Groeneveld2024}
Groeneveld C, van Weeren RJ, Osinga E, et~al (2024) Characterization of the decametre sky at subarcminute resolution. Nature Astronomy 8(6):786–795. \doi{10.1038/s41550-024-02266-z}, \urlprefix\url{http://dx.doi.org/10.1038/s41550-024-02266-z}

\bibitem[{{Hallinan} et~al(2021){Hallinan}, {Burns}, {Lux}, {Romero-Wolf}, {Teitelbaum}, {Chang}, {Kocz}, {Bowman}, {MacDowall}, {Kasper}, {Bradley}, {Anderson}, {Rapetti}, {Zhan}, {Wu}, {Keane}, {Panning}, {Klesh}, {Nesnas}, {Pober}, {Furlanetto}, and {Austin}}]{FarSide-2021BAAS}
{Hallinan} G, {Burns} J, {Lux} J, et~al (2021) {FARSIDE: A Low Radio Frequency Interferometric Array on the Lunar Farside}. In: Bulletin of the American Astronomical Society, p 379, \doi{10.3847/25c2cfeb.60683360}

\bibitem[{{Hess} and {Zarka}(2011)}]{hess_2011_mag_sig}
{Hess} SLG, {Zarka} P (2011) Modeling the radio signature of the orbital parameters, rotation, and magnetic field of exoplanets. \aap 531:A29. \doi{10.1051/0004-6361/201116510}, \urlprefix\url{https://doi.org/10.1051/0004-6361/201116510}

\bibitem[{Hibbard et~al(2025)Hibbard, Burns, MacDowall, Gopalswamy, Boardsen, Farrell, Bradley, Schulszas, Jones, Rapetti, and Turner}]{Hibbard+2025}
Hibbard JJ, Burns JO, MacDowall R, et~al (2025) Results from nasa's first radio telescope on the moon: Terrestrial technosignatures and the low-frequency galactic background observed by rolses-1 onboard the odysseus lander. ApJ submitted \urlprefix\url{https://arxiv.org/abs/2503.09842}, {\href{https://arxiv.org/abs/2503.09842}{{arXiv:2503.09842}}} {[astro-ph.IM]}

\bibitem[{{Hogan} and {Rees}(1979)}]{HoganRees1979}
{Hogan} CJ, {Rees} MJ (1979) {Spectral appearance of non-uniform gas at high z.} \mnras 188:791--798. \doi{10.1093/mnras/188.4.791}

\bibitem[{{IAU}(1994)}]{IAU-B15}
{IAU} (1994) {Resolution No. B 15 concerning the Bands to be used for Radiocommunications in the lunar envronment}. {Resolution B15}, {IAU XXIInd General Assembly, The Hague, The Netherlands, 1994}

\bibitem[{{ITU}(1979)}]{ITU-R-479}
{ITU} (1979) {Protection of Frequencies for Radioastronomical Measurements in the Shielded Zone of the Moon}. {Recommendation}, {International Telecommunications Union}, {ITU-R RA.479-5; latest edition confirmed by ITU in 2023 is available at https://www.itu.int/en/publications/ITU-R/pages/publications.aspx?parent=R-REG-RR-2020\&media=electronic}

\bibitem[{{ITU}(1999)}]{ITU-R-1417}
{ITU} (1999) { A radio-quiet zone in the vicinity of the L2 Sun-Earth Lagrange point}. {Recommendation}, {International Telecommunications Union}, {ITU-R RA.1417}

\bibitem[{{ITU}(2024)}]{ITU-RR}
{ITU} (2024) {Radio Regulations}. {Volume I, Articles}, {International Telecommunications Union}, {2024}

\bibitem[{{Jansky}(1933{\natexlab{a}})}]{Jansky-1933P}
{Jansky} KG (1933{\natexlab{a}}) {Electrical phenomena that apparently are of interstellar origin}. Popular Astronomy 41:548--555

\bibitem[{{Jansky}(1933{\natexlab{b}})}]{Jansky-1933N}
{Jansky} KG (1933{\natexlab{b}}) {Radio Waves from Outside the Solar System}. Nature 132(3323):66. \doi{10.1038/132066a0}

\bibitem[{{Jester} and {Falcke}(2009)}]{Jester-Falcke-2009}
{Jester} S, {Falcke} H (2009) {Science with a lunar low-frequency array: From the dark ages of the Universe to nearby exoplanets}. \nar 53(1-2):1--26. \doi{10.1016/j.newar.2009.02.001}, {\href{https://arxiv.org/abs/0902.0493}{{arXiv:0902.0493}}} {[astro-ph.CO]}

\bibitem[{Jia et~al(2018)Jia, Zou, Ping, Xue, Yan, and Ning}]{CE4science-2018}
Jia Y, Zou Y, Ping J, et~al (2018) The scientific objectives and payloads of chang’e-4 mission. Planetary and Space Science 162:207--215. \doi{https://doi.org/10.1016/j.pss.2018.02.011}, \urlprefix\url{https://www.sciencedirect.com/science/article/pii/S0032063317300211}, lunar Reconnaissance Orbiter – Seven Years of Exploration and Discovery

\bibitem[{{Jia} et~al(2018){Jia}, {Zou}, {Ping}, {Xue}, {Yan}, and {Ning}}]{NCLE-2018P&SS}
{Jia} Y, {Zou} Y, {Ping} J, et~al (2018) {The scientific objectives and payloads of Chang'E-4 mission}. Planetary Space Sci 162:207--215. \doi{10.1016/j.pss.2018.02.011}

\bibitem[{{Karapakula} et~al(2024){Karapakula}, {Brinkerink}, {Vecchio}, {Pourshaghaghi}, {Dolron}, {Jordans}, {Bertels}, {Aalbers}, {Ruiter}, {Boonstra}, {Bentum}, {Prinsloo}, {Arts}, {Bast}, {Damstra}, {van Duin}, {Ebbendorf}, {van der Marel}, {Morawietz}, {Witvers}, {Poiesz}, {van Dongen}, {Cecconi}, {Zarka}, {Dekkali}, {Chen}, {Wang}, {Zhang}, {Huang}, {Yan}, {Dong}, {Tan}, {Zhang}, {Xiong}, {Sun}, {Zhang}, {Ping}, {Wolt}, and {Falcke}}]{NCLE-2024}
{Karapakula} S, {Brinkerink} C, {Vecchio} A, et~al (2024) {Architecture Design and Ground Performance of Netherlands-China Low-Frequency Explorer}. Radio Science 59(8):e2023RS007906. \doi{10.1029/2023RS007906}

\bibitem[{{Keller} et~al(2023){Keller}, {Nikolic}, {Thyagarajan}, {Carilli}, {Bernardi}, {Charles}, {Bester}, {Smirnov}, {Kern}, {Dillon}, {Hazelton}, {Morales}, {Jacobs}, {Parsons}, {Abdurashidova}, {Adams}, {Aguirre}, {Alexander}, {Ali}, {Baartman}, {Balfour}, {Beardsley}, {Billings}, {Bowman}, {Bradley}, {Bull}, {Burba}, {Carey}, {Cheng}, {DeBoer}, {de Lera Acedo}, {Dexter}, {Eksteen}, {Ely}, {Ewall-Wice}, {Fagnoni}, {Fritz}, {Furlanetto}, {Gale-Sides}, {Glendenning}, {Gorthi}, {Greig}, {Grobbelaar}, {Halday}, {Hewitt}, {Hickish}, {Julius}, {Kariseb}, {Kerrigan}, {Kittiwisit}, {Kohn}, {Kolopanis}, {Lanman}, {Plante}, {Liu}, {Loots}, {Ma}, {MacMahon}, {Malan}, {Malgas}, {Malgas}, {Marero}, {Martinot}, {Mesinger}, {Molewa}, {Mosiane}, {Murray}, {Neben}, {Nuwegeld}, {Pascua}, {Patra}, {Pieterse}, {Pober}, {Razavi-Ghods}, {Robnett}, {Rosie}, {Santos}, {Sims}, {Smith}, {Swarts}, {Van Wyngaarden}, {Williams}, and {Zheng}}]{HERA-2023MNRAS}
{Keller} PM, {Nikolic} B, {Thyagarajan} N, et~al (2023) {Search for the Epoch of Reionization with HERA: upper limits on the closure phase delay power spectrum}. MNRAS 524(1):583--598. \doi{10.1093/mnras/stad371}, {\href{https://arxiv.org/abs/2302.07969}{{arXiv:2302.07969}}} {[astro-ph.CO]}

\bibitem[{Klein-Wolt et~al(2021)Klein-Wolt, Falcke, Brinkerink, Vecchio, Boonstra, Bentum, Koopmans, Rothkaehl, Ping, Chen, Huang, Burns, Cecconi, Zarka, Bergman, and Carpenter}]{KleinWolt2021}
Klein-Wolt M, Falcke H, Brinkerink C, et~al (2021) {Astronomical Lunar Observatory}. In: URSI GASS 2021, Rome, Italy, 28 August - 4 September 2021

\bibitem[{{Klessen} and {Glover}(2023)}]{Klessen2023}
{Klessen} RS, {Glover} SCO (2023) {The First Stars: Formation, Properties, and Impact}. \araa 61:65--130. \doi{10.1146/annurev-astro-071221-053453}, {\href{https://arxiv.org/abs/2303.12500}{{arXiv:2303.12500}}} {[astro-ph.CO]}

\bibitem[{{Kolopanis} et~al(2019){Kolopanis}, {Jacobs}, {Cheng}, {Parsons}, {Kohn}, {Pober}, {Aguirre}, {Ali}, {Bernardi}, {Bradley}, {Carilli}, {DeBoer}, {Dexter}, {Dillon}, {Kerrigan}, {Klima}, {Liu}, {MacMahon}, {Moore}, {Thyagarajan}, {Nunhokee}, {Walbrugh}, and {Walker}}]{2019ApJ...883..133K}
{Kolopanis} M, {Jacobs} DC, {Cheng} C, et~al (2019) {A Simplified, Lossless Reanalysis of PAPER-64}. The Astrophysical Journal 883(2):133. \doi{10.3847/1538-4357/ab3e3a}, {\href{https://arxiv.org/abs/1909.02085}{{arXiv:1909.02085}}} {[astro-ph.CO]}

\bibitem[{{Konovalenko} et~al(2016){Konovalenko}, {Sodin}, {Zakharenko}, {Zarka}, {Ulyanov}, {Sidorchuk}, {Stepkin}, {Tokarsky}, {Melnik}, {Kalinichenko}, {Stanislavsky}, {Koliadin}, {Shepelev}, {Dorovskyy}, {Ryabov}, {Koval}, {Bubnov}, {Yerin}, {Gridin}, {Kulishenko}, {Reznichenko}, {Bortsov}, {Lisachenko}, {Reznik}, {Kvasov}, {Mukha}, {Litvinenko}, {Khristenko}, {Shevchenko}, {Shevchenko}, {Belov}, {Rudavin}, {Vasylieva}, {Miroshnichenko}, {Vasilenko}, {Olyak}, {Mylostna}, {Skoryk}, {Shevtsova}, {Plakhov}, {Kravtsov}, {Volvach}, {Lytvinenko}, {Shevchuk}, {Zhouk}, {Bovkun}, {Antonov}, {Vavriv}, {Vinogradov}, {Kozhin}, {Kravtsov}, {Bulakh}, {Kuzin}, {Vasilyev}, {Brazhenko}, {Vashchishin}, {Pylaev}, {Koshovyy}, {Lozinsky}, {Ivantyshin}, {Rucker}, {Panchenko}, {Fischer}, {Lecacheux}, {Denis}, {Coffre}, {Grie{\ss}meier}, {Tagger}, {Girard}, {Charrier}, {Briand}, and {Mann}}]{Konovalenko+2016}
{Konovalenko} A, {Sodin} L, {Zakharenko} V, et~al (2016) {The modern radio astronomy network in Ukraine: UTR-2, URAN and GURT}. Experimental Astronomy 42(1):11--48. \doi{10.1007/s10686-016-9498-x}

\bibitem[{{Konovalenko} et~al(2024){Konovalenko}, {Zakharenko}, {Novosyadlyj}, {Gurvits}, {Stepkin}, {Vasylkivskyi}, {Tokarsky}, {Ulyanov}, {Stanislavsky}, and {Bubnov}}]{UTR-2-GURT2024}
{Konovalenko} A, {Zakharenko} V, {Novosyadlyj} B, et~al (2024) {On the Possibility of Detecting a Global Signal in the Line of the Hyperfine Structure of Hydrogen from the Dark Ages}. Journal of Physical Studies 28(1):id. 1902. \doi{10.48550/arXiv.2401.09096}, {\href{https://arxiv.org/abs/2401.09096}{{arXiv:2401.09096}}} {[astro-ph.CO]}

\bibitem[{{Koopmans} et~al(2015){Koopmans}, {Pritchard}, {Mellema}, {Aguirre}, {Ahn}, {Barkana}, {van Bemmel}, {Bernardi}, {Bonaldi}, {Briggs}, {de Bruyn}, {Chang}, {Chapman}, {Chen}, {Ciardi}, {Dayal}, {Ferrara}, {Fialkov}, {Fiore}, {Ichiki}, {Illiev}, {Inoue}, {Jelic}, {Jones}, {Lazio}, {Maio}, {Majumdar}, {Mack}, {Mesinger}, {Morales}, {Parsons}, {Pen}, {Santos}, {Schneider}, {Semelin}, {de Souza}, {Subrahmanyan}, {Takeuchi}, {Vedantham}, {Wagg}, {Webster}, {Wyithe}, {Datta}, and {Trott}}]{SKA-2015sci}
{Koopmans} L, {Pritchard} J, {Mellema} G, et~al (2015) {The Cosmic Dawn and Epoch of Reionisation with SKA}. In: Advancing Astrophysics with the Square Kilometre Array (AASKA14), p~1, \doi{10.22323/1.215.0001}, \eprint{1505.07568}

\bibitem[{{Koopmans} et~al(2021){Koopmans}, {Barkana}, {Bentum}, {Bernardi}, {Boonstra}, {Bowman}, {Burns}, {Chen}, {Datta}, {Falcke}, {Fialkov}, {Gehlot}, {Gurvits}, {Jeli{\'c}}, {Klein-Wolt}, {Lazio}, {Meerburg}, {Mellema}, {Mertens}, {Mesinger}, {Offringa}, {Pritchard}, {Semelin}, {Subrahmanyan}, {Silk}, {Trott}, {Vedantham}, {Verde}, {Zaroubi}, and {Zarka}}]{Koopmans+2021}
{Koopmans} LVE, {Barkana} R, {Bentum} M, et~al (2021) {Peering into the dark (ages) with low-frequency space interferometers}. Experimental Astronomy 51(3):1641--1676. \doi{10.1007/s10686-021-09743-7}, {\href{https://arxiv.org/abs/1908.04296}{{arXiv:1908.04296}}} {[astro-ph.IM]}

\bibitem[{{Lamy}(2020)}]{lamy2020auroral}
{Lamy} L (2020) {Auroral emissions from Uranus and Neptune}. Philosophical Transactions of the Royal Society of London Series A 378(2187):20190481. \doi{10.1098/rsta.2019.0481}

\bibitem[{{Liu} et~al(2025){Liu}, {Wan}, {Wu}, {Wang}, {Zhang}, {Guo}, {Zhang}, and {Jiang}}]{Chinese-ULWSKA}
{Liu} J, {Wan} J, {Wu} C, et~al (2025) {Design of lunar based square kilometer array radio telescope antenna}. Chinese Space Science and Technology (in Chinese) 45:88--93. \doi{10.16708/j.cnki.1000-758X.2025.0026}

\bibitem[{{Louarn} and {Le Qu{\'e}au}(1996)}]{1996P&SS...44..211L}
{Louarn} P, {Le Qu{\'e}au} D (1996) {Generation of the Auroral Kilometric Radiation in plasma cavities - II. The cyclotron maser instability in small size sources}. \planss 44(3):211--224. \doi{10.1016/0032-0633(95)00122-0}

\bibitem[{{Macquart} et~al(2020){Macquart}, {Prochaska}, {McQuinn}, {Bannister}, {Bhandari}, {Day}, {Deller}, {Ekers}, {James}, {Marnoch}, {Os{\l}owski}, {Phillips}, {Ryder}, {Scott}, {Shannon}, and {Tejos}}]{Macquart+2020Nature}
{Macquart} JP, {Prochaska} JX, {McQuinn} M, et~al (2020) {A census of baryons in the Universe from localized fast radio bursts}. Nature 581(7809):391--395. \doi{10.1038/s41586-020-2300-2}, {\href{https://arxiv.org/abs/2005.13161}{{arXiv:2005.13161}}} {[astro-ph.CO]}

\bibitem[{{Madau} et~al(1997){Madau}, {Meiksin}, and {Rees}}]{Madau1997}
{Madau} P, {Meiksin} A, {Rees} MJ (1997) {21 Centimeter Tomography of the Intergalactic Medium at High Redshift}. \apj 475(2):429--444. \doi{10.1086/303549}, {\href{https://arxiv.org/abs/astro-ph/9608010}{{arXiv:astro-ph/9608010}}} {[astro-ph]}

\bibitem[{{Maia} et~al(2000){Maia}, {Pick}, {Vourlidas}, and {Howard}}]{Maia2000}
{Maia} D, {Pick} M, {Vourlidas} A, et~al (2000) Development of coronal mass ejections: Radio shock signatures. The Astrophysical Journal Letters 528:L49--L51

\bibitem[{{Mertens} et~al(2021){Mertens}, {Semelin}, and {Koopmans}}]{NenuFAR-2021}
{Mertens} FG, {Semelin} B, {Koopmans} LVE (2021) {Exploring the Cosmic Dawn with NenuFAR}. In: {Siebert} A, {Bailli{\'e}} K, {Lagadec} E, et~al (eds) SF2A-2021: Proceedings of the Annual meeting of the French Society of Astronomy and Astrophysics, pp 211--214, \doi{10.48550/arXiv.2109.10055}, \eprint{2109.10055}

\bibitem[{{Mimoun} et~al(2012){Mimoun}, {Wieczorek}, {Alkalai}, {Banerdt}, {Baratoux}, {Bougeret}, {Bouley}, {Cecconi}, {Falcke}, {Flohrer}, {Garcia}, {Grimm}, {Grott}, {Gurvits}, {Jaumann}, {Johnson}, {Knapmeyer}, {Kobayashi}, {Konovalenko}, {Lawrence}, {Le Feuvre}, {Lognonn{\'e}}, {Neal}, {Oberst}, {Olsen}, {R{\"o}ttgering}, {Spohn}, {Vennerstrom}, {Woan}, and {Zarka}}]{Mimoun+2012}
{Mimoun} D, {Wieczorek} MA, {Alkalai} L, et~al (2012) {Farside explorer: unique science from a mission to the farside of the moon}. Experimental Astronomy 33(2-3):529--585. \doi{10.1007/s10686-011-9252-3}

\bibitem[{{Mondal} et~al(2020){Mondal}, {Oberoi}, and {Vourlidas}}]{Mondal2020}
{Mondal} S, {Oberoi} D, {Vourlidas} A (2020) Estimation of the physical parameters of a cme at high coronal heights using low-frequency radio observations. The Astrophysical Journal 893:28

\bibitem[{{Monsalve} et~al(2017){Monsalve}, {Rogers}, {Bowman}, and {Mozdzen}}]{Monsalve2017}
{Monsalve} RA, {Rogers} AEE, {Bowman} JD, et~al (2017) {Results from EDGES High-band. I. Constraints on Phenomenological Models for the Global 21 cm Signal}. \apj 847(1):64. \doi{10.3847/1538-4357/aa88d1}, {\href{https://arxiv.org/abs/1708.05817}{{arXiv:1708.05817}}} {[astro-ph.CO]}

\bibitem[{{Morosan} et~al(2019){Morosan}, {Carley}, {Hayes}, {Murray}, {Zucca}, {Fallows}, {McCauley}, {Kilpua}, {Mann}, {Vocks}, and {Gallagher}}]{Morosan2019}
{Morosan} DE, {Carley} EP, {Hayes} LA, et~al (2019) Multiple regions of shock-accelerated particles during a solar coronal mass ejection. Nature Astronomy 3:452--461

\bibitem[{{Mortlock} et~al(2011){Mortlock}, {Warren}, {Venemans}, {Patel}, {Hewett}, {McMahon}, {Simpson}, {Theuns}, {Gonz{\'a}les-Solares}, {Adamson}, {Dye}, {Hambly}, {Hirst}, {Irwin}, {Kuiper}, {Lawrence}, and {R{\"o}ttgering}}]{Mortlock2011}
{Mortlock} DJ, {Warren} SJ, {Venemans} BP, et~al (2011) {A luminous quasar at a redshift of z = 7.085}. \nat 474(7353):616--619. \doi{10.1038/nature10159}, {\href{https://arxiv.org/abs/1106.6088}{{arXiv:1106.6088}}} {[astro-ph.CO]}

\bibitem[{Mozdzen et~al(2016)Mozdzen, Bowman, Monsalve, and Rogers}]{Mozdzen2016}
Mozdzen TJ, Bowman JD, Monsalve RA, et~al (2016) {Limits on foreground subtraction from chromatic beam effects in global redshifted 21 cm measurements}. Monthly Notices of the Royal Astronomical Society 455(4):3890--3900. \doi{10.1093/MNRAS/STV2601}

\bibitem[{{Nichols}(2011)}]{Nichols2011}
{Nichols} JD (2011) {Magnetosphere-ionosphere coupling at Jupiter-like exoplanets with internal plasma sources: implications for detectability of auroral radio emissions}. \mnras 414(3):2125--2138. \doi{10.1111/j.1365-2966.2011.18528.x}, {\href{https://arxiv.org/abs/1102.2737}{{arXiv:1102.2737}}} {[physics.space-ph]}

\bibitem[{{Novaco} and {Brown}(1978)}]{Novaco+Brown-1978}
{Novaco} JC, {Brown} LW (1978) {Nonthermal galactic emission below 10 megahertz.} ApJ 221:114--123. \doi{10.1086/156009}

\bibitem[{{Paciga} et~al(2013){Paciga}, {Albert}, {Bandura}, {Chang}, {Gupta}, {Hirata}, {Odegova}, {Pen}, {Peterson}, {Roy}, {Shaw}, {Sigurdson}, and {Voytek}}]{2013MNRAS.433..639P}
{Paciga} G, {Albert} JG, {Bandura} K, et~al (2013) {A simulation-calibrated limit on the H I power spectrum from the GMRT Epoch of Reionization experiment}. MNRAS 433(1):639--647. \doi{10.1093/mnras/stt753}, {\href{https://arxiv.org/abs/1301.5906}{{arXiv:1301.5906}}} {[astro-ph.CO]}

\bibitem[{{Petroff} et~al(2022){Petroff}, {Hessels}, and {Lorimer}}]{Petroff+2022A&ARv}
{Petroff} E, {Hessels} JWT, {Lorimer} DR (2022) {Fast radio bursts at the dawn of the 2020s}. {Astron \& Astrophys Rev} 30(1):2. \doi{10.1007/s00159-022-00139-w}, {\href{https://arxiv.org/abs/2107.10113}{{arXiv:2107.10113}}} {[astro-ph.HE]}

\bibitem[{{Planck Collaboration} et~al(2020){Planck Collaboration}, {Aghanim, N.}, {Akrami, Y.}, {Arroja, F.}, {Ashdown, M.}, {Aumont, J.}, {Baccigalupi, C.}, {Ballardini, M.}, {Banday, A. J.}, {Barreiro, R. B.}, {Bartolo, N.}, {Basak, S.}, {Battye, R.}, {Benabed, K.}, {Bernard, J.-P.}, {Bersanelli, M.}, {Bielewicz, P.}, {Bock, J. J.}, {Bond, J. R.}, {Borrill, J.}, {Bouchet, F. R.}, {Boulanger, F.}, {Bucher, M.}, {Burigana, C.}, {Butler, R. C.}, {Calabrese, E.}, {Cardoso, J.-F.}, {Carron, J.}, {Casaponsa, B.}, {Challinor, A.}, {Chiang, H. C.}, {Colombo, L. P. L.}, {Combet, C.}, {Contreras, D.}, {Crill, B. P.}, {Cuttaia, F.}, {de Bernardis, P.}, {de Zotti, G.}, {Delabrouille, J.}, {Delouis, J.-M.}, {Désert, F.-X.}, {Di Valentino, E.}, {Dickinson, C.}, {Diego, J. M.}, {Donzelli, S.}, {Doré, O.}, {Douspis, M.}, {Ducout, A.}, {Dupac, X.}, {Efstathiou, G.}, {Elsner, F.}, {Enßlin, T. A.}, {Eriksen, H. K.}, {Falgarone, E.}, {Fantaye, Y.}, {Fergusson, J.}, {Fernandez-Cobos, R.}, {Finelli, F.}, {Forastieri, F.},
  {Frailis, M.}, {Franceschi, E.}, {Frolov, A.}, {Galeotta, S.}, {Galli, S.}, {Ganga, K.}, {Génova-Santos, R. T.}, {Gerbino, M.}, {Ghosh, T.}, {González-Nuevo, J.}, {Górski, K. M.}, {Gratton, S.}, {Gruppuso, A.}, {Gudmundsson, J. E.}, {Hamann, J.}, {Handley, W.}, {Hansen, F. K.}, {Helou, G.}, {Herranz, D.}, {Hildebrandt, S. R.}, {Hivon, E.}, {Huang, Z.}, {Jaffe, A. H.}, {Jones, W. C.}, {Karakci, A.}, {Keihänen, E.}, {Keskitalo, R.}, {Kiiveri, K.}, {Kim, J.}, {Kisner, T. S.}, {Knox, L.}, {Krachmalnicoff, N.}, {Kunz, M.}, {Kurki-Suonio, H.}, {Lagache, G.}, {Lamarre, J.-M.}, {Langer, M.}, {Lasenby, A.}, {Lattanzi, M.}, {Lawrence, C. R.}, {Le Jeune, M.}, {Leahy, J. P.}, {Lesgourgues, J.}, {Levrier, F.}, {Lewis, A.}, {Liguori, M.}, {Lilje, P. B.}, {Lilley, M.}, {Lindholm, V.}, {López-Caniego, M.}, {Lubin, P. M.}, {Ma, Y.-Z.}, {Macías-Pérez, J. F.}, {Maggio, G.}, {Maino, D.}, {Mandolesi, N.}, {Mangilli, A.}, {Marcos-Caballero, A.}, {Maris, M.}, {Martin, P. G.}, {Martinelli, M.}, {Martínez-González, E.},
  {Matarrese, S.}, {Mauri, N.}, {McEwen, J. D.}, {Meerburg, P. D.}, {Meinhold, P. R.}, {Melchiorri, A.}, {Mennella, A.}, {Migliaccio, M.}, {Millea, M.}, {Mitra, S.}, {Miville-Deschênes, M.-A.}, {Molinari, D.}, {Moneti, A.}, {Montier, L.}, {Morgante, G.}, {Moss, A.}, {Mottet, S.}, {Münchmeyer, M.}, {Natoli, P.}, {Nørgaard-Nielsen, H. U.}, {Oxborrow, C. A.}, {Pagano, L.}, {Paoletti, D.}, {Partridge, B.}, {Patanchon, G.}, {Pearson, T. J.}, {Peel, M.}, {Peiris, H. V.}, {Perrotta, F.}, {Pettorino, V.}, {Piacentini, F.}, {Polastri, L.}, {Polenta, G.}, {Puget, J.-L.}, {Rachen, J. P.}, {Reinecke, M.}, {Remazeilles, M.}, {Renault, C.}, {Renzi, A.}, {Rocha, G.}, {Rosset, C.}, {Roudier, G.}, {Rubiño-Martín, J. A.}, {Ruiz-Granados, B.}, {Salvati, L.}, {Sandri, M.}, {Savelainen, M.}, {Scott, D.}, {Shellard, E. P. S.}, {Shiraishi, M.}, {Sirignano, C.}, {Sirri, G.}, {Spencer, L. D.}, {Sunyaev, R.}, {Suur-Uski, A.-S.}, {Tauber, J. A.}, {Tavagnacco, D.}, {Tenti, M.}, {Terenzi, L.}, {Toffolatti, L.}, {Tomasi, M.},
  {Trombetti, T.}, {Valiviita, J.}, {Van Tent, B.}, {Vibert, L.}, {Vielva, P.}, {Villa, F.}, {Vittorio, N.}, {Wandelt, B. D.}, {Wehus, I. K.}, {White, M.}, {White, S. D. M.}, {Zacchei, A.}, and {Zonca, A.}}]{Planck2020-I}
{Planck Collaboration}, {Aghanim, N.}, {Akrami, Y.}, et~al (2020) Planck 2018 results - i. overview and the cosmological legacy of planck. \aap 641:A1. \doi{10.1051/0004-6361/201833880}, \urlprefix\url{https://doi.org/10.1051/0004-6361/201833880}

\bibitem[{{Pleunis} et~al(2021){Pleunis}, {Michilli}, {Bassa}, {Hessels}, {Naidu}, {Andersen}, {Chawla}, {Fonseca}, {Gopinath}, {Kaspi}, {Kondratiev}, {Li}, {Bhardwaj}, {Boyle}, {Brar}, {Cassanelli}, {Gupta}, {Josephy}, {Karuppusamy}, {Keimpema}, {Kirsten}, {Leung}, {Marcote}, {Masui}, {Mckinven}, {Meyers}, {Ng}, {Nimmo}, {Paragi}, {Rahman}, {Scholz}, {Shin}, {Smith}, {Stairs}, and {Tendulkar}}]{Pleunis+2021ApJ}
{Pleunis} Z, {Michilli} D, {Bassa} CG, et~al (2021) {LOFAR Detection of 110-188 MHz Emission and Frequency-dependent Activity from FRB 20180916B}. {ApJL} 911(1):L3. \doi{10.3847/2041-8213/abec72}, {\href{https://arxiv.org/abs/2012.08372}{{arXiv:2012.08372}}} {[astro-ph.HE]}

\bibitem[{{Polidan} et~al(2024){Polidan}, {Burns}, {Ignatiev}, {Hegedus}, {Pober}, {Mahesh}, {Chang}, {Hallinan}, {Ning}, and {Bowman}}]{FarView-2024arXiv}
{Polidan} RS, {Burns} JO, {Ignatiev} A, et~al (2024) {FarView: An In-Situ Manufactured Lunar Far Side Radio Array Concept for 21-cm Dark Ages Cosmology}. arXiv e-prints arXiv:2404.03840. \doi{10.48550/arXiv.2404.03840}, {\href{https://arxiv.org/abs/2404.03840}{{arXiv:2404.03840}}} {[astro-ph.IM]}

\bibitem[{Pritchard and Loeb(2010)}]{Pritchard2010}
Pritchard JR, Loeb A (2010) Constraining the unexplored period between the dark ages and reionization with observations of the global 21 cm signal. Physical Review D 82(2). \doi{10.1103/physrevd.82.023006}, \urlprefix\url{http://dx.doi.org/10.1103/PhysRevD.82.023006}

\bibitem[{Pritchard and Loeb(2012)}]{Pritchard2012}
Pritchard JR, Loeb A (2012) 21 cm cosmology in the 21st century. Reports on Progress in Physics 75(8):086901. \doi{10.1088/0034-4885/75/8/086901}, \urlprefix\url{http://dx.doi.org/10.1088/0034-4885/75/8/086901}

\bibitem[{Rajan et~al(2016)Rajan, Boonstra, Bentum, Klein-Wolt, Belien, Arts, Saks, and van~der Veen}]{rajan2016}
Rajan R, Boonstra A, Bentum M, et~al (2016) Space-based aperture array for ultra-long wavelength radio astronomy. Experimental Astronomy 41:271–306. \doi{10.1007/s10686-015-9486-6}

\bibitem[{Rajan et~al(2011)Rajan, Engelen, Bentum, and Verhoeven}]{rajan2011AE}
Rajan RT, Engelen S, Bentum M, et~al (2011) Orbiting low frequency array for radio astronomy. In: 2011 Aerospace Conference, pp 1--11, \doi{10.1109/AERO.2011.5747222}

\bibitem[{{Roger} et~al(1999){Roger}, {Costain}, {Landecker}, and {Swerdlyk}}]{Roger+1999}
{Roger} RS, {Costain} CH, {Landecker} TL, et~al (1999) {The radio emission from the Galaxy at 22 MHz}. \aaps 137:7--19. \doi{10.1051/aas:1999239}, {\href{https://arxiv.org/abs/astro-ph/9902213}{{arXiv:astro-ph/9902213}}} {[astro-ph]}

\bibitem[{{Sathyanarayana Rao} et~al(2023){Sathyanarayana Rao}, {Singh}, {K.~S.}, {B.~S.}, {Sathish}, {Somashekar}, {Agaram}, {Kavitha}, {Vishwapriya}, {Anand}, {Udaya Shankar}, and {Seetha}}]{PRATUSH-2023}
{Sathyanarayana Rao} M, {Singh} S, {K.~S.} S, et~al (2023) {PRATUSH experiment concept and design overview}. Experimental Astronomy 56(2-3):741--778. \doi{10.1007/s10686-023-09909-5}

\bibitem[{{Scott} and {Rees}(1990)}]{ScottRees1990}
{Scott} D, {Rees} MJ (1990) {The 21-cm line at high redshift: a diagnostic for the origin of large scale structure}. \mnras 247:510

\bibitem[{{Turnpenney} et~al(2020){Turnpenney}, {Nichols}, {Wynn}, and {Jia}}]{Turnpenney2020}
{Turnpenney} S, {Nichols} JD, {Wynn} GA, et~al (2020) {Magnetohydrodynamic modelling of star-planet interaction and associated auroral radio emission}. \mnras 494(4):5044--5055. \doi{10.1093/mnras/staa824}, {\href{https://arxiv.org/abs/2003.09991}{{arXiv:2003.09991}}} {[astro-ph.EP]}

\bibitem[{{Vallat} et~al(2018){Vallat}, {Roatsch}, {Dougherty}, {Coustenis}, {Kasaba}, {Wahlund}, {Santolik}, {Barabash}, {Cremonese}, {Palumbo}, {Rothkaehl}, {Iess}, {Piccioni}, {Brandt}, {Hussmann}, {Langevin}, {Poulet}, {Plaut}, {Lorente}, {Jaumann}, {Altobelli}, {Fletcher}, {Retherford}, {Bunce}, {Van Hoolst}, {Hoffmann}, {Tobie}, {Mueller-Wodarg}, {Cecconi}, {Accomazzo}, {Wurz}, {Gurvits}, {Tosi}, {Gladstone}, {Erd}, {Cimo}, {Krupp}, {Hartogh}, {Witasse}, {Bruzzone}, {Grasset}, {Kaspi}, {Masters}, {Cavalie}, {Stevenson}, {Boutonnet}, {Munoz Crego}, and {Tanco}}]{JUICE-2018cospar}
{Vallat} C, {Roatsch} T, {Dougherty} MK, et~al (2018) {JUICE: A European Mission to Jupiter and its Icy Moons}. In: 42nd COSPAR Scientific Assembly, pp B5.3--31--18

\bibitem[{{van Haarlem} et~al(2013){van Haarlem}, {Wise}, {Gunst}, {Heald}, {McKean}, {Hessels}, {de Bruyn}, {Nijboer}, {Swinbank}, {Fallows}, {Brentjens}, {Nelles}, {Beck}, {Falcke}, {Fender}, {H{\"o}randel}, {Koopmans}, {Mann}, {Miley}, {R{\"o}ttgering}, {Stappers}, {Wijers}, {Zaroubi}, {van den Akker}, {Alexov}, {Anderson}, {Anderson}, {van Ardenne}, {Arts}, {Asgekar}, {Avruch}, {Batejat}, {B{\"a}hren}, {Bell}, {Bell}, {van Bemmel}, {Bennema}, {Bentum}, {Bernardi}, {Best}, {B{\^\i}rzan}, {Bonafede}, {Boonstra}, {Braun}, {Bregman}, {Breitling}, {van de Brink}, {Broderick}, {Broekema}, {Brouw}, {Br{\"u}ggen}, {Butcher}, {van Cappellen}, {Ciardi}, {Coenen}, {Conway}, {Coolen}, {Corstanje}, {Damstra}, {Davies}, {Deller}, {Dettmar}, {van Diepen}, {Dijkstra}, {Donker}, {Doorduin}, {Dromer}, {Drost}, {van Duin}, {Eisl{\"o}ffel}, {van Enst}, {Ferrari}, {Frieswijk}, {Gankema}, {Garrett}, {de Gasperin}, {Gerbers}, {de Geus}, {Grie{\ss}meier}, {Grit}, {Gruppen}, {Hamaker}, {Hassall}, {Hoeft}, {Holties}, {Horneffer},
  {van der Horst}, {van Houwelingen}, {Huijgen}, {Iacobelli}, {Intema}, {Jackson}, {Jelic}, {de Jong}, {Juette}, {Kant}, {Karastergiou}, {Koers}, {Kollen}, {Kondratiev}, {Kooistra}, {Koopman}, {Koster}, {Kuniyoshi}, {Kramer}, {Kuper}, {Lambropoulos}, {Law}, {van Leeuwen}, {Lemaitre}, {Loose}, {Maat}, {Macario}, {Markoff}, {Masters}, {McFadden}, {McKay-Bukowski}, {Meijering}, {Meulman}, {Mevius}, {Middelberg}, {Millenaar}, {Miller-Jones}, {Mohan}, {Mol}, {Morawietz}, {Morganti}, {Mulcahy}, {Mulder}, {Munk}, {Nieuwenhuis}, {van Nieuwpoort}, {Noordam}, {Norden}, {Noutsos}, {Offringa}, {Olofsson}, {Omar}, {Orr{\'u}}, {Overeem}, {Paas}, {Pandey-Pommier}, {Pandey}, {Pizzo}, {Polatidis}, {Rafferty}, {Rawlings}, {Reich}, {de Reijer}, {Reitsma}, {Renting}, {Riemers}, {Rol}, {Romein}, {Roosjen}, {Ruiter}, {Scaife}, {van der Schaaf}, {Scheers}, {Schellart}, {Schoenmakers}, {Schoonderbeek}, {Serylak}, {Shulevski}, {Sluman}, {Smirnov}, {Sobey}, {Spreeuw}, {Steinmetz}, {Sterks}, {Stiepel}, {Stuurwold}, {Tagger}, {Tang},
  {Tasse}, {Thomas}, {Thoudam}, {Toribio}, {van der Tol}, {Usov}, {van Veelen}, {van der Veen}, {ter Veen}, {Verbiest}, {Vermeulen}, {Vermaas}, {Vocks}, {Vogt}, {de Vos}, {van der Wal}, {van Weeren}, {Weggemans}, {Weltevrede}, {White}, {Wijnholds}, {Wilhelmsson}, {Wucknitz}, {Yatawatta}, {Zarka}, {Zensus}, and {van Zwieten}}]{LOFAR-2013A&A}
{van Haarlem} MP, {Wise} MW, {Gunst} AW, et~al (2013) {LOFAR: The LOw-Frequency ARray}. Astron Astrophys 556:A2. \doi{10.1051/0004-6361/201220873}, {\href{https://arxiv.org/abs/1305.3550}{{arXiv:1305.3550}}} {[astro-ph.IM]}

\bibitem[{Vertegaal et~al(2021)Vertegaal, Bentum, and Pourshaghaghi}]{Vertegaal2021}
Vertegaal CJC, Bentum MJ, Pourshaghaghi HR (2021) Using shape memory alloy for cubesat antenna design in space. In: 2021 15th European Conference on Antennas and Propagation (EuCAP), pp 1--5, \doi{10.23919/EuCAP51087.2021.9411188}

\bibitem[{{Whitney} et~al(2011){Whitney}, {Booler}, {Bowman}, {Emrich}, {Goeke}, and {Remillard}}]{MWA-2011AAS}
{Whitney} A, {Booler} T, {Bowman} J, et~al (2011) {The Murchison Widefield Array (MWA): Current Status and Plans}. In: American Astronomical Society Meeting Abstracts \#218, p 132.07

\bibitem[{{Wouthuysen}(1952)}]{Wouthuysen1952}
{Wouthuysen} SA (1952) {On the excitation mechanism of the 21-cm (radio-frequency) interstellar hydrogen emission line.} \aj 57:31--32. \doi{10.1086/106661}

\bibitem[{{Wu} and {Lee}(1979)}]{1979ApJ...230..621W}
{Wu} CS, {Lee} LC (1979) {A theory of the terrestrial kilometric radiation.} \apj 230:621--626. \doi{10.1086/157120}

\bibitem[{{Yan} et~al(2023){Yan}, {Wu}, {Gurvits}, {Wu}, {Deng}, {Zhao}, {Zhou}, {Lan}, {Fan}, {Yi}, {Yang}, {Yang}, {Wei}, {Guo}, {Qiu}, {Wu}, {Hu}, {Chen}, {Rothkaehl}, and {Morawski}}]{Yan+2023ExpA}
{Yan} J, {Wu} J, {Gurvits} LI, et~al (2023) {Ultra-low-frequency radio astronomy observations from a Seleno-centric orbit}. Experimental Astronomy 56(1):333--353. \doi{10.1007/s10686-022-09887-0}, {\href{https://arxiv.org/abs/2212.09590}{{arXiv:2212.09590}}} {[astro-ph.IM]}

\bibitem[{Zandboer(2023)}]{Zandboer2023}
Zandboer JCF (2023) Active antenna design for lunar detection of global 21cm-signals from the dark ages. Master's thesis, Eindhoven University of Technology (TU/e), Eindhoven, The Netherlands, \urlprefix\url{https://research.tue.nl/nl/studentTheses/active-antenna-design-for-lunar-detection-of-global-21cm-signals-}

\bibitem[{Zandboer et~al(2024)Zandboer, Prinsloo, Johannsen, and Bentum}]{Zandboer2024}
Zandboer JCF, Prinsloo DS, Johannsen U, et~al (2024) Active antenna design for lunar-based detection of global 21cm-signals from the dark ages. In: 2024 18th European Conference on Antennas and Propagation (EuCAP), pp 1--5, \doi{10.23919/EuCAP60739.2024.10501339}

\bibitem[{{Zarka}(2000)}]{zarka_GM119_00}
{Zarka} P (2000) {Radio Emissions from the Planets and their Moons}. Geophysical Monograph Series 119:167--178. \doi{10.1029/GM119p0167}

\bibitem[{{Zarka}(2007)}]{Zarka2007}
{Zarka} P (2007) {Plasma interactions of exoplanets with their parent star and associated radio emissions}. \planss 55(5):598--617. \doi{10.1016/j.pss.2006.05.045}

\bibitem[{Zarka and Cecconi(2022)}]{MASER_zarka2022solar}
Zarka P, Cecconi B (2022) Solar system low frequency radio spectra (version 1.0).[dataset]. PADC/MASER \doi{10.25935/YAWF-AF18}

\bibitem[{{Zarka} et~al(2012){Zarka}, {Bougeret}, {Briand}, {Cecconi}, {Falcke}, {Girard}, {Grie{\ss}meier}, {Hess}, {Klein-Wolt}, {Konovalenko}, {Lamy}, {Mimoun}, and {Aminaei}}]{2012P&SS...74..156Z}
{Zarka} P, {Bougeret} JL, {Briand} C, et~al (2012) {Planetary and exoplanetary low frequency radio observations from the Moon}. \planss 74(1):156--166. \doi{10.1016/j.pss.2012.08.004}

\bibitem[{{Zarka} et~al(2018){Zarka}, {Marques}, {Louis}, {Ryabov}, {Lamy}, {Echer}, and {Cecconi}}]{Zarka2018}
{Zarka} P, {Marques} MS, {Louis} C, et~al (2018) {Jupiter radio emission induced by Ganymede and consequences for the radio detection of exoplanets}. \aap 618:A84. \doi{10.1051/0004-6361/201833586}, {\href{https://arxiv.org/abs/1808.08055}{{arXiv:1808.08055}}} {[astro-ph.EP]}

\bibitem[{{Zarka} et~al(2019){Zarka}, {Li}, {Grie{\ss}meier}, {Lamy}, {Girard}, {Hess}, {Lazio}, and {Hallinan}}]{2019RAA....19...23Z}
{Zarka} P, {Li} D, {Grie{\ss}meier} JM, et~al (2019) {Detecting exoplanets with FAST?} Research in Astronomy and Astrophysics 19(2):023. \doi{10.1088/1674-4527/19/2/23}, {\href{https://arxiv.org/abs/1904.01239}{{arXiv:1904.01239}}} {[astro-ph.IM]}

\bibitem[{Zarka et~al(2020)Zarka, Denis, Tagger, Girard, Coffre, Viou, Taffoureau, Charrier, Bondonneau, Briand, Casoli, Cecconi, Cognard, Corbel, Dallier, Ferrari, Grießmeier, Loh, Martin, and Zakharenko}]{Zarka2020}
Zarka P, Denis L, Tagger M, et~al (2020) The low-frequency radio telescope nenufar

\bibitem[{{Zhang}(2020)}]{Zhang-2020Nature}
{Zhang} B (2020) {The physical mechanisms of fast radio bursts}. {Nature} 587(7832):45--53. \doi{10.1038/s41586-020-2828-1}, {\href{https://arxiv.org/abs/2011.03500}{{arXiv:2011.03500}}} {[astro-ph.HE]}

\bibitem[{{Zhang} et~al(2020){Zhang}, {Zucca}, {Sridhar}, {Wang}, {Bisi}, {Dabrowski}, {Krankowski}, {Mann}, {Magdalenic}, {Morosan}, and {Vocks}}]{Zhang2020}
{Zhang} P, {Zucca} P, {Sridhar} SS, et~al (2020) Interferometric imaging with lofar remote baselines of the fine structures of a solar type-iiib radio burst. Astronomy \& Astrophysics 639:A115

\bibitem[{{Zheng} et~al(2014){Zheng}, {Tegmark}, {Buza}, {Dillon}, {Gharibyan}, {Hickish}, {Kunz}, {Liu}, {Losh}, {Lutomirski}, {Morrison}, {Narayanan}, {Perko}, {Rosner}, {Sanchez}, {Schutz}, {Tribiano}, {Valdez}, {Yang}, {Adami}, {Zelko}, {Zheng}, {Armstrong}, {Bradley}, {Dexter}, {Ewall-Wice}, {Magro}, {Matejek}, {Morgan}, {Neben}, {Pan}, {Penna}, {Peterson}, {Su}, {Villasenor}, {Williams}, and {Zhu}}]{Zheng2014}
{Zheng} H, {Tegmark} M, {Buza} V, et~al (2014) {MITEoR: a scalable interferometer for precision 21 cm cosmology}. \mnras 445(2):1084--1103. \doi{10.1093/mnras/stu1773}, {\href{https://arxiv.org/abs/1405.5527}{{arXiv:1405.5527}}} {[astro-ph.IM]}

\bibitem[{{Zucca} et~al(2018){Zucca}, {Morosan}, {Rouillard}, {Fallows}, {Gallagher}, {Magdalenic}, {Klein}, {Mann}, {Vocks}, {Carley} et~al}]{Zucca2018}
{Zucca} P, {Morosan} DE, {Rouillard} AP, et~al (2018) Shock location and cme 3d reconstruction of a solar type ii radio burst with lofar. Astronomy \& Astrophysics 615:A89

\end{thebibliography}

\backmatter

\bmhead{Supplementary information}

\bmhead{Acknowledgments}
We are grateful to the CRAF frequency manager, Waleed Markour, for providing input for the lunar far-side electromagenetic environment section.

\section*{Declarations}

\begin{itemize}
\item Funding for the ALO-DEX CDF Study Astrophysical Lunar Observatory, CDF-219(A), October 2021, was provided by ESA. 

\item Author's contributions to be listed at the separate title page: M. Klein-Wolt is PI of the ESA ALO-DEX Topical Team and, together with L. Koopmans, guided the scientific scope of the DEX concept during the CDF study phase. M. Klein-Wolt, M.J. Arts, A.J.~Boonstra, C.D.~Brinkerink,
J.~Garcia Guti\'errez, J.~Grenouilleau, L.I.~Gurvits, L.V.E. Koopmans, D. Prinsloo, M.~Ruiter, J.A.~Tauber, H.K.~Vedantham, and the ESA ESTEC Concurrent Design Facility team at ESTEC conducted the CDF study that forms the basis of this paper. A.J. Boonstra, C.D. Brinkerink, S.~Ghosh, L.I.~Gurvits, L.V.E.~Koopmans, Z.~Paragi, J.A.~Tauber, H.K.~Vedantham, J.C.F. Zandboer and P. Zucca wrote the initial version of this paper. All authors commented on previous versions of the manuscript. All authors read and approved the final manuscript.
\item The authors declare no competing interests.
\end{itemize} 


\end{document}